\documentclass[oneside,11pt]{article}

\usepackage{url,mathrsfs,algorithm}
\usepackage{amsmath,wrapfig,color,bigfoot}
\usepackage{amsfonts}
\usepackage{graphicx}
\usepackage{subfigure}

\usepackage[utf8]{inputenc} 
\usepackage[T1]{fontenc}    
\usepackage{url}            
\usepackage{booktabs}       
\usepackage{amsfonts}       
\usepackage{nicefrac}       
\usepackage{microtype}      

\usepackage{latexsym}
\usepackage{amssymb}
\usepackage{mathrsfs}
\usepackage{verbatim}

\newcommand{\bel}{\begin{eqnarray}\label}
\newcommand{\eel}{\end{eqnarray}}
\newcommand{\bes}{\begin{eqnarray*}}
\newcommand{\ees}{\end{eqnarray*}}
\newcommand{\bei}{\begin{itemize}}
\newcommand{\eei}{\end{itemize}}
\newcommand{\beiftnt}{\begin{itemize}\footnotesize}

\newcommand{\convd}{\stackrel{{\rm D}}{\longrightarrow}}
\def\toD{\convd}

\def\benu{\begin{enumerate}}
\def\eenu{\end{enumerate}}

\def\real{{\mathbb{R}}}

\def\E{{\mathbb{E}}}
\def\P{{\mathbb{P}}}

\def\complex{\mathop{{\rm I}\kern-.58em\hbox{\rm C}}\nolimits}

\def\Var{\hbox{\rm Var}}

\def\eps{\epsilon}

\def\hrho{\widehat{\rho}}

\usepackage{amsmath,amsthm,amssymb}
\usepackage{latexsym}
\usepackage{mathrsfs}
\usepackage{amsfonts}

\usepackage{mathrsfs,algorithm}
\usepackage{amsmath,wrapfig,color,bigfoot}
\usepackage{graphicx}
\usepackage{subfigure}
\newtheorem{theorem}{Theorem}

\def\real{\mathop{{\rm I}\kern-.2em\hbox{\rm R}}\nolimits}

\title{\Huge Theory of the GMM Kernel}

\author{
\textbf{\Large Ping Li} \vspace{0.05in}\\
         Department of Statistics and Biostatistics\\
         Department of Computer Science\\
       Rutgers University\\
          Piscataway, NJ 08854, USA\\
       \texttt{pingli@stat.rutgers.edu}
\and
\textbf{\Large Cun-Hui Zhang} \vspace{0.05in}\\
         Department of Statistics and Biostatistics\\
       Rutgers University\\
          Piscataway, NJ 08854, USA\\
       \texttt{cunhui@stat.rutgers.edu}
}

\date{}
\begin{document}

\maketitle

\begin{abstract}

\noindent We\footnote{Presented at  ICSA Conference on Data Science  and Stanford Statistics Seminar.} develop  some theoretical results for a robust similarity measure named  ``generalized min-max'' (GMM). This similarity has direct applications in machine learning as a positive definite kernel and can be efficiently computed via probabilistic hashing. Owing to the discrete nature, the hashed values  can also be used for efficient near neighbor search. We prove the theoretical limit of GMM and the consistency result, assuming that the data follow an elliptical distribution, which is a  very general family of  distributions and includes the  multivariate $t$-distribution as a special case. The consistency result holds as long as the data have bounded first moment (an assumption which essentially holds for  datasets commonly encountered in practice). Furthermore, we establish the asymptotic normality of GMM. Compared to the ``cosine'' similarity which is routinely adopted in current practice in statistics and machine learning, the consistency of GMM requires much weaker conditions. Interestingly, when the data follow the $t$-distribution with $\nu$ degrees of freedom, GMM typically provides a better measure of similarity than  ``cosine'' roughly when $\nu<8$ (which is already very close to normal).  These theoretical results will help explain the recent success of GMM~\cite{Report:Li_GMM16,Report:Li_GMM_Nys16} in   learning tasks.

\end{abstract}

\section{Introduction}

In statistics and machine learning, it is often crucial to choose, either explicitly or implicitly, some measure of data similarity. The most commonly  adopted measure might be the ``cosine''  similarity:
\begin{align}\label{eqn_cosine}
Cos(x,y) =\frac{\sum_{i=1}^n x_iy_i}{\sqrt{\sum_{i=1}^n x_i^2 \sum_{i=1}^n y_i^2}}
\end{align}
where $x$ and $y$ are $n$-dimensional data vectors. This measure implicitly assumes that the data have bounded second moment otherwise it will not converge to a fixed limit as the sample size increases.  The data encountered in the real-world, however, are virtually always heavy-tailed~\cite{Article:Leland_power-law,Article:Crovella_power-law,Proc:Faloutsos_Sigcomm99}. \cite{Article:Newman_05} argued that the many natural datasets follow the power law with exponent (denote by $\nu$)  varying between 1 and 2. For example, $\nu = 1.2$ for the frequency of use of words, $\nu=2.04$ for the number of citations to papers, $\nu=1.4$ for the number of hits on the web sites, etc. Basically, $\nu>2$ means that data have bounded second moment. The cosine similarity (\ref{eqn_cosine}) will not converge (as $n\rightarrow\infty$) to a fixed constant if the data do not have bounded second moment.

In this study, we analyze the ``generalized min-max'' (GMM) similarity. First, we define
\begin{align}\notag
x_{i+} =\left\{\begin{array}{cc}
 x_i &\text{ if } x_i \geq 0\\
 0 &\text{ otherwise }
 \end{array}
 \right. ,
 \hspace{0.2in}
x_{i-} =\left\{\begin{array}{cc}
 -x_i &\text{ if } x_i < 0,\\
 0 &\text{ otherwise }
 \end{array}
 \right. ,
 \hspace{0.2in} x_i = x_{i+} - x_{i-}
\end{align}
Then we  compute GMM as follows:
\begin{align}\label{eqn_gmm}
GMM(x,y) =  \frac{\sum_{i=1}^n\left[\min(x_{i+},y_{i+}) + \min(x_{i-},y_{i-})\right]}{\sum_{i=1}^n\left[\max(x_{i+},y_{i+}) + \max(x_{i-},y_{i-})\right]} \overset{\triangle}{=}g_n(x,y)
\end{align}
Note that for nonnative data, GMM becomes the original ``min-max'' kernel, which has been studied in the literature~\cite{Proc:Kleinberg_FOCS99,Proc:Charikar,Report:Manasse_CWS10,Proc:Ioffe_ICDM10,Proc:Li_KDD15}. This paper focuses on analyzing theoretical properties of GMM. In particular, we are interested in the limit of $g_n(x,y)$ as $n\rightarrow\infty$ and how fast $g_n$ converges to the limit. The convergency and speed of convergence are important. For example, the cosine similarity (\ref{eqn_cosine}) is popular largely because, as long as the data have bounded second moments, $Cos(x,y)$ converges to a fixed limit which is believed to be a good characterization of the similarity between $x$ and $y$.

\vspace{0.1in}

To proceed with the analysis, we will have to make   assumptions on the data. In this paper, we adopt the ``elliptical distribution'' model~\cite{Book:Anderson03} which is very broad and includes many common distributions (such as Gaussian and Cauchy) as special cases. We first provide a simulation study.

\section{Simulations Based on  $t$-Distribution}

\begin{figure}[h!]
\begin{center}
\mbox{
\includegraphics[width=1.7in]{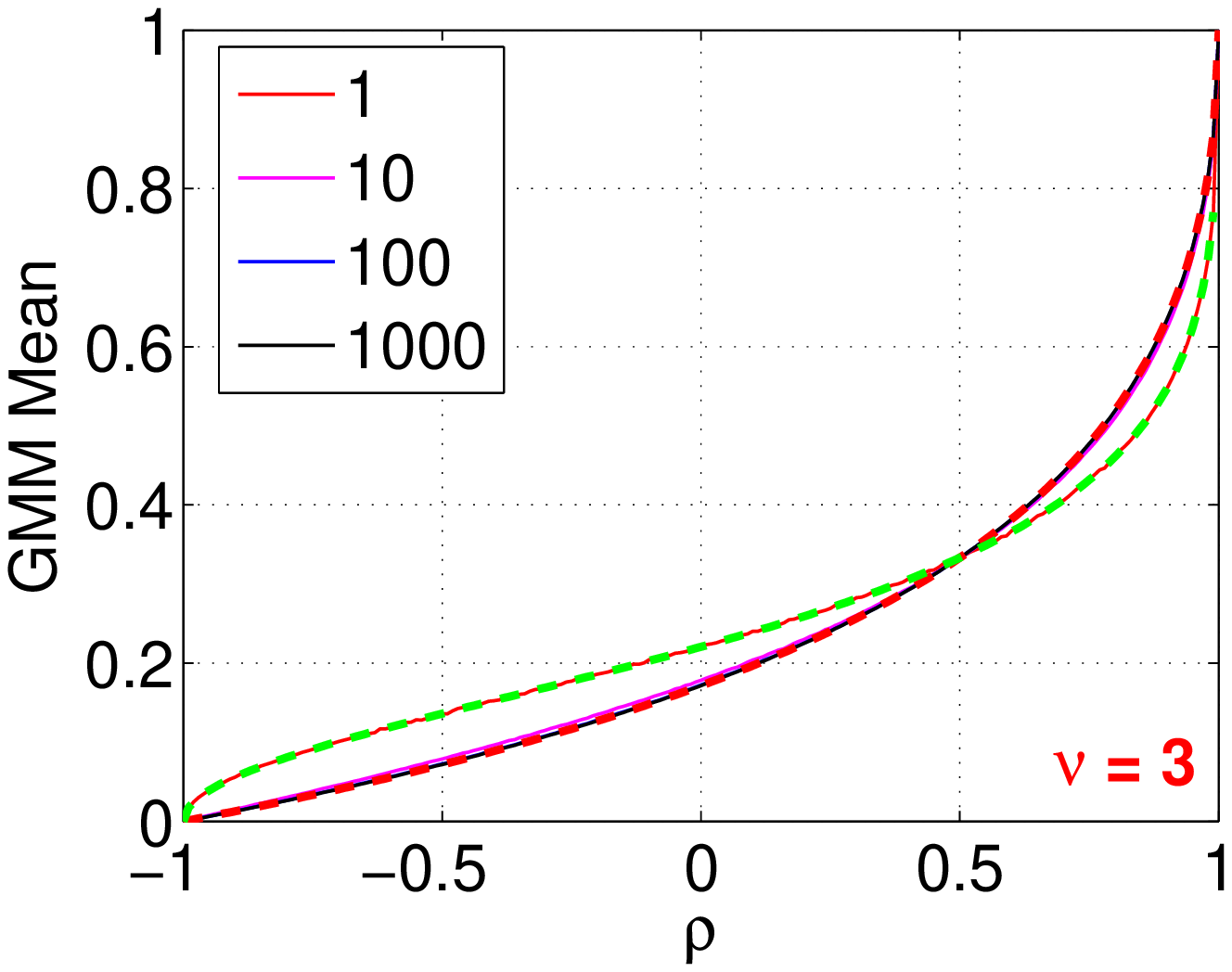}\hspace{-0.1in}
\includegraphics[width=1.7in]{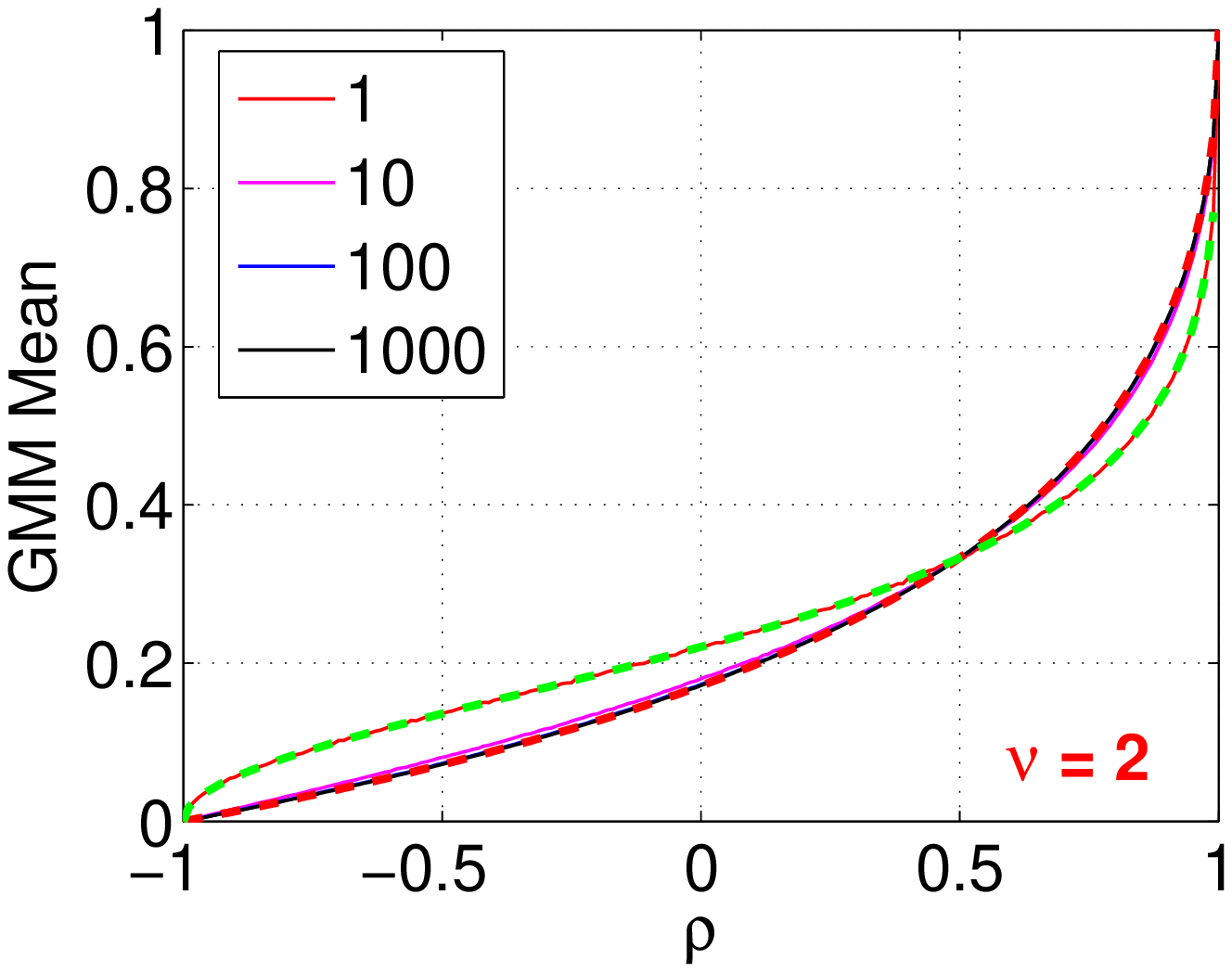}\hspace{-0.1in}
\includegraphics[width=1.7in]{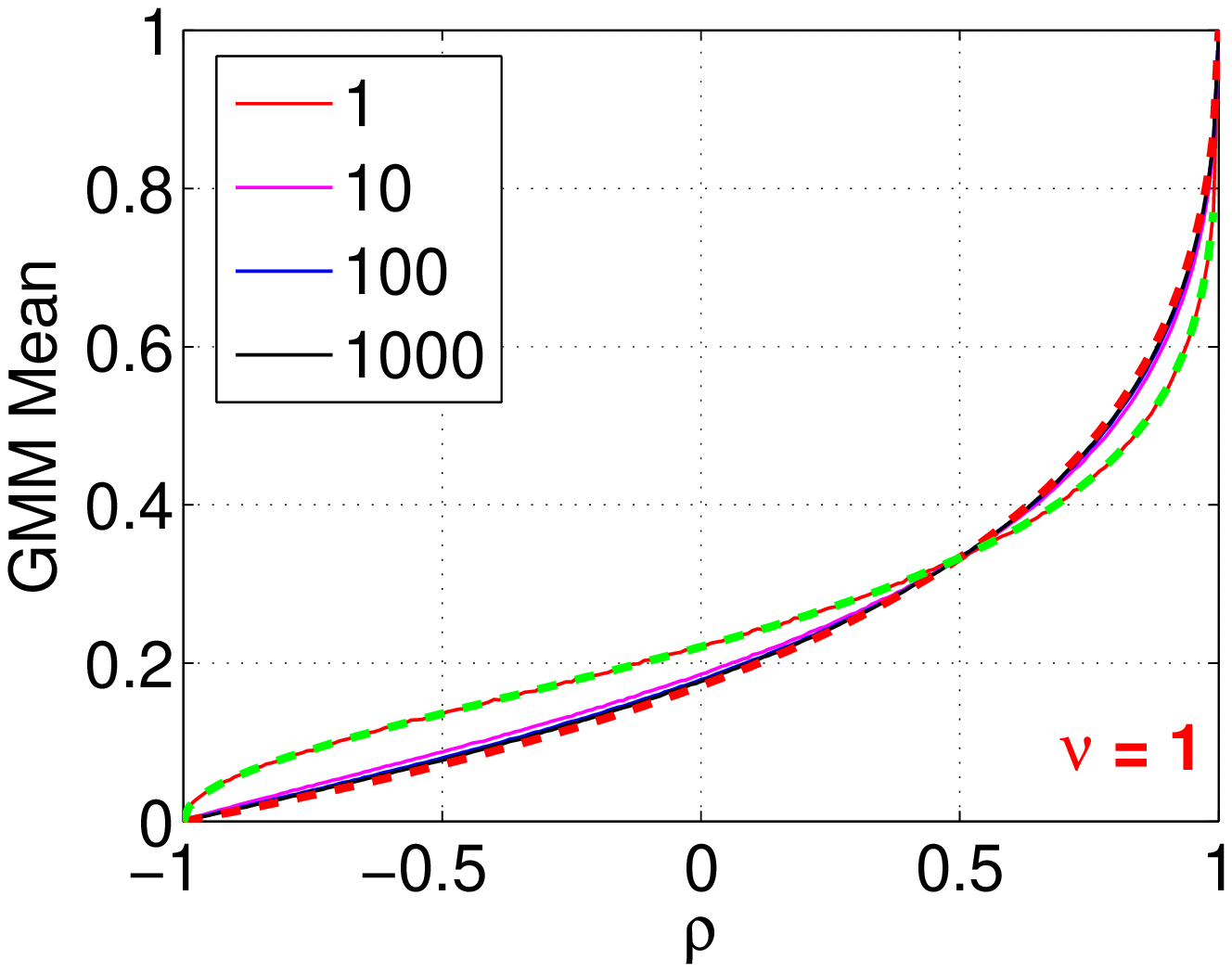}\hspace{-0.1in}
\includegraphics[width=1.7in]{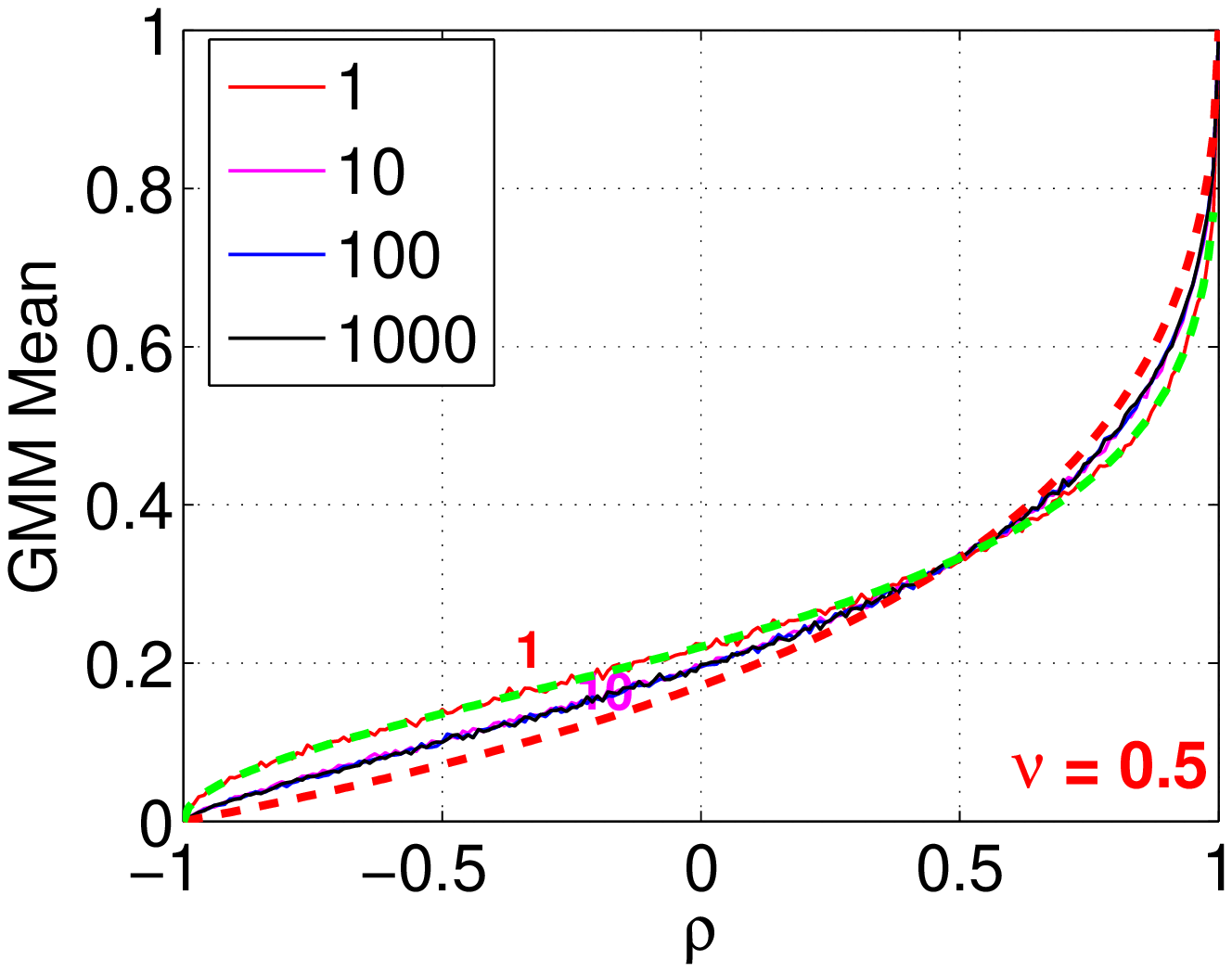}
}
\mbox{
\includegraphics[width=1.7in]{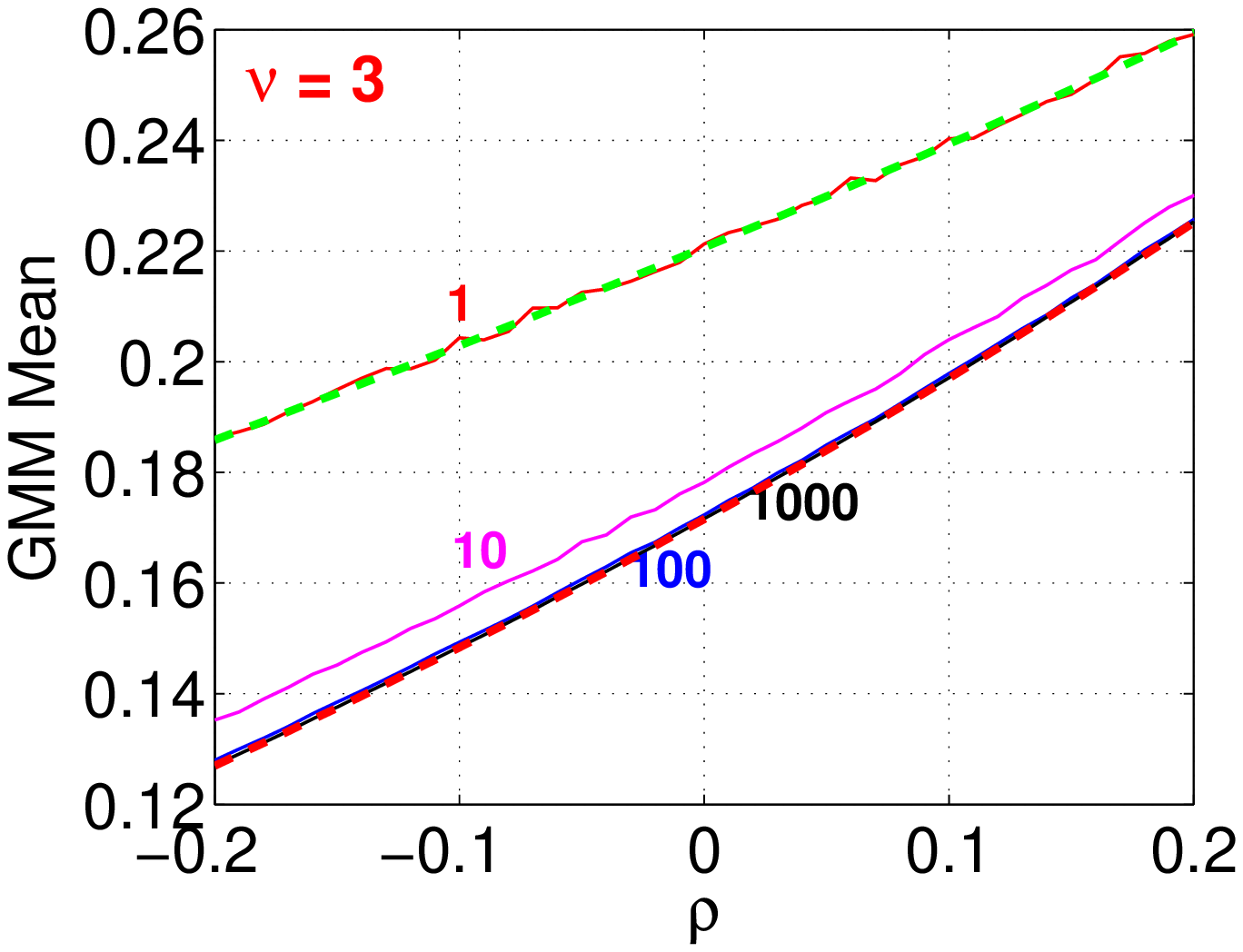}\hspace{-0.1in}
\includegraphics[width=1.7in]{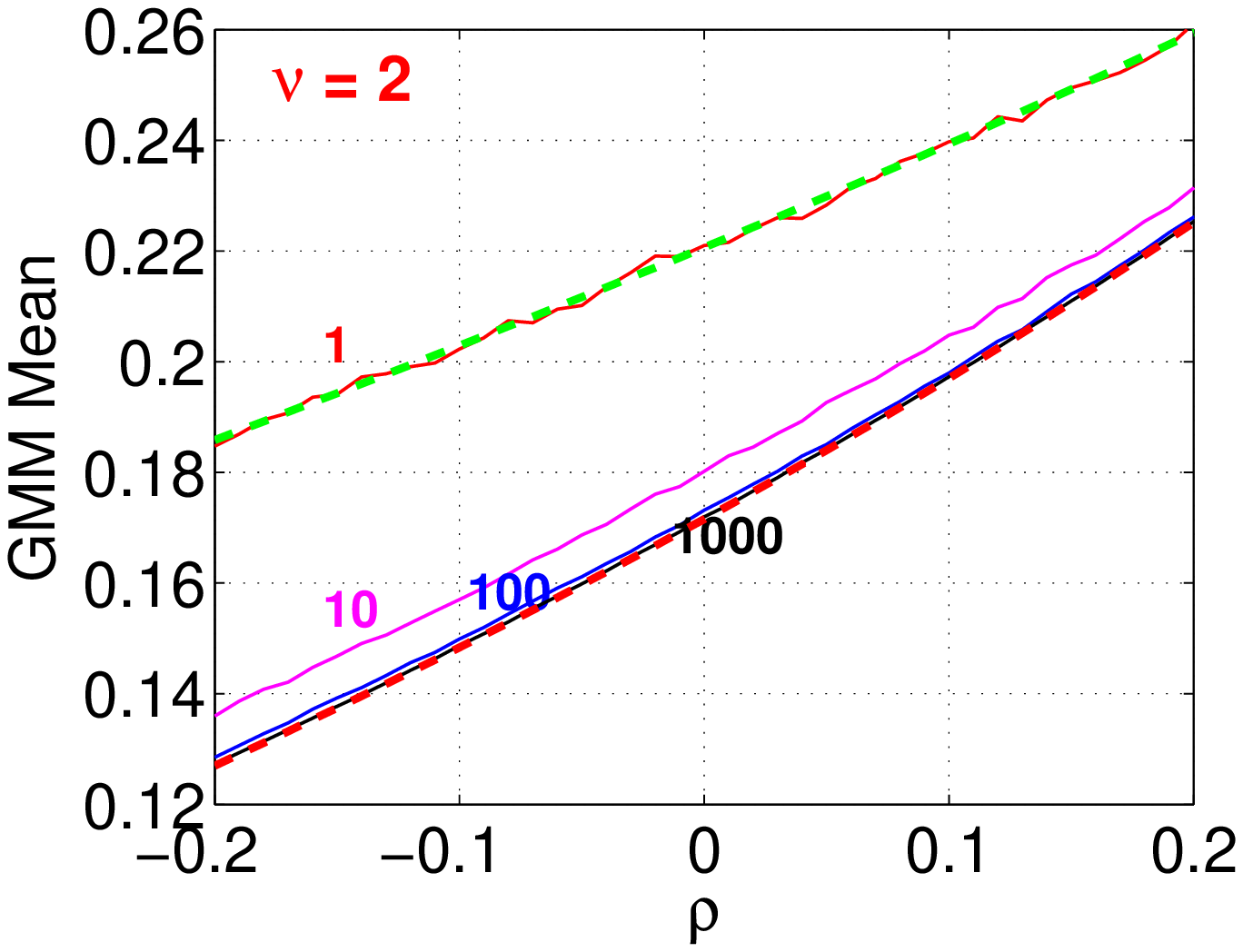}\hspace{-0.1in}
\includegraphics[width=1.7in]{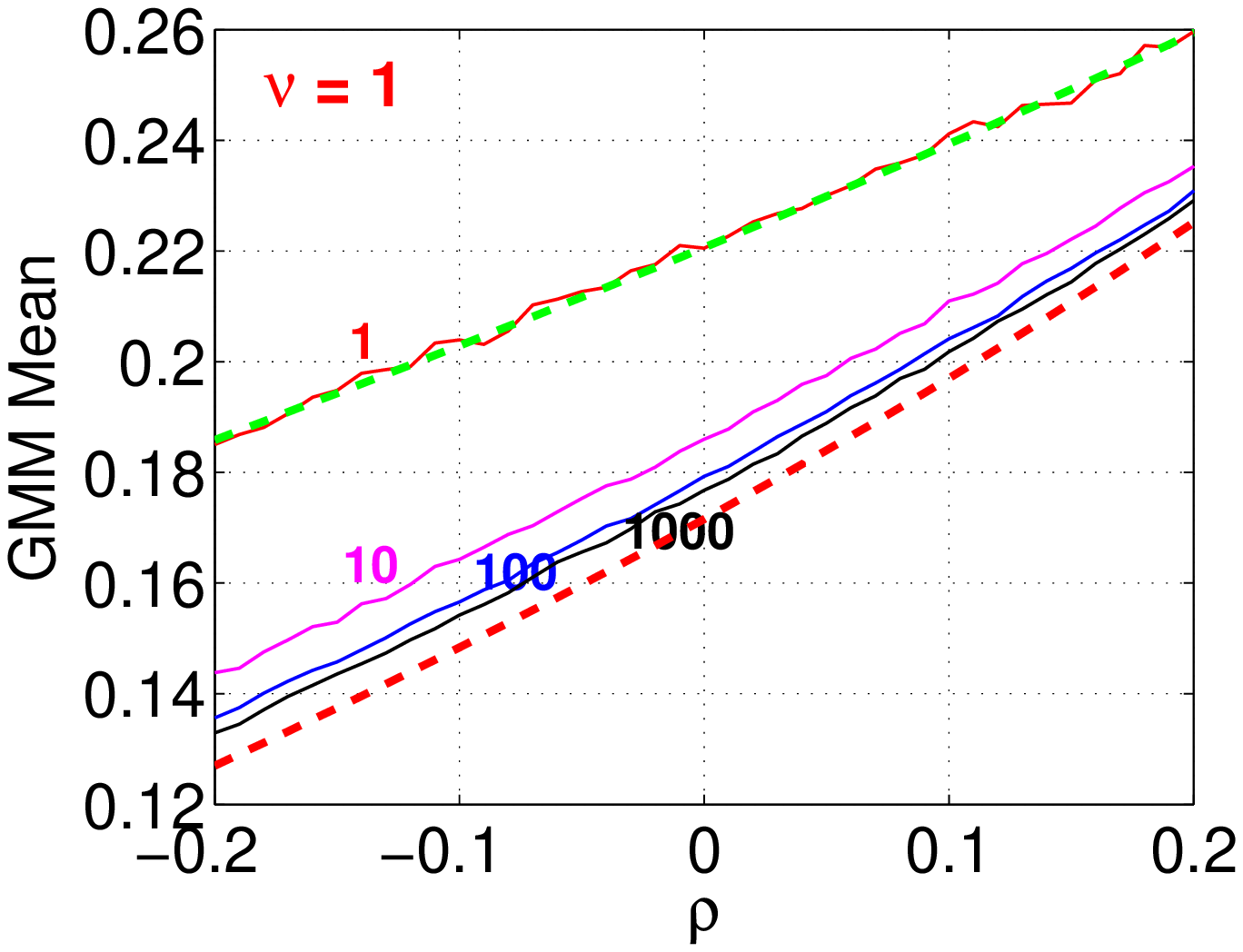}\hspace{-0.1in}
\includegraphics[width=1.7in]{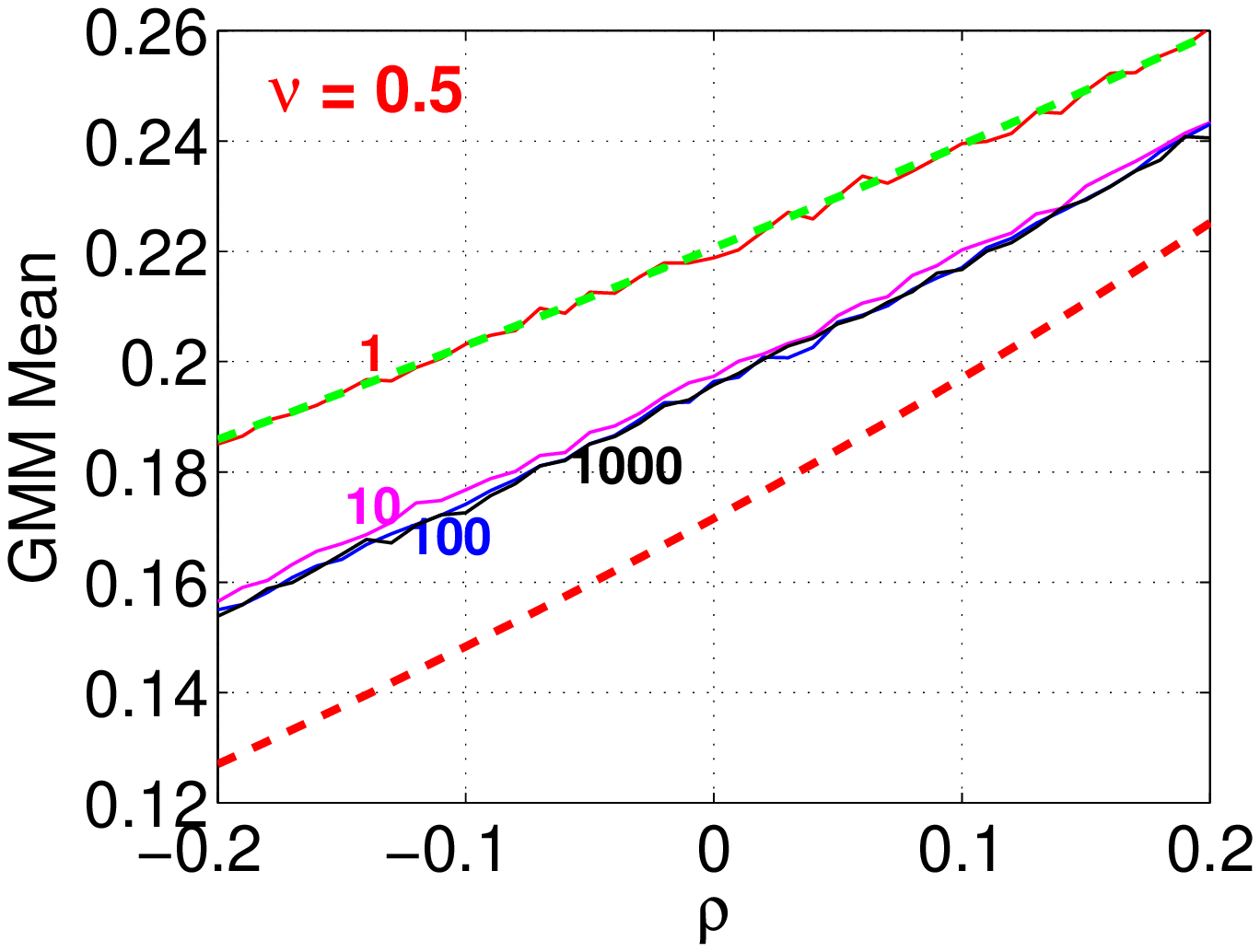}
}

\mbox{
\includegraphics[width=1.7in]{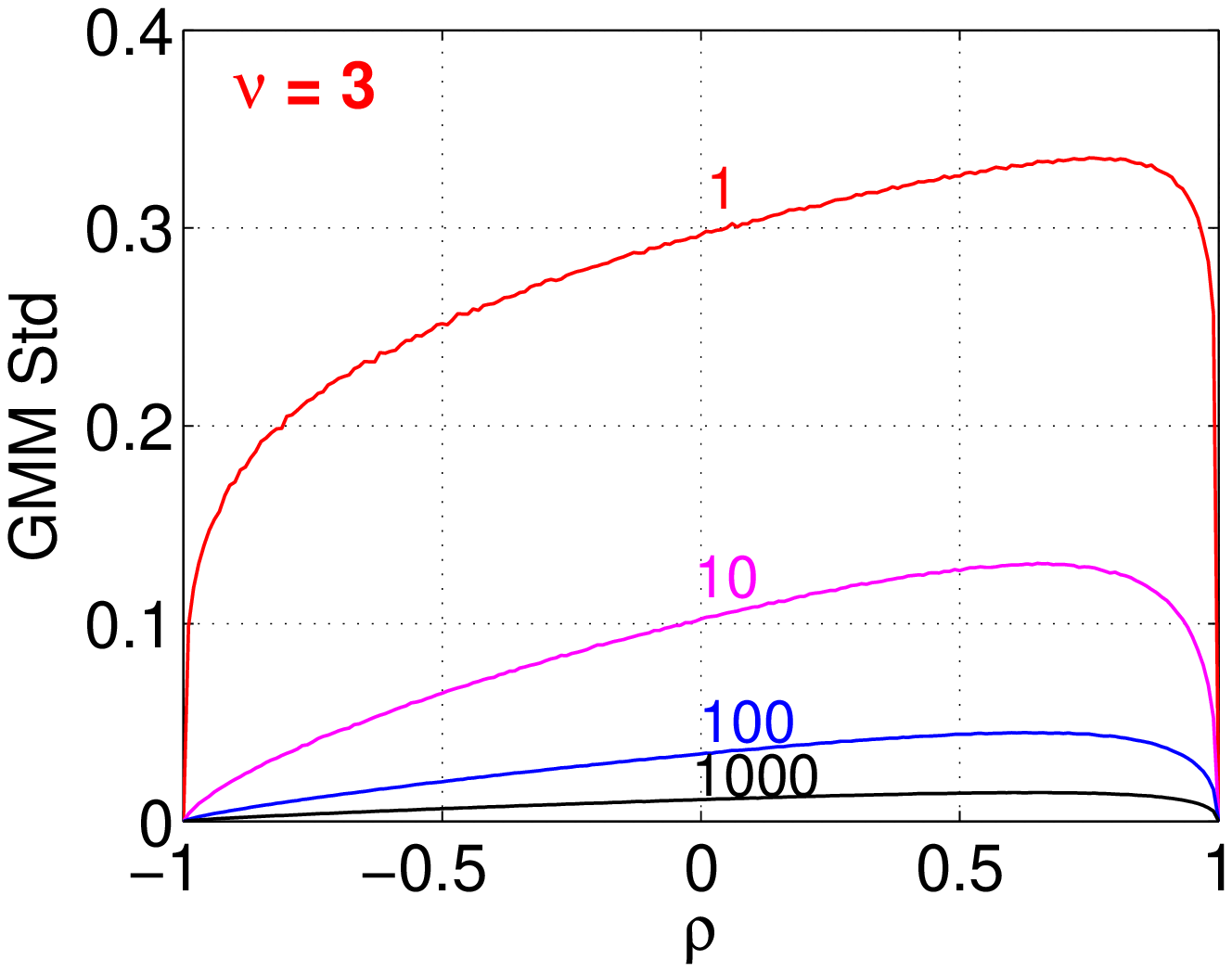}\hspace{-0.1in}
\includegraphics[width=1.7in]{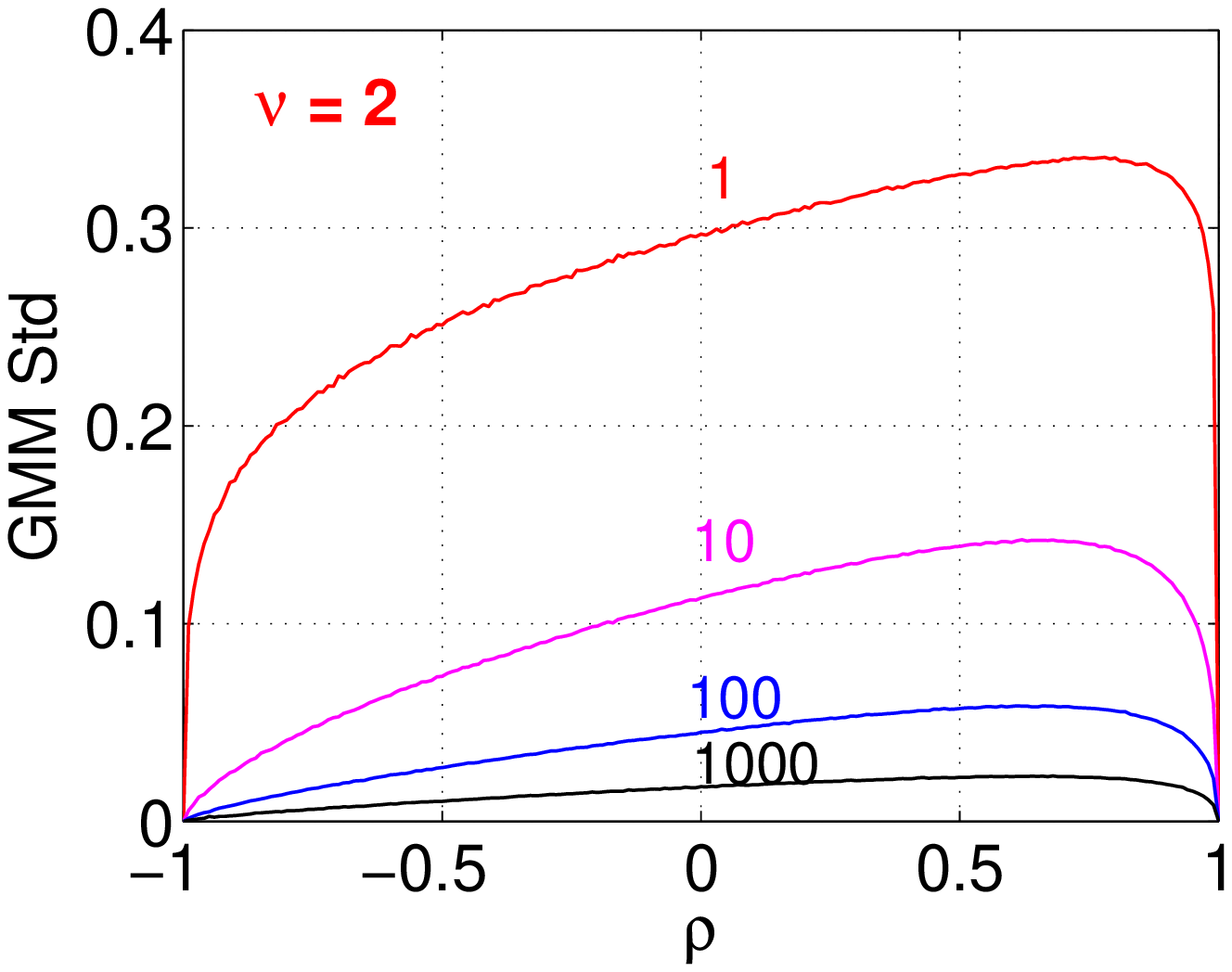}\hspace{-0.1in}
\includegraphics[width=1.7in]{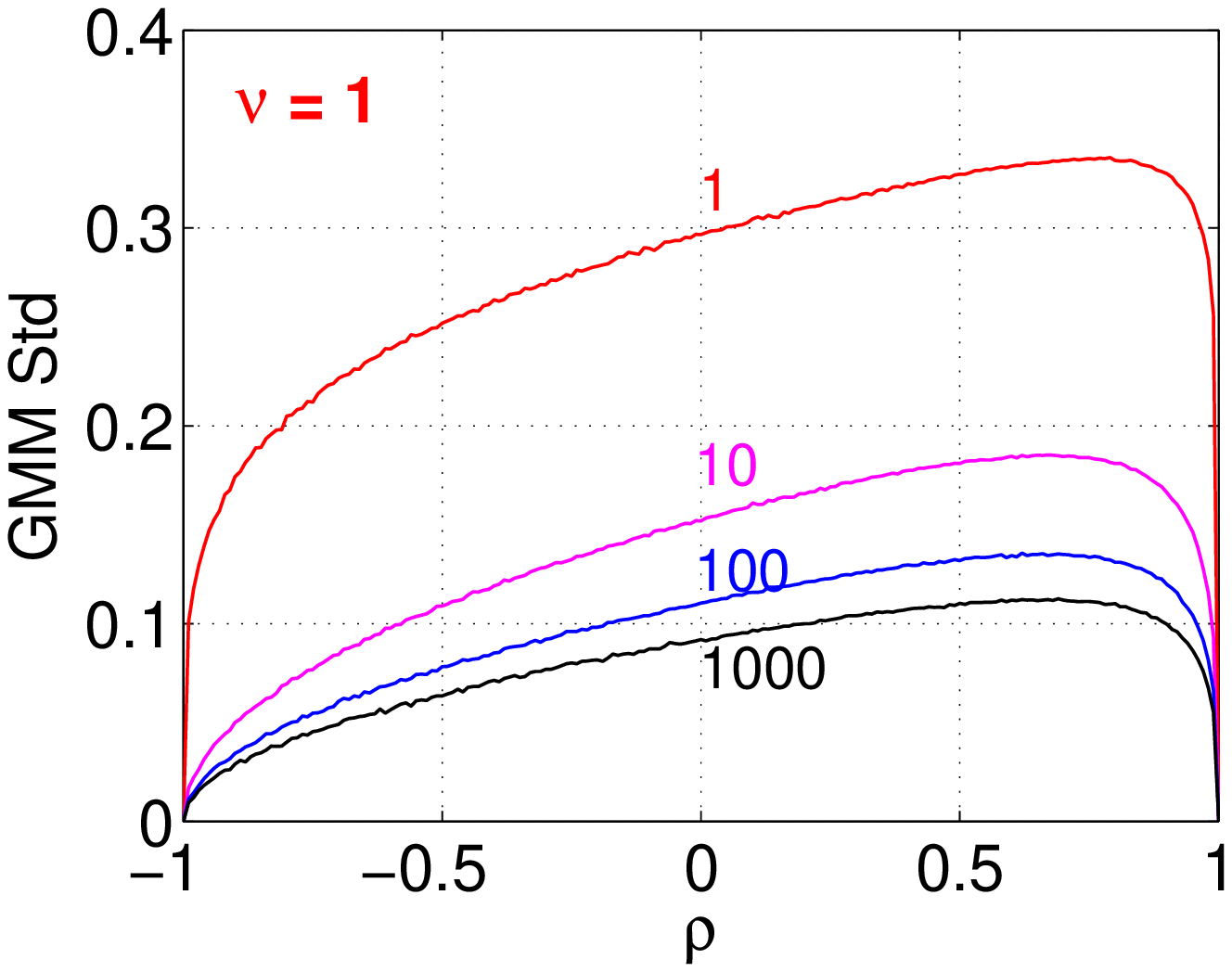}\hspace{-0.1in}
\includegraphics[width=1.7in]{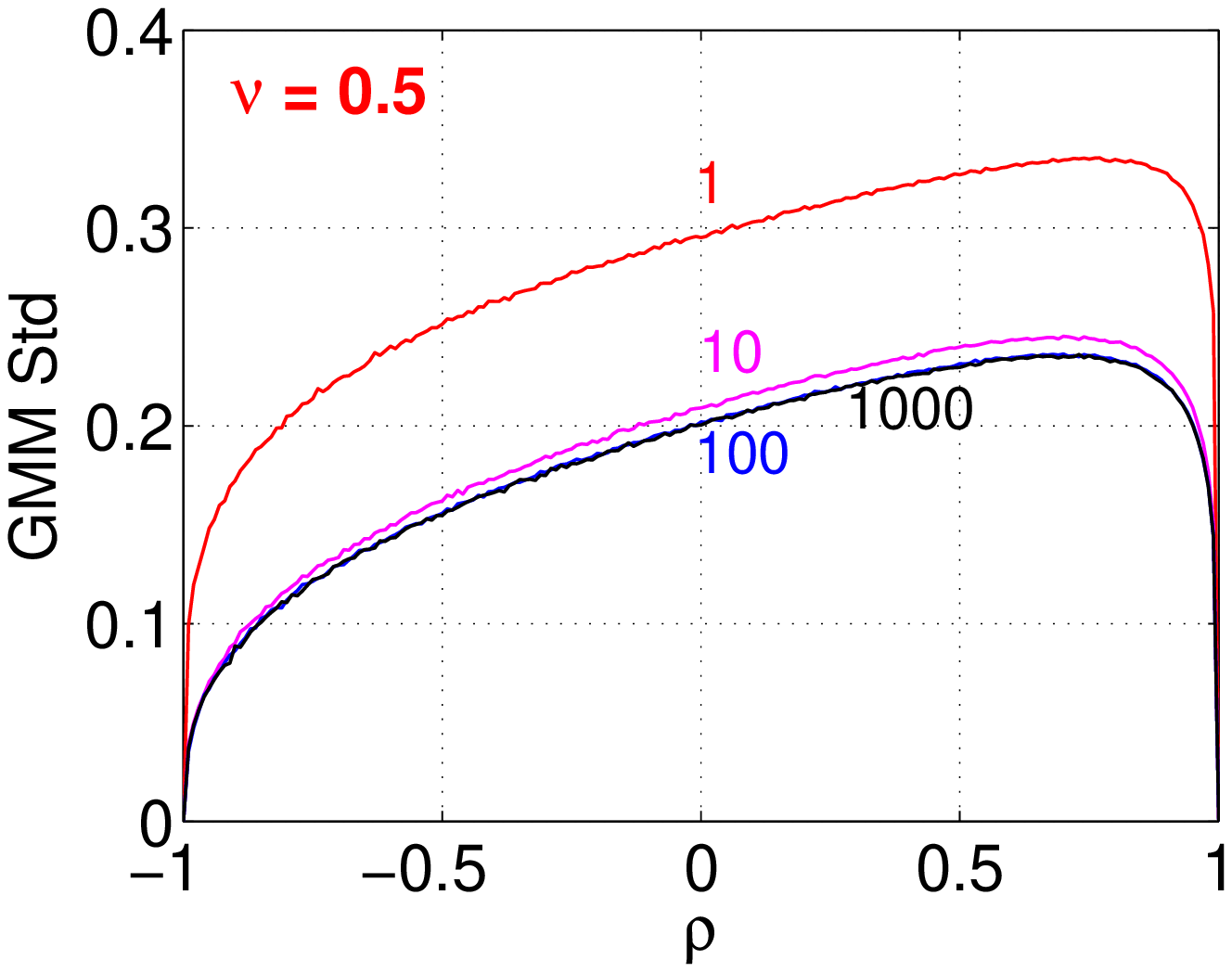}
}

\end{center}
\vspace{-0.2in}
\caption{We simulate $GMM = g_n$ defined in (\ref{eqn_gmm}) from the bivariate $t$-distribution with $\nu=0.5, 1, 2, 3$ degrees of freedom, and $n=1, 10, 100, 1000$, for 10000 repetitions. In the panels of the first two rows, we plot the mean curves together with two fixed (dashed) curves $f_1$ and $f_\infty$ defined in (\ref{eqn_mu_t}). The panels in the second row are the zoomed-in version of the panels in the first row. The bottom panels plot the empirical standard deviation of $g_n$.
}\label{fig_simulation_t}
\end{figure}

The bivariate $t$-distribution has an explicit density and is a special case of the elliptical distribution.  Denote by $t_{\Sigma, \nu}$ the bivariate $t$-distribution with covariance matrix $\Sigma$ and  $\nu$ degrees of freedom. Basically, if
two independent variables  $Z\sim N(0,\Sigma)$ and $u \sim \chi^2_\nu$, then we have  $Z\sqrt{\nu/u} \sim t_{\Sigma, \nu}$. Here, we let
$\Sigma = \left[\begin{array}{cc}
1 &\rho\\
\rho &1\end{array}
\right]$, where $-1\leq \rho \leq 1$. We consider $n$ iid samples $(x_i,y_i)\sim t_{\Sigma,\nu}$ and  compute $GMM(x,y) = g_n(x,y)$ according to (\ref{eqn_gmm}). We are interested in the mean an standard deviation of GMM for $n\in\{1,10,100,1000,10000\}$ and $\nu\in\{3,2,1,0.5\}$, as shown in Figure~\ref{fig_simulation_t}.

The panels in the first (top) row present the mean of GMM ($g_n$). The curves of GMM lie between two fixed curves, $f_1$ and $f_\infty$, which we will calculate to be the following expressions:
\begin{align}\label{eqn_mu_t}
f_1 =& \rho +\frac{1}{\pi}\left[\sqrt{1-\rho^2}
\log(2-2\rho)-2\rho \sin^{-1}\big(\sqrt{(1 - \rho)/2}\big)\right],\hspace{0.15in}
f_\infty = \frac{1-\sqrt{(1-\rho)/2}}{1+\sqrt{(1-\rho)/2}}
\end{align}
For better clarity, the panels in the second (middle) row plot the magnified portion. In each panel, the top (dashed and green if color is available) curve represent $f_1$ and the bottom (dashed and red) curve represent $f_\infty$. We can see that for $\nu=3$ and $\nu=2$, $g_n$ converges to $f_\infty$  fast. For $\nu=1$, $g_n$ also converges to $f_\infty$ but much slower. With $\nu=0.5$, $g_n$ does not converge to $f_\infty$.

The panels in the third (bottom) row plot the standard deviation (std). For $\nu\geq 1$, the std curves converge to 0, although at $\nu=1$ the convergence is much slower. When $\nu=0.5$, the standard deviation does not converge to 0.

Basically, the simulations suggest that $g_n$ converges to $f_\infty$ as long as the data have bounded first moment (i.e., $\nu>1$) and the convergence still holds for the boundary case (i.e., $\nu=1$).  We will provide thorough theoretical analysis on $g_n$ for the general elliptical distribution.

Because $\rho$ measures data similarity, the fact that  $g_n\rightarrow f_\infty$ as long as $\nu\geq1$ is important because it means we have a robust measure of $\rho$ as long as the data are ``reasonably'' distributed. As shown by~\cite{Article:Newman_05}, most natural datasets have the equivalent $\nu>1$.


\section{Analysis Based on  Elliptical Distributions}

We consider $(x_i,y_i)$, $i=1$ to $n$, are iid copy of $(X,Y)$. Our goal is to analyze the statistical behavior of  GMM, especially as $n\rightarrow \infty$,
\begin{align}\notag
GMM(x,y) =g_n(x,y) =  \frac{\sum_{i=1}^n\left[\min(x_{i+},y_{i+}) + \min(x_{i-},y_{i-})\right]}{\sum_{i=1}^n\left[\max(x_{i+},y_{i+}) + \max(x_{i-},y_{i-})\right]}
\end{align}

To proceed with the theoretical analysis, we make a very general  distributional assumption on the data. We say the vector $(X,Y)$ has an elliptical distribution if
\bel{elliptical}
(X,Y)^T = AU T = {a_1^TU T\choose a_2^TUT}
\eel
where $A = (a_1,a_2)^T$ is a deterministic $2\times 2$ matrix,
$U$ is a vector uniformly distribution in the unit circle and $T$ is a positive random variable
independent of $U$. See~\cite{Book:Anderson03} for an introduction.

In the family of elliptical distributions, there are  two important special cases:
\begin{enumerate}
\item \textbf{Gaussian distribution}: In this case, we have  $T^2  \sim \chi^2_2$ and
\begin{align}
(X,Y)^T \sim N(0,\Sigma)\sim AU\sqrt{\chi^2_2},\hspace{0.3in} \text{where } \ \Sigma = AA^T = \begin{pmatrix} 1 & \sigma \rho \cr \sigma\rho & \sigma^2 \end{pmatrix}.
\end{align}
Note that for analyzing $g_n$, it suffices to set $\Var(X)=1$, due to cancelation in GMM.

\item \textbf{$t$-distribution}: In this case, we have $T \sim \sqrt{\chi^2_2 \nu /\chi^2_\nu}$ and
\begin{align}
(X,Y)^T \sim N(0,\Sigma)\sqrt{\nu/\chi^2_\nu}.
\end{align}

\end{enumerate}

Note that in $\Sigma$ we consider $\sigma \neq 1$ to allow the situation that two vectors have different scales. For the convenience of presenting our theoretical results, we summarize the notations:
\begin{itemize}
\item $\Sigma = \begin{pmatrix} 1 & \sigma \rho \cr \sigma\rho & \sigma^2 \end{pmatrix}$, where $\rho \in [-1,1]$ and $\sigma>0$.
\item $\alpha = \sin^{-1}\big(\sqrt{1/2 - \rho/2}\big)\in [0,\pi/2]$.
\item $\tau \in [-\pi/2+2\alpha,\pi/2]$ is the solution of $\cos(\tau - 2\alpha)/\cos\tau = \sigma$, i.e., $\tau = \arctan(\sigma/\sin(2\alpha) - \cot(2\alpha))$. Note that $\tau = \alpha$ if $\sigma=1$.
\end{itemize}
In addition, we need the following  definitions of $f_1(\rho,\sigma)$ and $f_\infty(\rho,\sigma)$, for general $\sigma$ as well as $\sigma=1$:
\begin{align}\notag
f_1(\rho,\sigma) =& \frac{1}{\sigma\pi}\Big((\tau+\pi/2-2\alpha)\cos(2\alpha)+\sin(2\alpha)
\log\frac{\cos(2\alpha-\pi/2)}{\cos\tau}\Big)\\
 &+ \frac{\sigma}{\pi} \Big((\pi/2-\tau)\cos(2\alpha)+\sin(2\alpha)
\log\frac{\cos(2\alpha-\pi/2)}{\cos(2\alpha-\tau)}\Big),\\
\overset{ \sigma =1}{=}& \rho +\frac{1}{\pi}\left[\sqrt{1-\rho^2}
\log(2-2\rho)-2\rho \sin^{-1}\big(\sqrt{(1 - \rho)/2}\big)\right] \\\notag \\\label{f_infty}
f_\infty(\rho,\sigma) =&
\frac{1-\sin(2\alpha-\tau) +\sigma(1 - \sin\tau)}
{\sigma(1+\sin\tau) + 1+\sin(2\alpha-\tau)}\\
\overset{ \sigma =1}{=}&\frac{1-\sqrt{(1-\rho)/2}}{1+\sqrt{(1-\rho)/2}}
\end{align}

Theorem~\ref{thm_mean}  presents the results  for consistency.
\begin{theorem}\label{thm_mean}\textbf{(Consistency)}\hspace{0.2in}
Assume $(X,Y)$ has an elliptical distribution with $(X,Y)^T = AU T$ and $\Sigma = AA^T = \begin{pmatrix} 1 & \sigma \rho \cr \sigma\rho & \sigma^2 \end{pmatrix}$. Let $(x_i,y_i)$, $i=1$ to $n$, be iid copies of $(X,Y)$, and $GMM(x,y) = g_n(x,y)$ as defined in (\ref{eqn_gmm}). Then the following statements hold:
 \begin{itemize}
 \item $g_1 = f_1(\rho,\sigma)$
 \item If\ \  $\E T  <\infty$, then $g_n \rightarrow f_\infty(\rho,\sigma)$, almost surely.
 \item If  we have
\bel{cond}
\lim_{t\to\infty} \frac{t\,\P(T>t)}{\E\min(T,t)} = 0,
\eel
 then $g_n \rightarrow f_\infty(\rho,\sigma)$, in probability.
\item If $(X,Y)$ has a  $t$-distribution with $\nu$ degrees of freedom, then $g_n\rightarrow f_\infty(\rho,\sigma)$ almost surely if $\nu>1$ and $g_n\rightarrow f_\infty(\rho,\sigma)$ in probability if $\nu=1$.
 \end{itemize}
\end{theorem}

\newpage\clearpage

Theorem~\ref{thm_normality}  presents the results  for asymptotic normality.

\begin{theorem}\label{thm_normality}\textbf{(Asymptotic Normality)} \hspace{0.2in} \\ With the same notation and definitions as in Theorem~\ref{thm_mean}, the following statements hold:
\begin{itemize}
\item If $\E T^2< \infty$, then
\begin{align}
n^{1/2}\left(g_n(x,y) - f_\infty(\rho,\sigma)\right) \overset{D}{\longrightarrow} N\left(0, \frac{V}{H^4} \frac{\E T^2}{\E^2 T}\right)
\end{align}
where
\begin{align}
&V\\\notag
=&\frac{1}{4\pi^3}\left\{2\tau+\pi-4\alpha+\sin(2\tau-4\alpha)+\sigma^2\left(\pi-2\tau-\sin(2\tau)\right)\right\}\left\{\sigma(1+\sin\tau) + 1+\sin(2\alpha-\tau)\right\}^2\\\notag
+&\frac{1}{4\pi^3}\left\{\sigma^2 \left(2{\tau}+\sin(2\tau) + {\pi}\right)
+ \left({\pi}+4\alpha-2{\tau}- \sin(2\tau - 4\alpha)\right) +4{\sigma} \left(\sin2\alpha-2\alpha\cos2\alpha\right)\right\}\\\notag
&\hspace{0.2in}\times\left\{1-\sin(2\alpha-\tau) +\sigma(1 - \sin\tau)\right\}^2\\\notag
-&\frac{\sigma}{\pi^3} \left((\pi-2\alpha)\cos2\alpha + \sin2\alpha\right)\left\{1-\sin(2\alpha-\tau) +\sigma(1 - \sin\tau)\right\}\left\{\sigma(1+\sin\tau) + 1+\sin(2\alpha-\tau)\right\}\\
\overset{\sigma=1}{=}& \frac{4}{\pi^3}\sin^2\alpha\left(3\pi-8\cos\alpha+2\sin2\alpha+\pi\cos2\alpha
-8\alpha\sin\alpha-4\alpha\cos2\alpha\right)
\end{align}
and
\begin{align}
H =& \frac{1}{\pi}\left\{\sigma(1+\sin\tau) + 1+\sin(2\alpha-\tau)\right\}  \overset{\sigma=1}{=} \frac{2}{\pi}(1+\sin\alpha)
\end{align}
\item If $(X,Y)$ has a  $t$-distribution with $\nu$ degrees of freedom and $\nu>2$, then
\bel{normality-1}
n^{1/2}\left(g_n(x,y) - f_\infty(\rho,\sigma)\right)
\toD N\left(0, \frac{V}{H^4}\frac{\E T^2}{\E^2T}\right).
\eel
where $\E T^2 = \frac{2\nu}{\nu-2}$ and $\E T
= \frac{\sqrt{\nu}\,\Gamma(\nu/2-1/2)\Gamma(1/2)}{2\,\Gamma(\nu/2)}$
\item
If $(X,Y)$ has a  $t$-distribution with $\nu=2$ degrees of freedom, then
\begin{align}
\left(\frac{n}{\log n}\right)^{1/2}\left(g_n(x,y) - f_\infty(\rho,\sigma)\right)
\toD N\left(0, \frac{V}{H^4} \frac{4}{\pi^2}\right).
\end{align}

\end{itemize}

\end{theorem}

\vspace{0.4in}

\begin{figure}
\begin{center}
\mbox{
\includegraphics[width=2.7in]{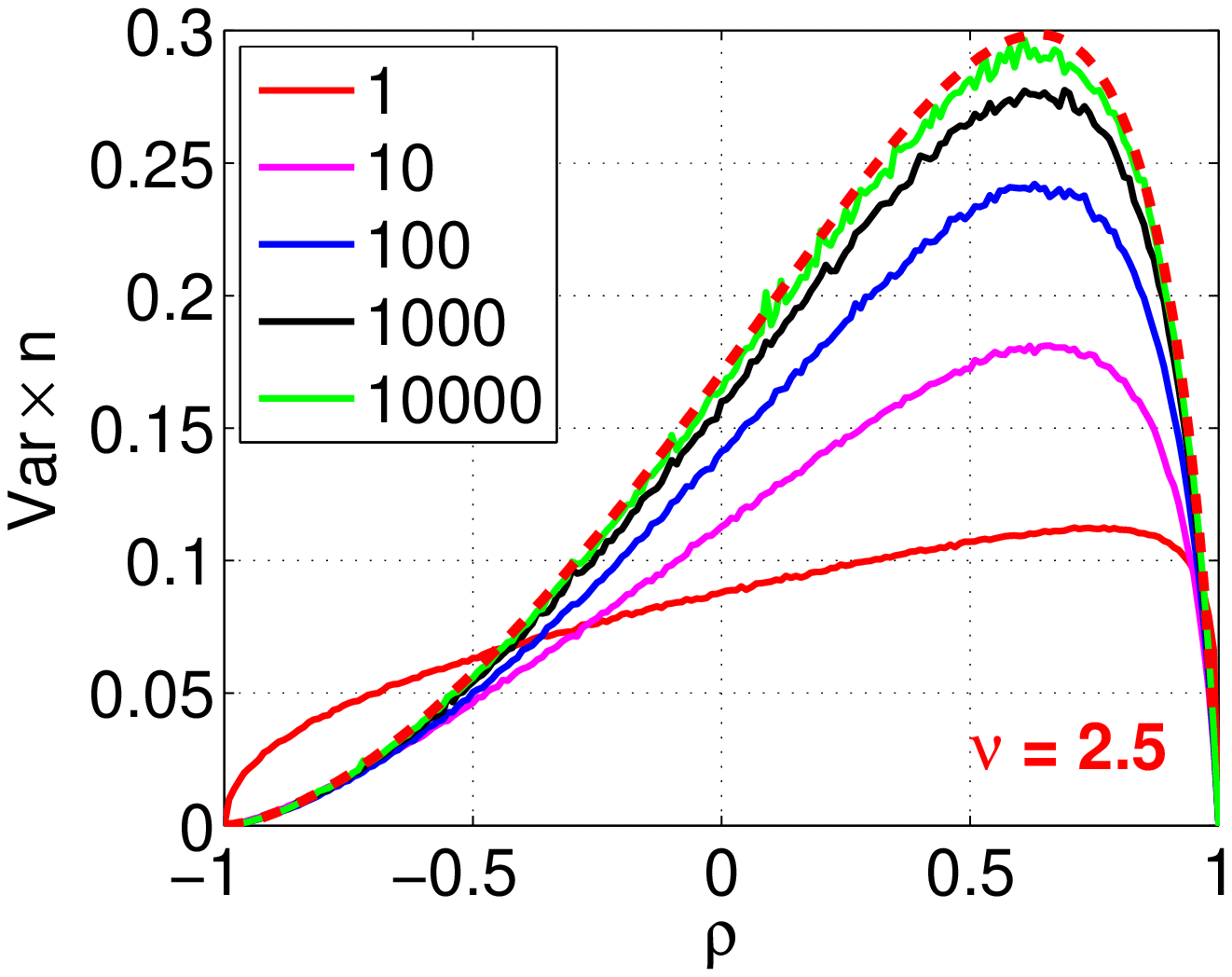}
\includegraphics[width=2.7in]{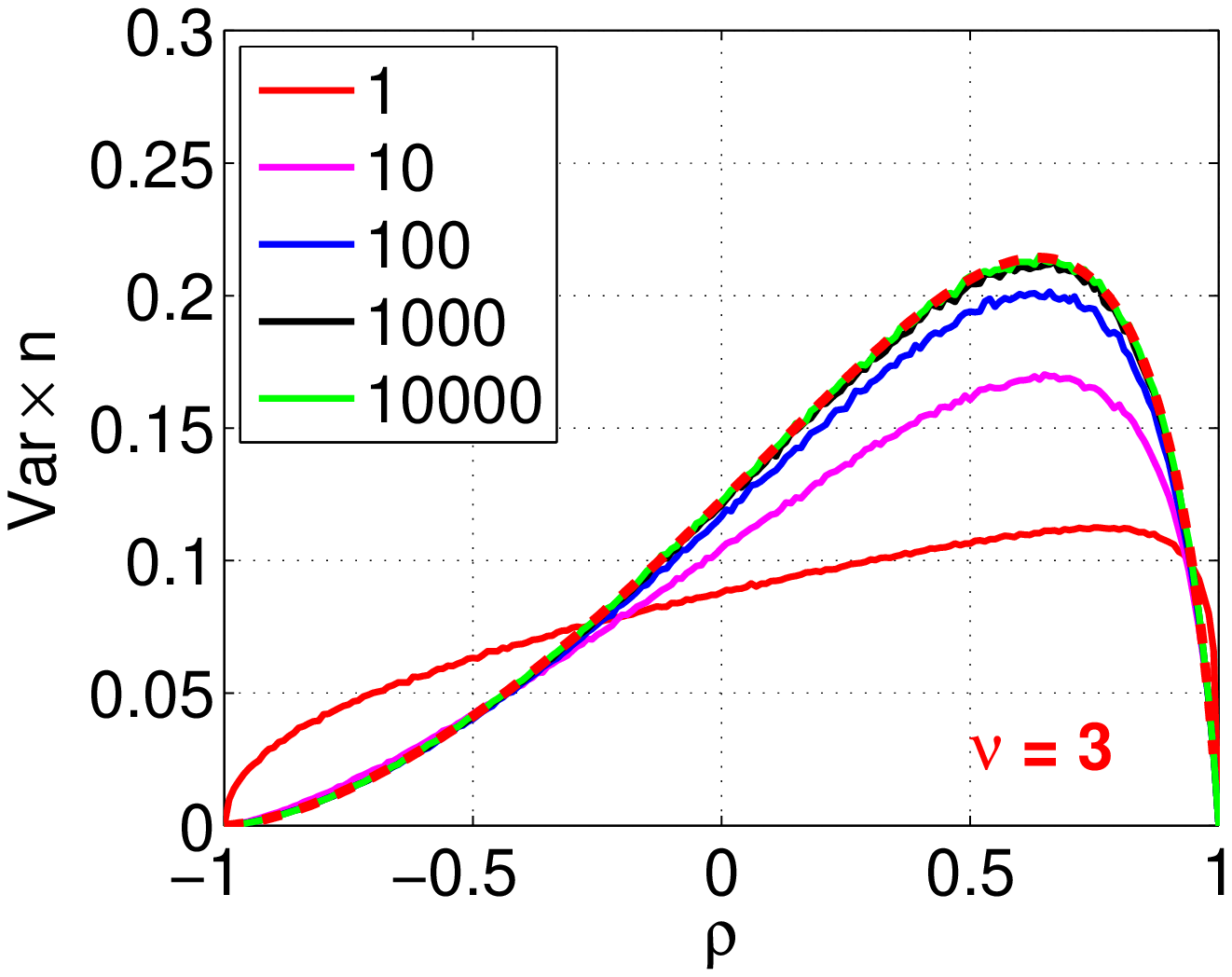}
}

\mbox{
\includegraphics[width=2.7in]{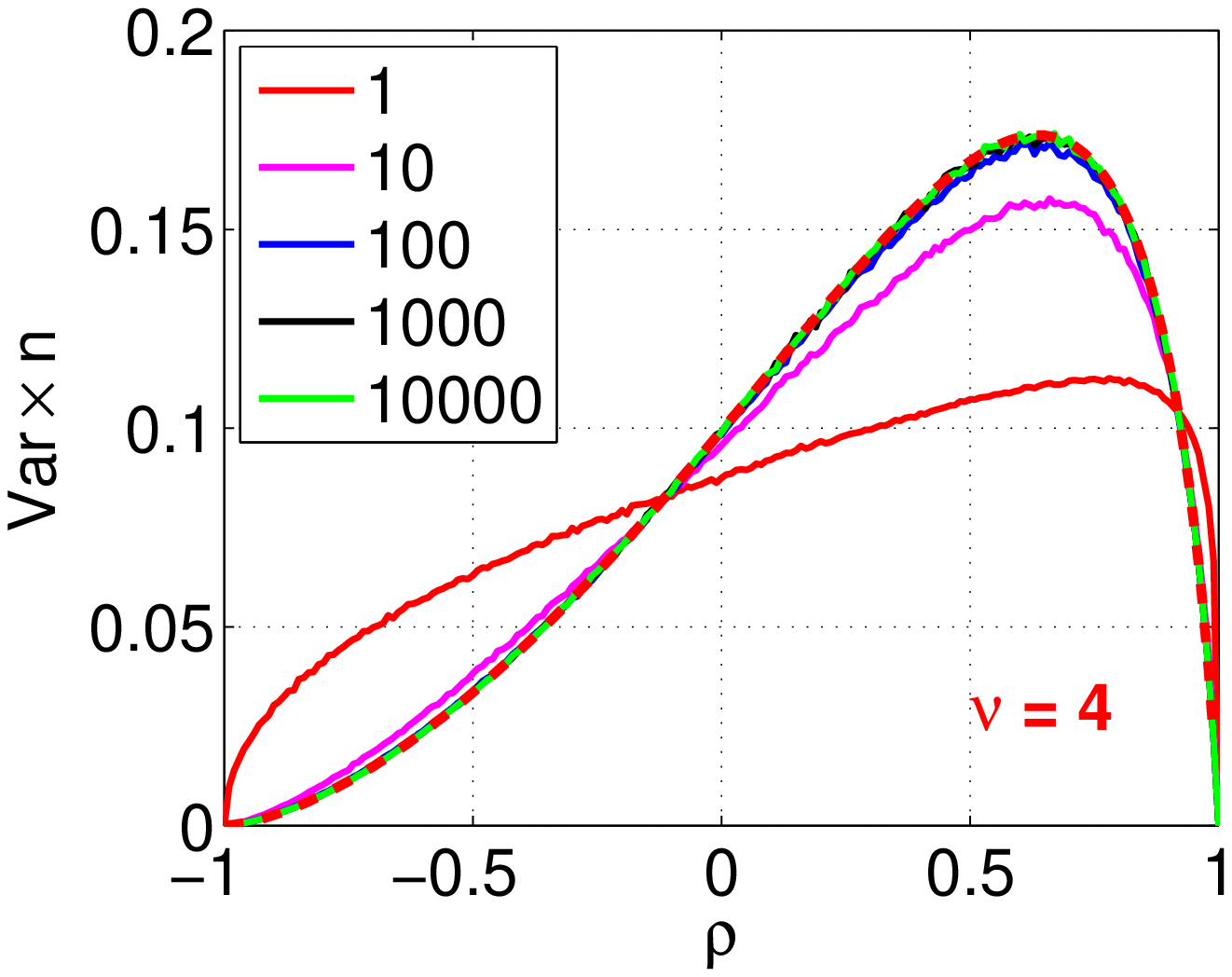}
\includegraphics[width=2.7in]{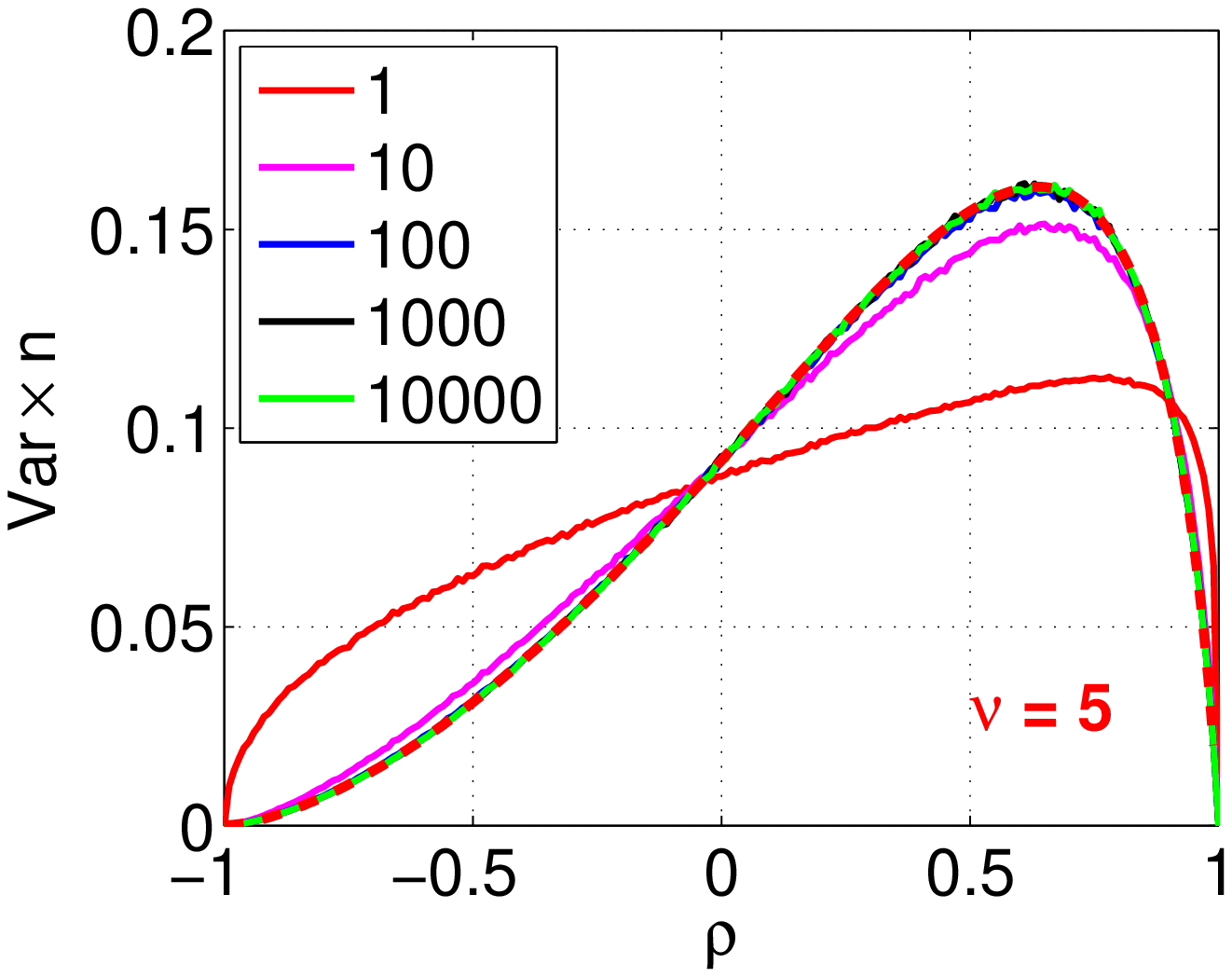}
}
\end{center}
\vspace{-0.2in}
\caption{Simulations for verifying the asymptotic variance formula (\ref{eqn_VarGMM}) based on $t$-distribution with $\nu$ degrees of freedom where $\nu \in \{2.5, 3, 4, 5\}$ and $\rho\in[-1,1]$ spaced at 0.01. For each case, we repeat the simulation 10000 times. We report the empirical $Var(g_n)\times n$ with the theoretical asymptotic value $\frac{V}{H^4} \frac{\E T^2}{\E^2 T}$ plotted as dashed curves. For $n$ large enough, the asymptotic variance formula (\ref{eqn_VarGMM}) becomes accurate. For small $n$ values, however, the formula can be quite conservative.    }\label{fig_VarGMM}
\end{figure}

Figure~\ref{fig_VarGMM} presents a simulation study to verify the asymptotic normality, in particular, the asymptotic variance formula
\begin{align}\label{eqn_VarGMM}
Var\left(g_n\right) =  \frac{1}{n} \frac{V}{H^4} \frac{\E T^2}{\E^2 T} + O\left(\frac{1}{n^2}\right)
\end{align}
by considering that the data follow a $t$-distribution with $\nu$ degrees of freedom and $\nu = 2.5,\ 3,\ 4,\ 5$.  The simulation results confirm the asymptotic variance formula at large enough sample size $n$. When $n$ is not too large, the asymptotic variance formula (\ref{eqn_VarGMM}) can be conservative.
\newpage

\section{Estimation of $\rho$}

The fact that $g_n(x,y) \rightarrow f_\infty(\rho,\sigma)$ also provides a robust and convenient  way to estimate the similarity between data vectors. Here, for convenience we consider $\sigma =1$. For this case, we have $f_\infty = \frac{1-\sqrt{(1-\rho)/2}}{1+\sqrt{(1-\rho)/2}}$. This suggests an estimator of $\rho$:
\begin{align}
\hat{\rho}_g = 1- 2\left(\frac{1-g_n}{1+g_n}\right)^2
\end{align}
As $n\rightarrow\infty$, $g_n \rightarrow f_\infty$ and $\hat{\rho}_g\rightarrow \rho$. In other words, the estimator $\hat{\rho}_g$ is asymptotically unbiased.  The asymptotic variance of $\hat{\rho}_g$ can be computed using ``delta method'':
\begin{align}\notag
Var\left(\hat{\rho}_g\right) =& \left[8\frac{1-f_\infty}{(1+f_\infty)^3}\right]^2Var\left(g_n\right) + O\left(\frac{1}{n^2}\right)\\
 =& \frac{1}{n}2\left(1-\rho\right)\left(1+\sqrt{(1-\rho)/2}\right)^4 \frac{V}{H^4} \frac{\E T^2}{\E^2 T} + O\left(\frac{1}{n^2}\right)
\end{align}
See (\ref{eqn_VarGMM}) and Theorem~\ref{thm_normality} for more details. Again, we emphasize that this estimator $\hat{\rho}_g$ is meaningful as long as $\E T<\infty$  and $Var\left(\hat{\rho}_g\right) <\infty$ as long as $\E T^2<\infty$. \\

It is interesting to compare this estimator with the commonly used estimator based on the ``cosine'' similarity:
\begin{align}\notag
Cos(x,y) =\frac{\sum_{i=1}^n x_iy_i}{\sqrt{\sum_{i=1}^n x_i^2 \sum_{i=1}^n y_i^2}} \overset{\triangle} {=} c_n(x,y)
\end{align}
When the data are bivariate normal, it is a known result~\cite{Book:Anderson03} that $c_n(x,y)$, when appropriately normalized, converges in distribution to a normal
\begin{align}
n^{1/2}\left(c_n - \rho\right) \overset{D}{\longrightarrow} N\left(0, (1-\rho^2)^2\right)
\end{align}
This asymptotic normality (with difference in the variance term) holds as long as the data have bounded fourth moment. Here, we present the generalization as a theorem.
\begin{theorem}\label{thm_cosine}
If $\E T^4<\infty$, then
\begin{align}
n^{1/2}\left(c_n - \rho\right) \overset{D}{\longrightarrow} N\left(0, \frac{\E T^4}{2\E^2T^2} (1-\rho^2)^2\right)
\end{align}
\end{theorem}

\vspace{0.4in}

Based on Theorem~\ref{thm_cosine},  a natural estimator of $\rho$ and its asymptotic variance would be
\begin{align}
\hat{\rho}_c = c_n,\hspace{0.2in} Var\left(\hat{\rho}_c\right) = \frac{1}{n}\frac{\E T^4}{2\E^2T^2}\left(1-\rho^2\right)^2 + O\left(\frac{1}{n^2}\right)
\end{align}
When the data follow a $t$-distribution with $\nu$ degrees of freedom, we have
\begin{align}\notag
\E T^2 = \frac{2\nu}{\nu-2},\hspace{0.3in} \E T^4 = \frac{4\nu^3}{(\nu-2)^2(\nu-4)} + \frac{4\nu^2}{(\nu-2)^2}
\end{align}

\vspace{0.4in}

Figure~\ref{fig_Mse1} and Figure~\ref{fig_Mse2} provide a simulation study for comparing two estimators $\hat{\rho}_g$ and $\hat{\rho}_c$. We assume $t$-distribution with $\nu$ degrees of freedom, where $\nu \in\{2.5, 3, 4, 4.5, 5, 6, 8, 10\}$ as well $\nu=\infty$ (i.e., normal distribution). In each panel, we plot the empirical mean square errors (MSEs): $MSE(\hat{\rho}_g)$ and $MSE(\hat{\rho}_c)$ (computed from 10000 repetitions), along with the (asymptotic) theoretical variance of $\hat{\rho}_g$: $\frac{1}{n}2\left(1-\rho\right)\left(1+\sqrt{(1-\rho)/2}\right)^4 \frac{V}{H^4} \frac{\E T^2}{\E^2 T}$.  For clarity, we did not plot the theoretical variance of $\hat{\rho}_c$, which is fairly simple and more straightforward to be verified. \\

The results in Figure~\ref{fig_Mse1} and Figure~\ref{fig_Mse2} confirm that $\hat{\rho}_g$, the estimator based on GMM, is substantially more accurate than $\hat{\rho}_c$, the commonly used estimator based on cosine. Roughly speaking, when $\nu<8$, $\hat{\rho}_g$ is more preferable. Even when the data are perfectly Gaussian (the bottom row in Figure~\ref{fig_Mse2}), the use of $\hat{\rho}_g$ does not result in much loss of accuracy compared to $\hat{\rho}_c$.

\begin{figure}
\begin{center}
\mbox{
\includegraphics[width=2.2in]{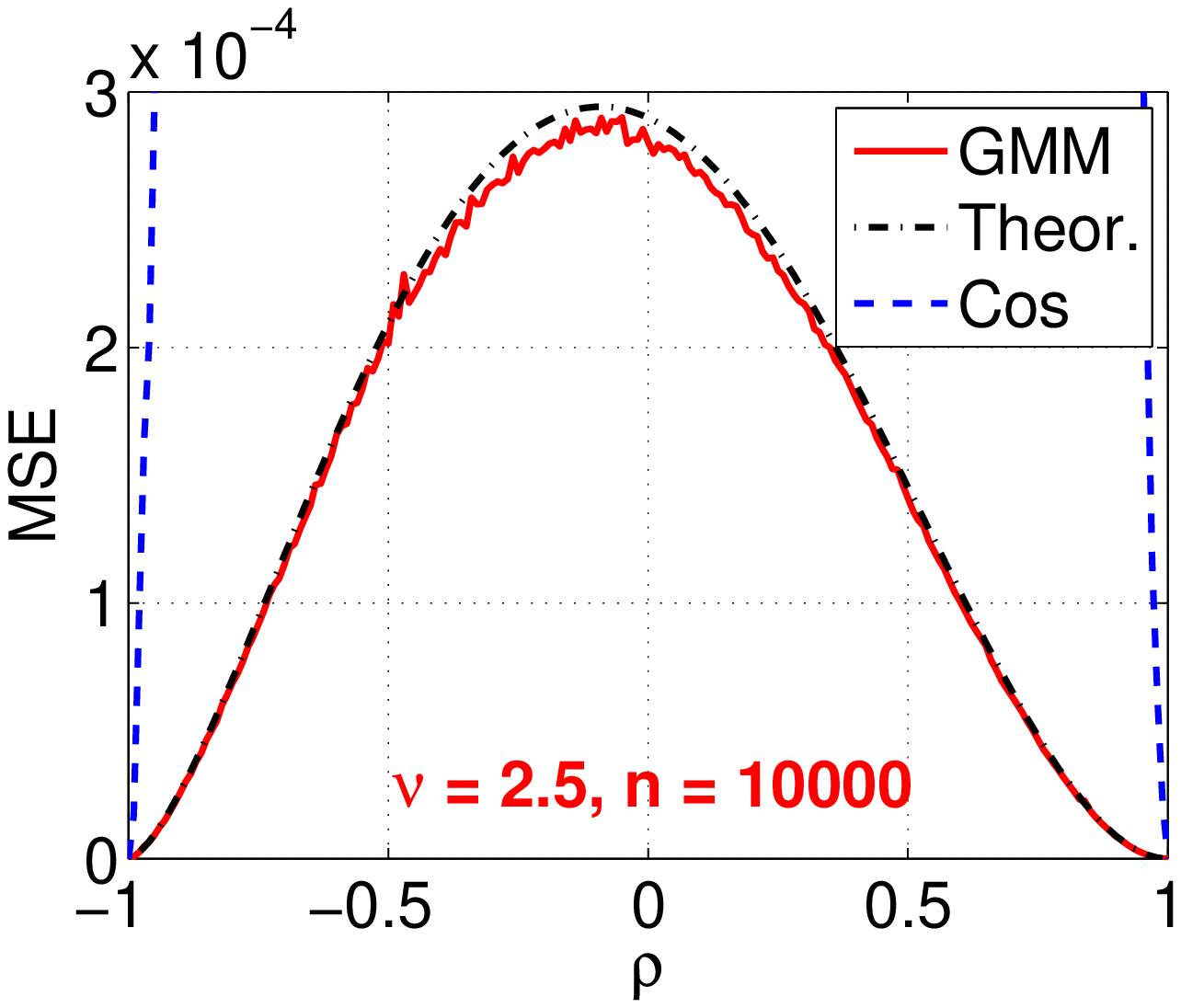}
\includegraphics[width=2.2in]{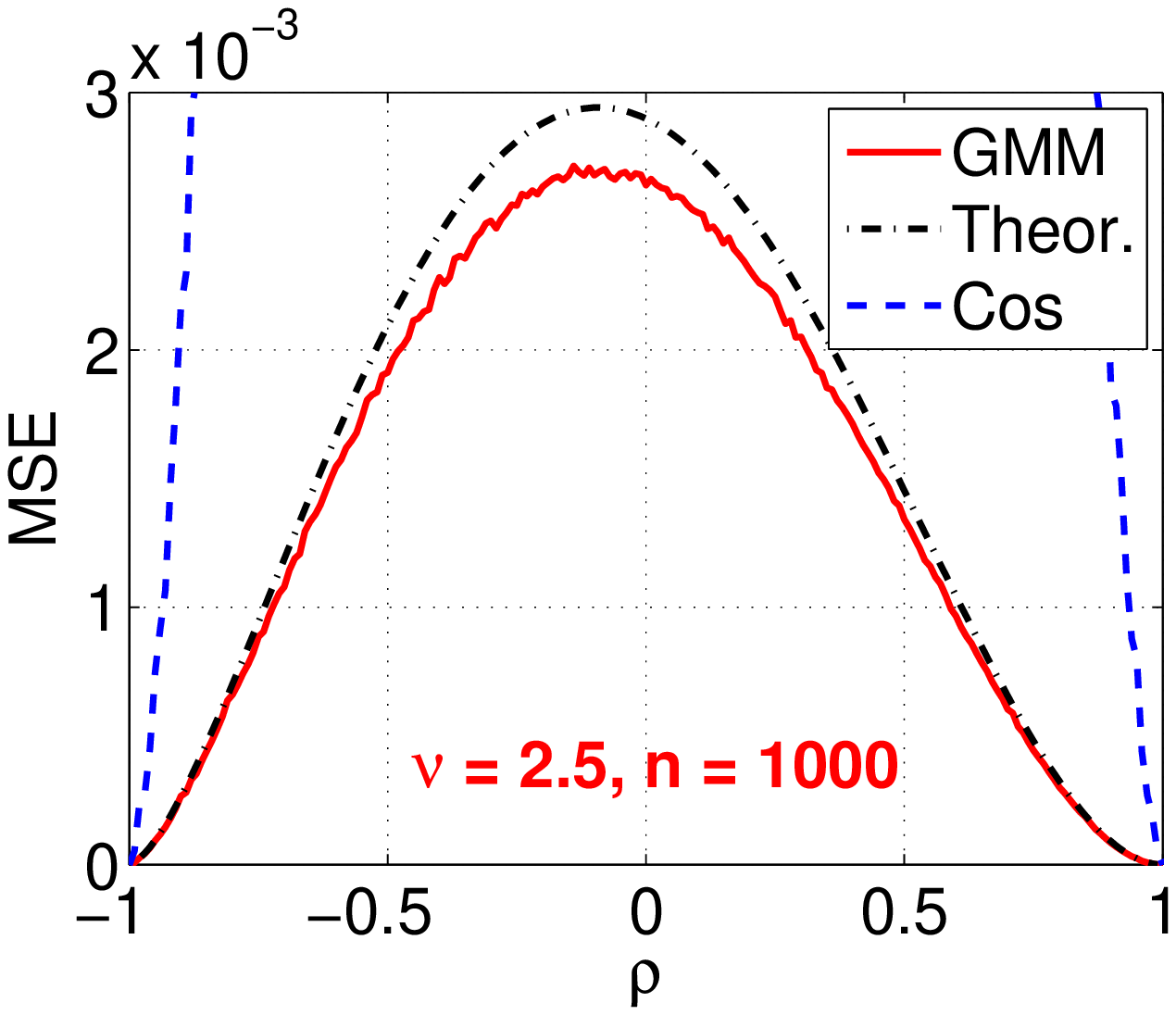}
\includegraphics[width=2.2in]{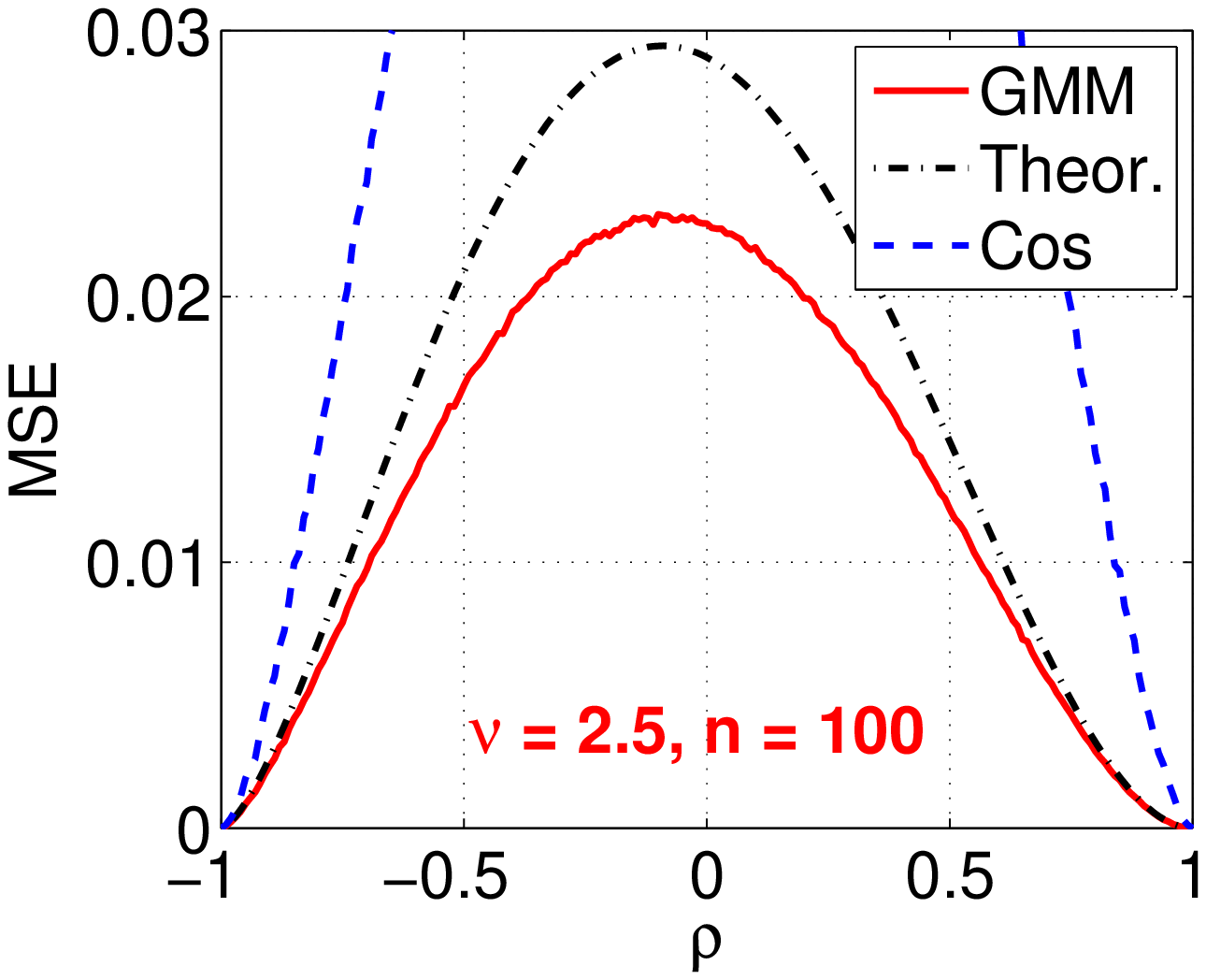}
}

\mbox{
\includegraphics[width=2.2in]{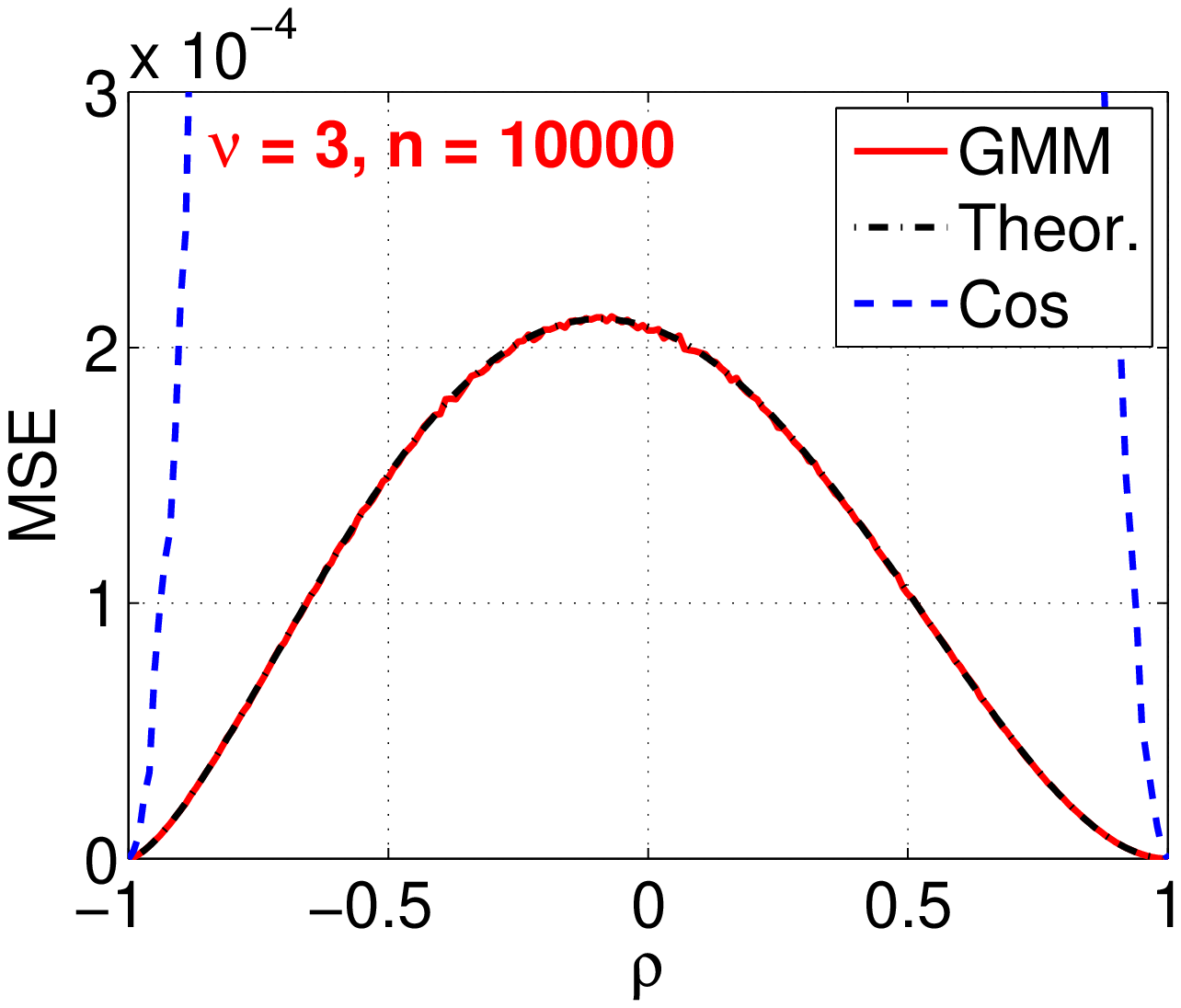}
\includegraphics[width=2.2in]{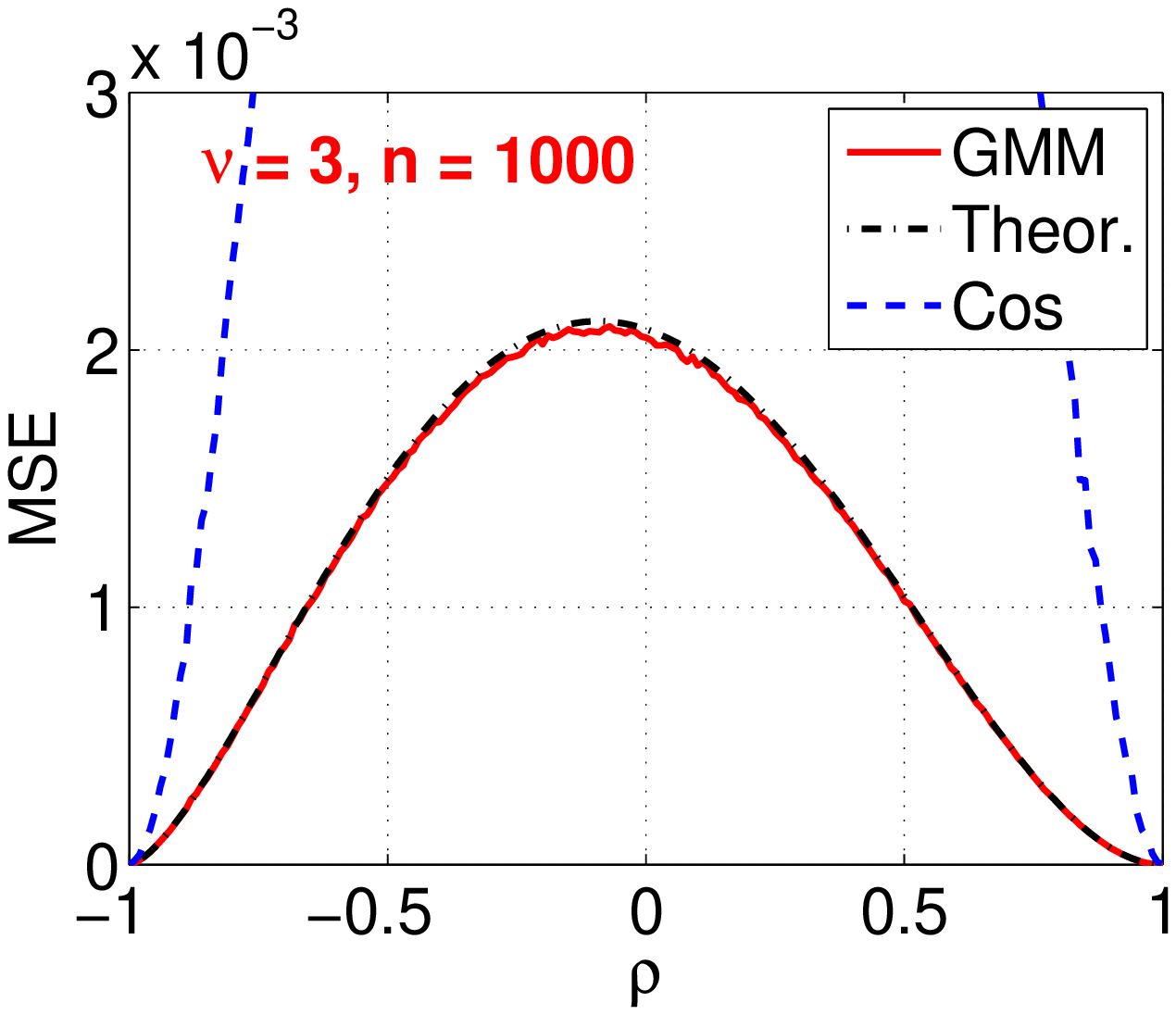}
\includegraphics[width=2.2in]{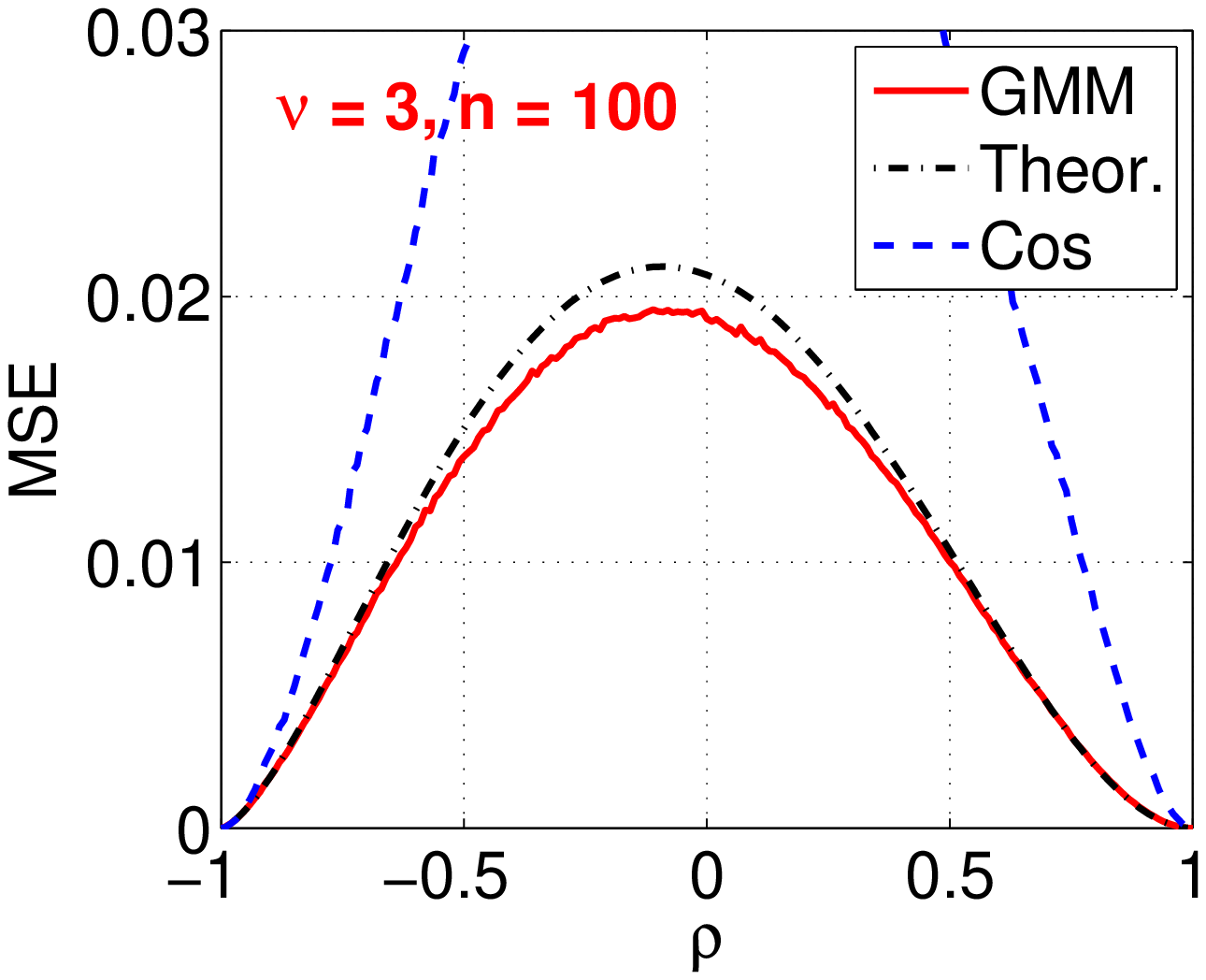}
}

\mbox{
\includegraphics[width=2.2in]{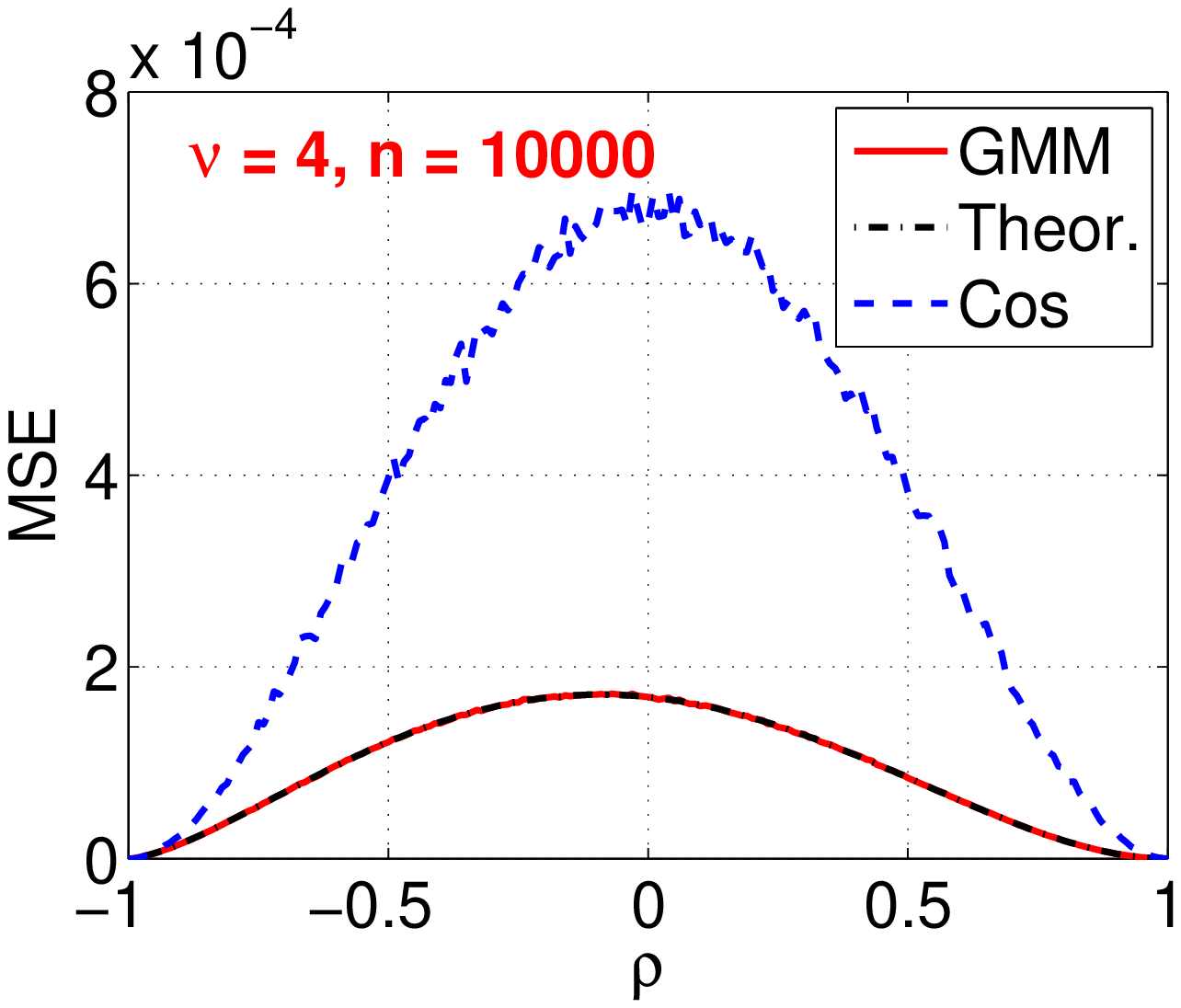}
\includegraphics[width=2.2in]{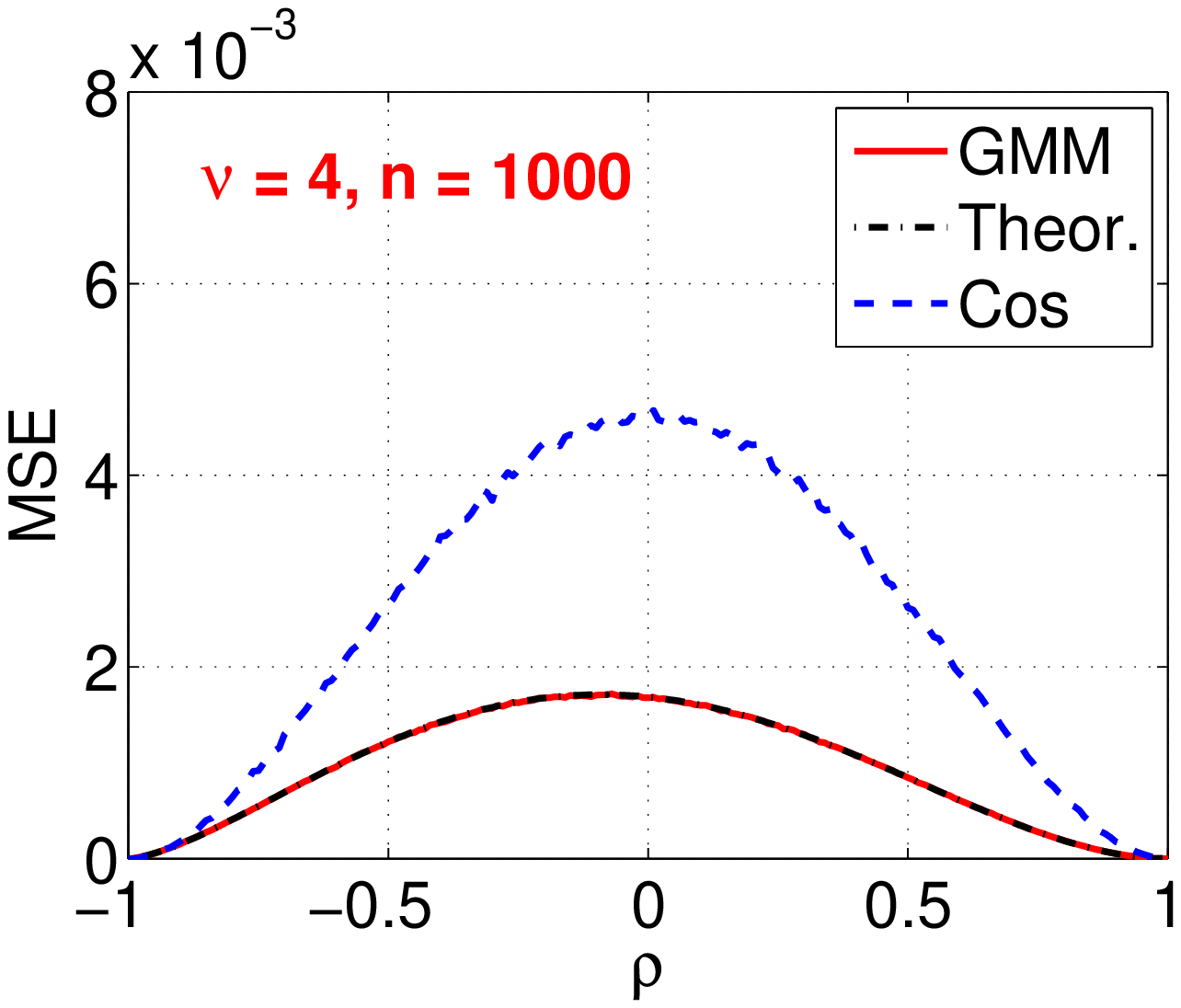}
\includegraphics[width=2.2in]{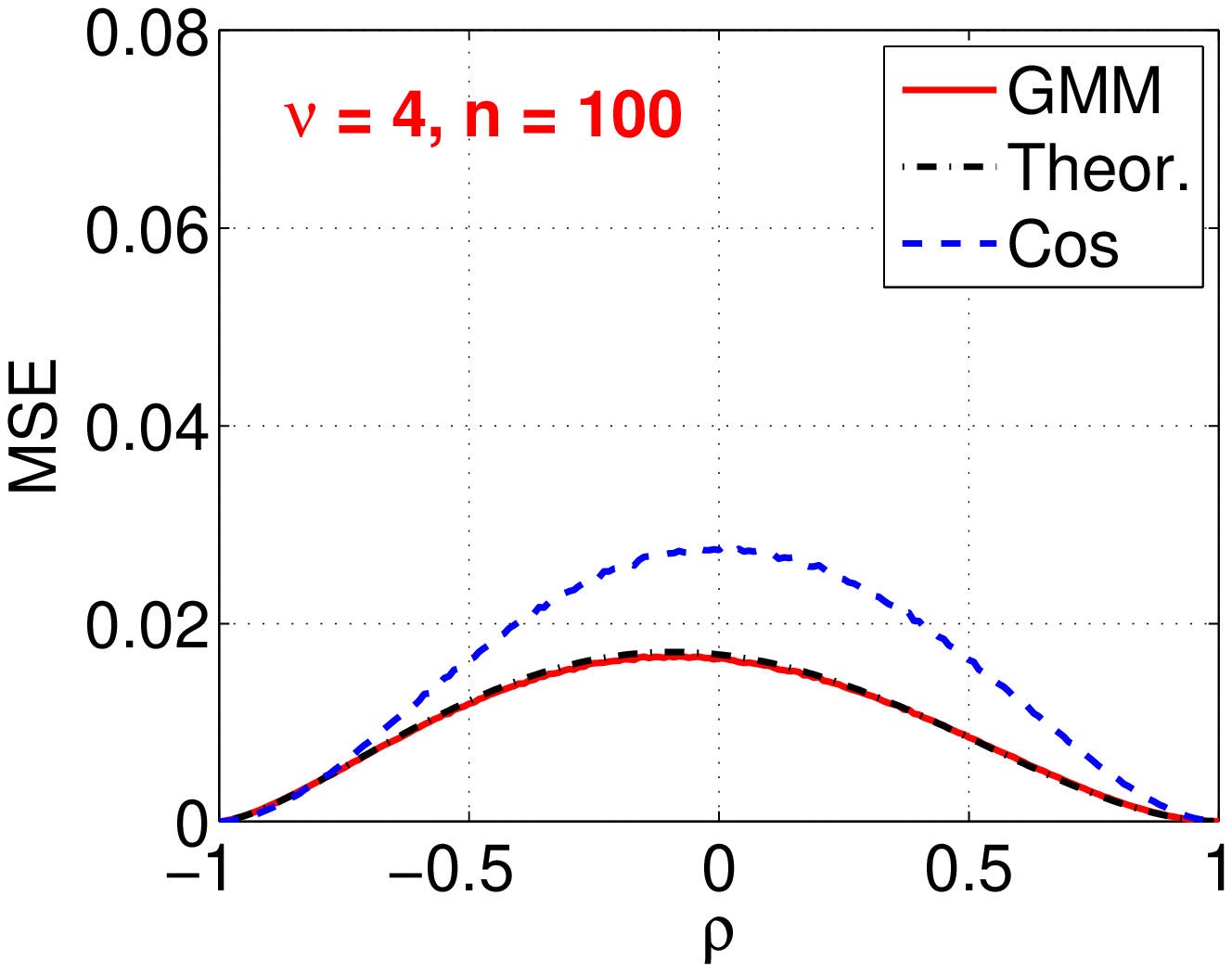}
}

\mbox{
\includegraphics[width=2.2in]{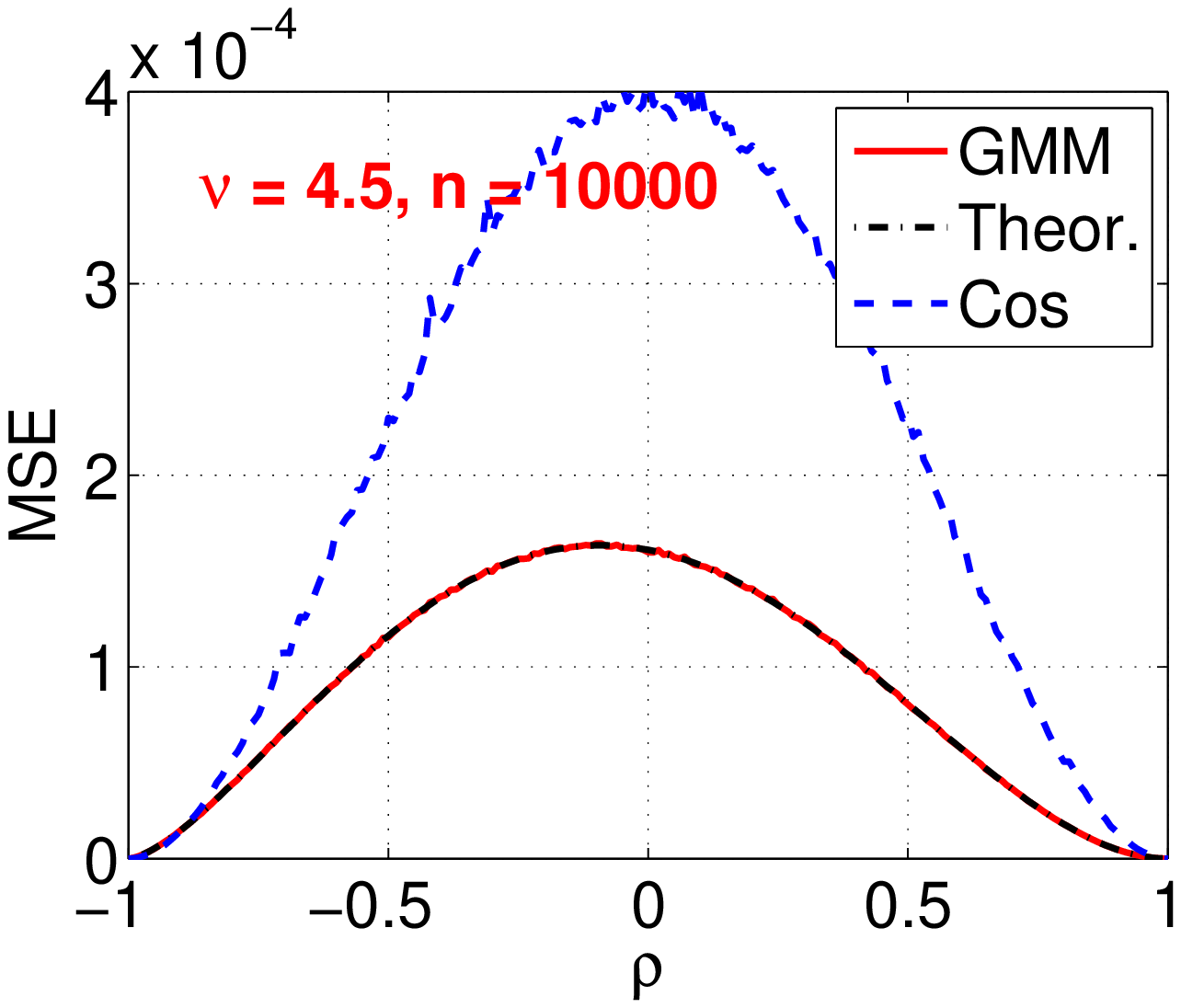}
\includegraphics[width=2.2in]{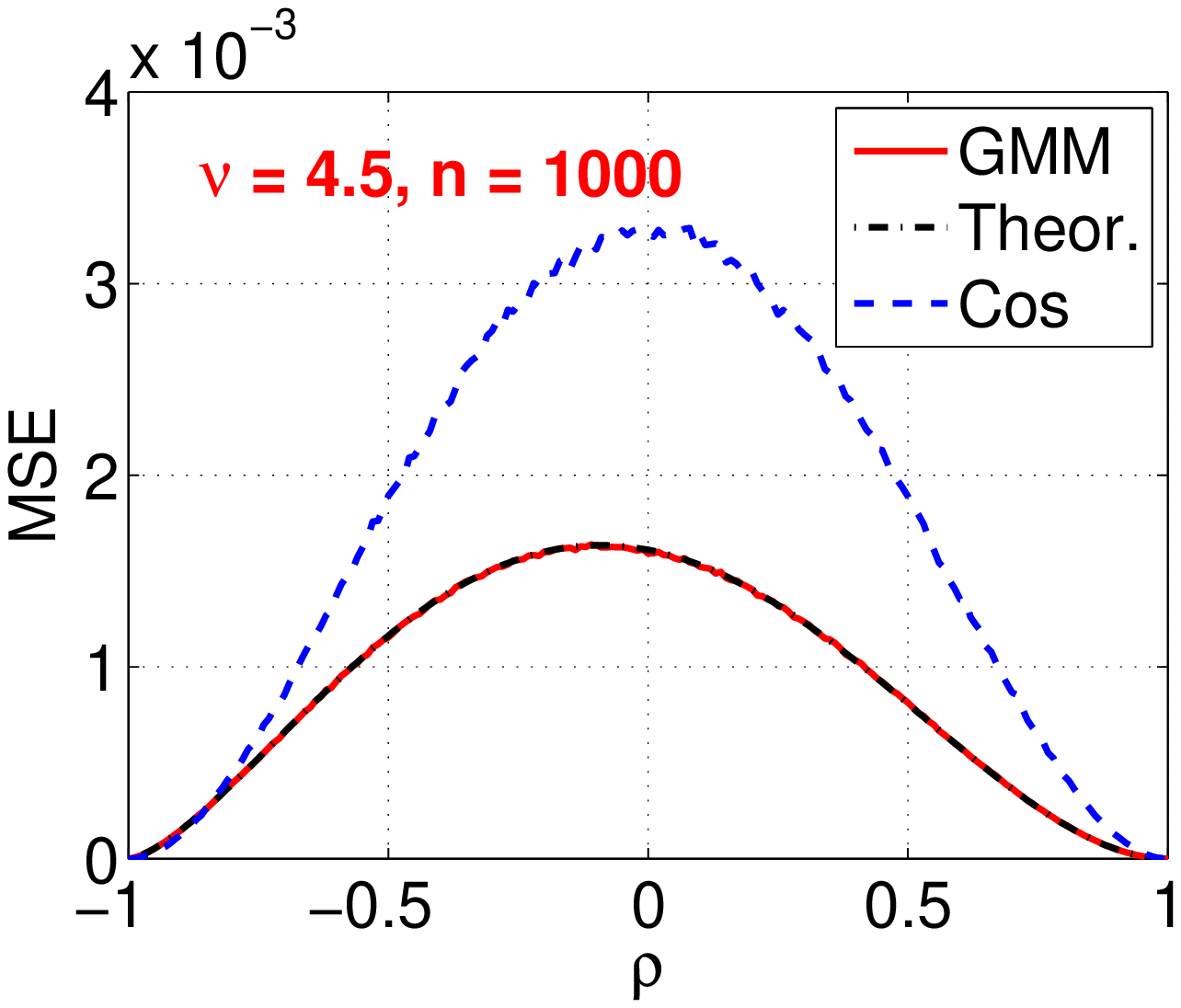}
\includegraphics[width=2.2in]{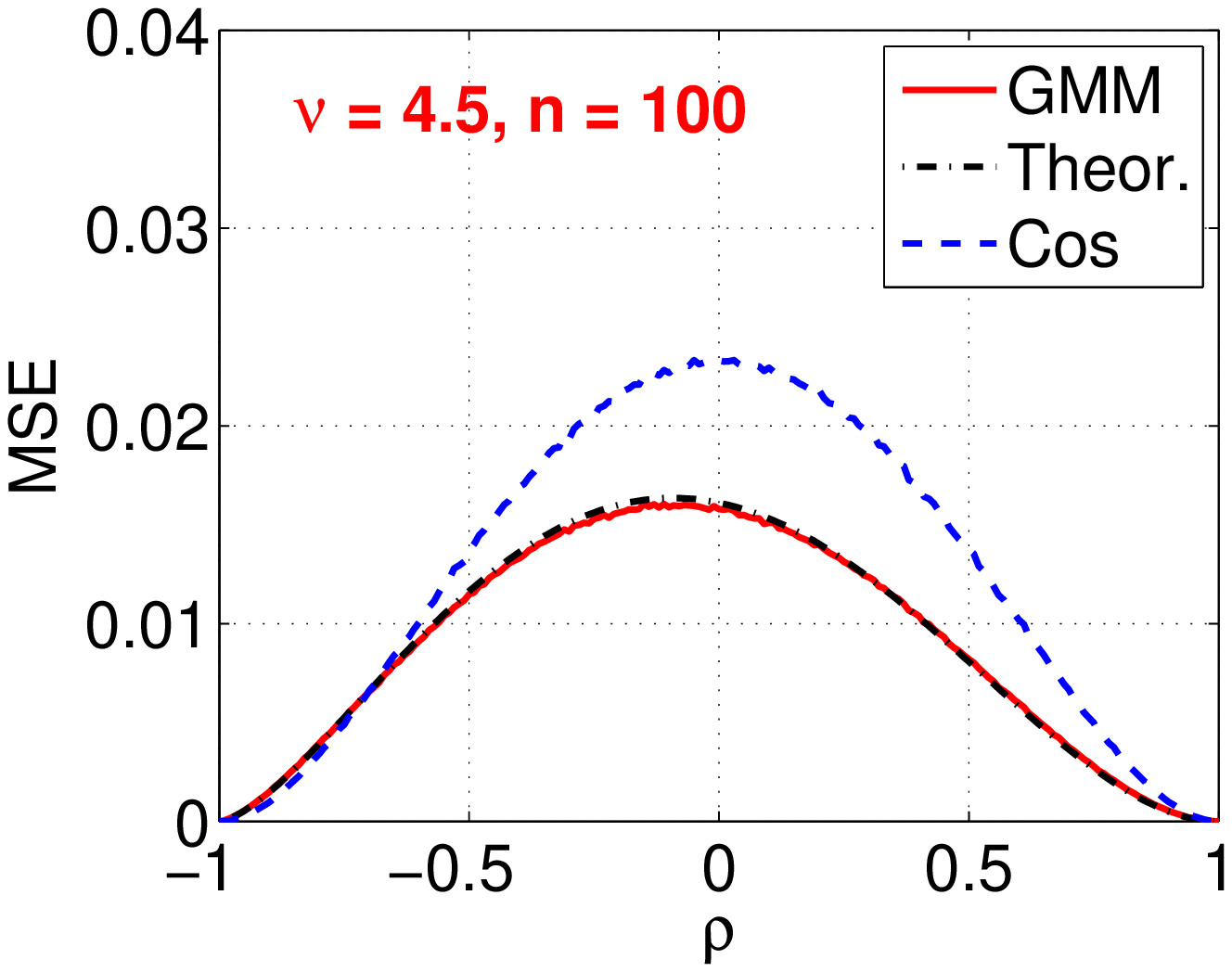}
}

\end{center}
\vspace{-0.2in}
\caption{Simulations for comparing two estimators of data similarity $\rho$: 1) $\hat{\rho}_g$, the estimator based on GMM, and 2) $\hat{\rho}_c$, the estimator based on cosine. We assume the data follow a $t$ distribution with $\nu$ degrees of freedom. In each panel (for each $\nu$), we plot the empirical MSE($\hat{\rho}_g$) and MSE($\hat{\rho}_c$) as well as the theoretical asymptotic variance of $\hat{\rho}_g$: $\frac{1}{n}2\left(1-\rho\right)\left(1+\sqrt{(1-\rho)/2}\right)^4 \frac{V}{H^4} \frac{\E T^2}{\E^2 T}$. It is clear from the results that $\hat{\rho}_g$ is substantially more accurate than $\hat{\rho}_c$. The theoretical asymptotic variance formula, despite the complexity of its expression, is accurate when $\nu$ is not too close to 2.   }\label{fig_Mse1}
\end{figure}

\begin{figure}
\begin{center}

\mbox{
\includegraphics[width=2.2in]{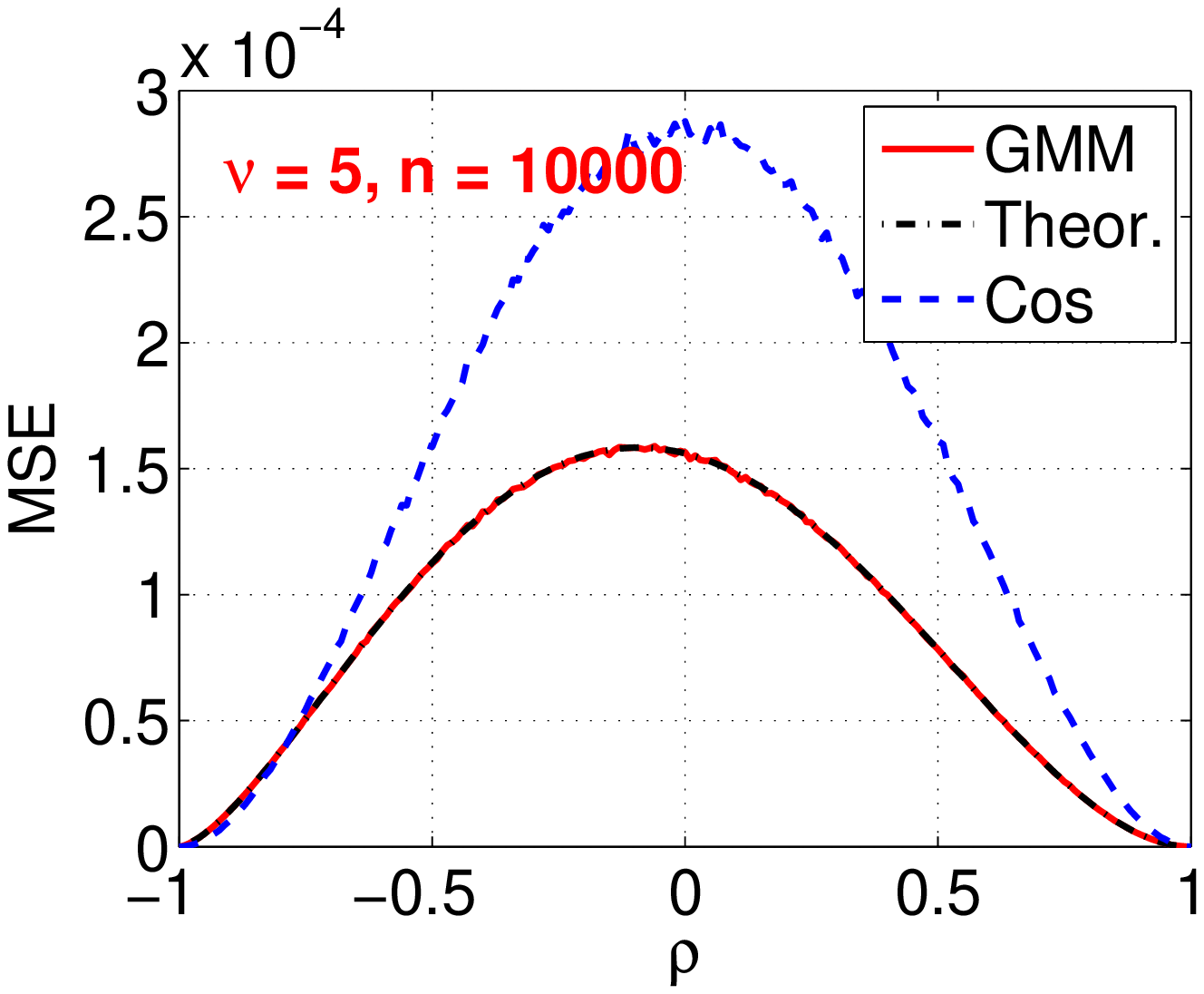}
\includegraphics[width=2.2in]{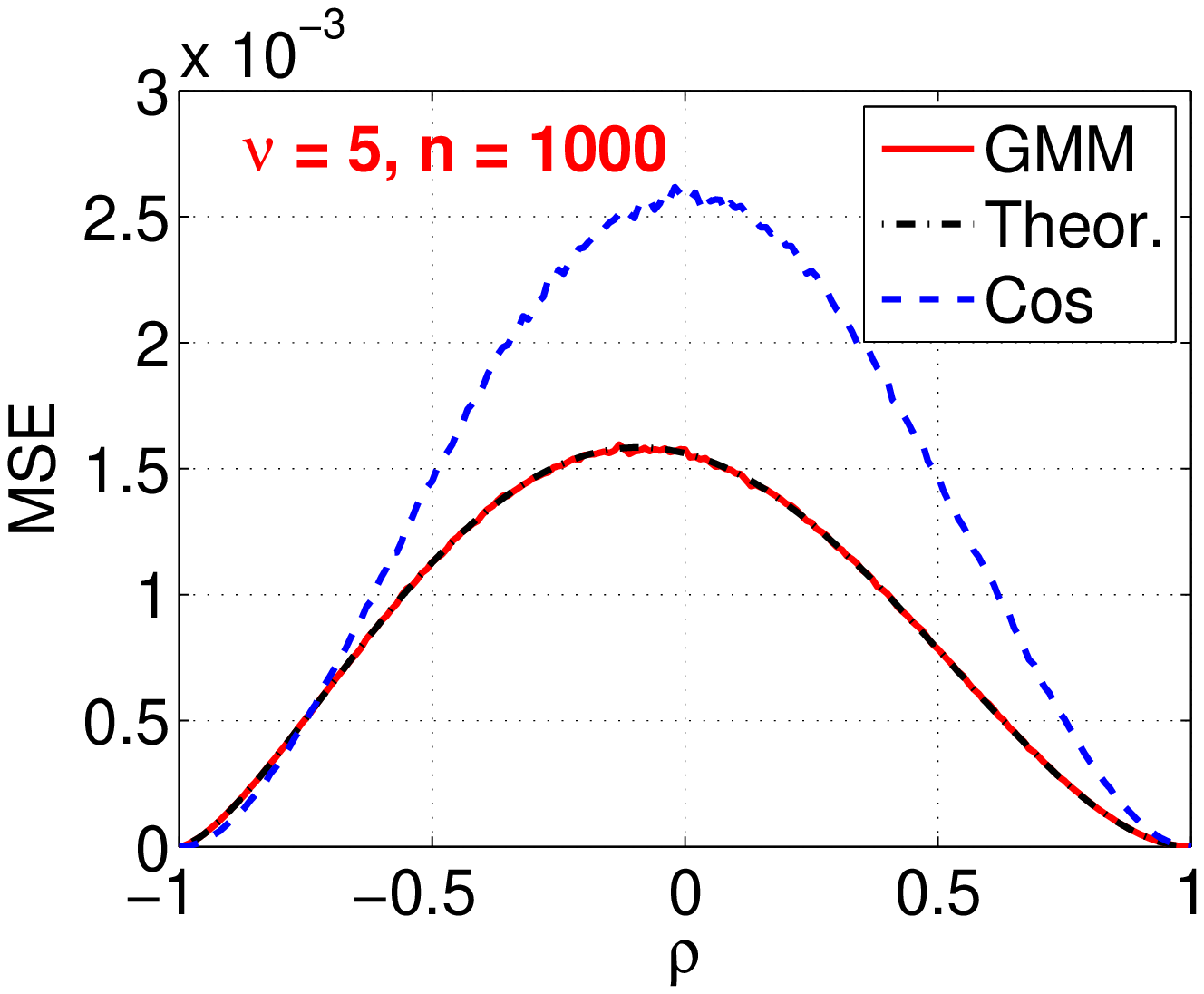}
\includegraphics[width=2.2in]{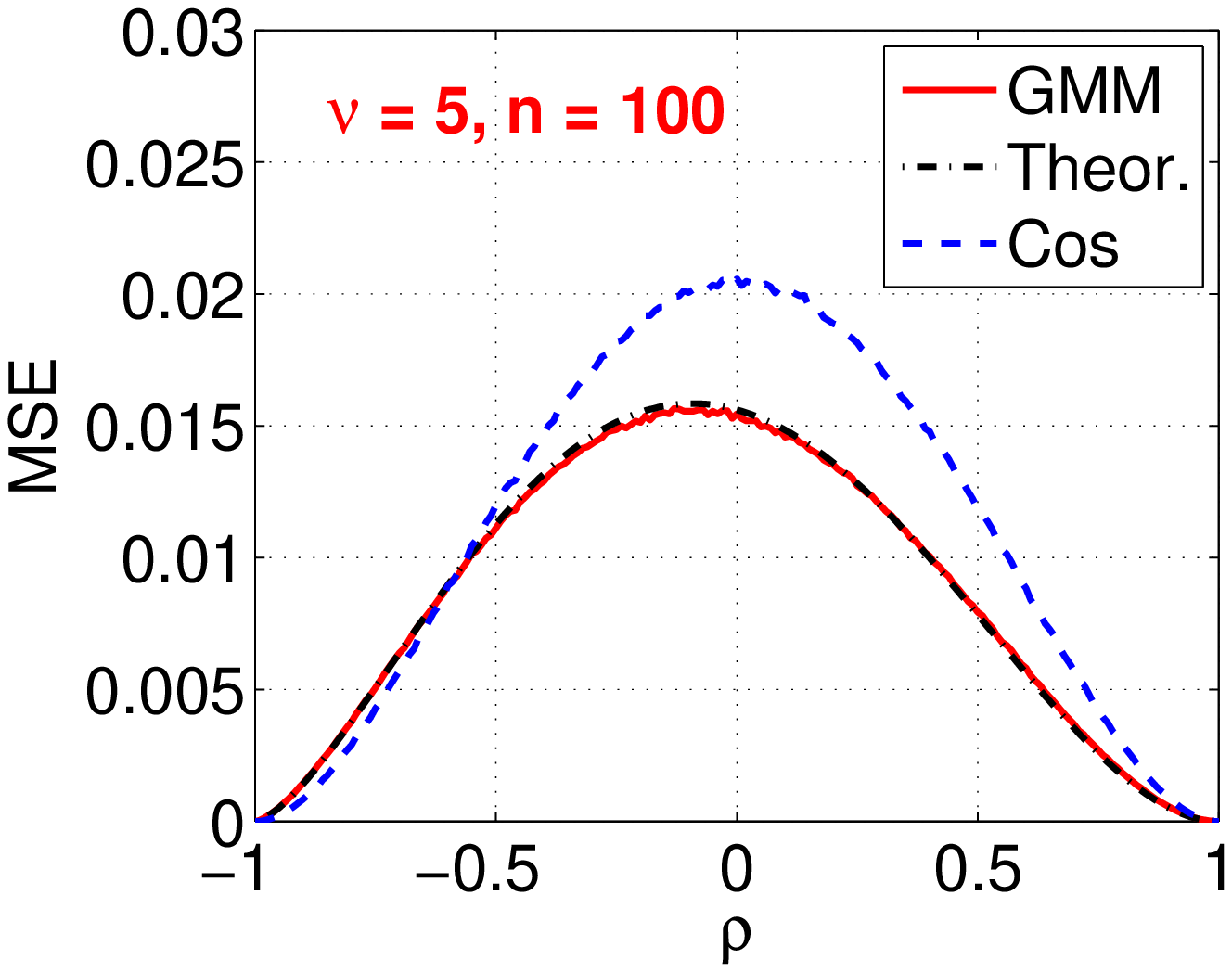}
}

\mbox{
\includegraphics[width=2.2in]{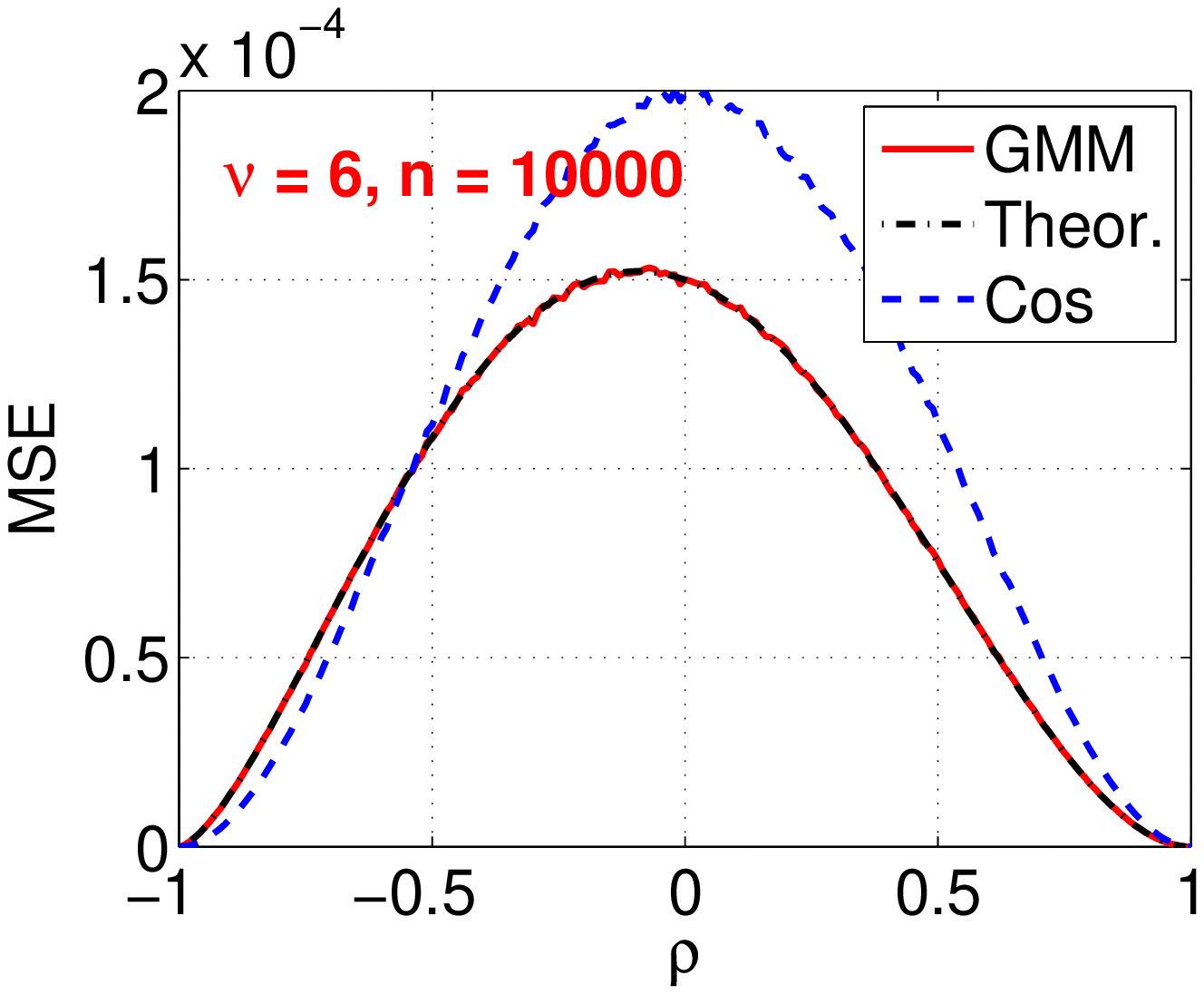}
\includegraphics[width=2.2in]{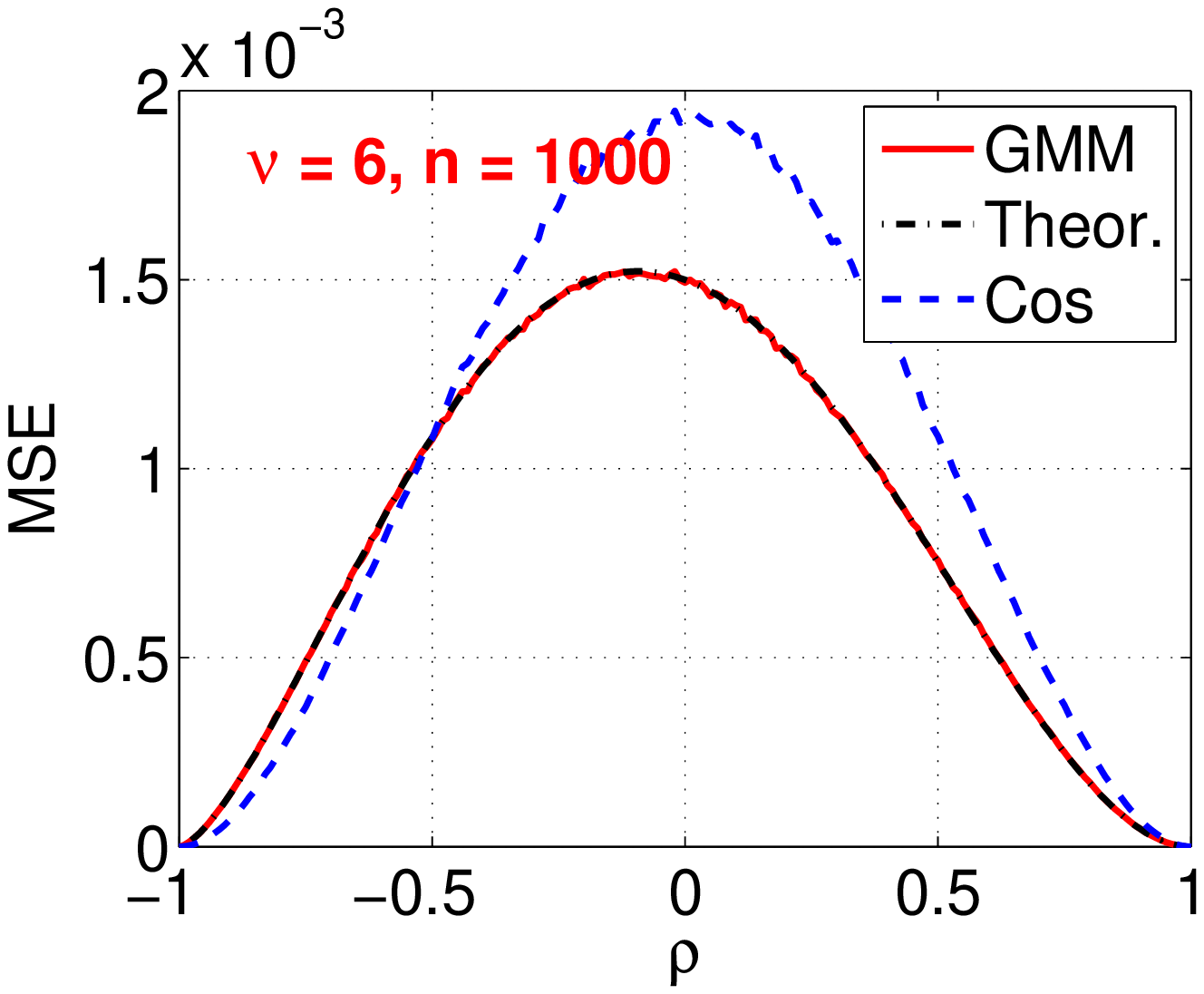}
\includegraphics[width=2.2in]{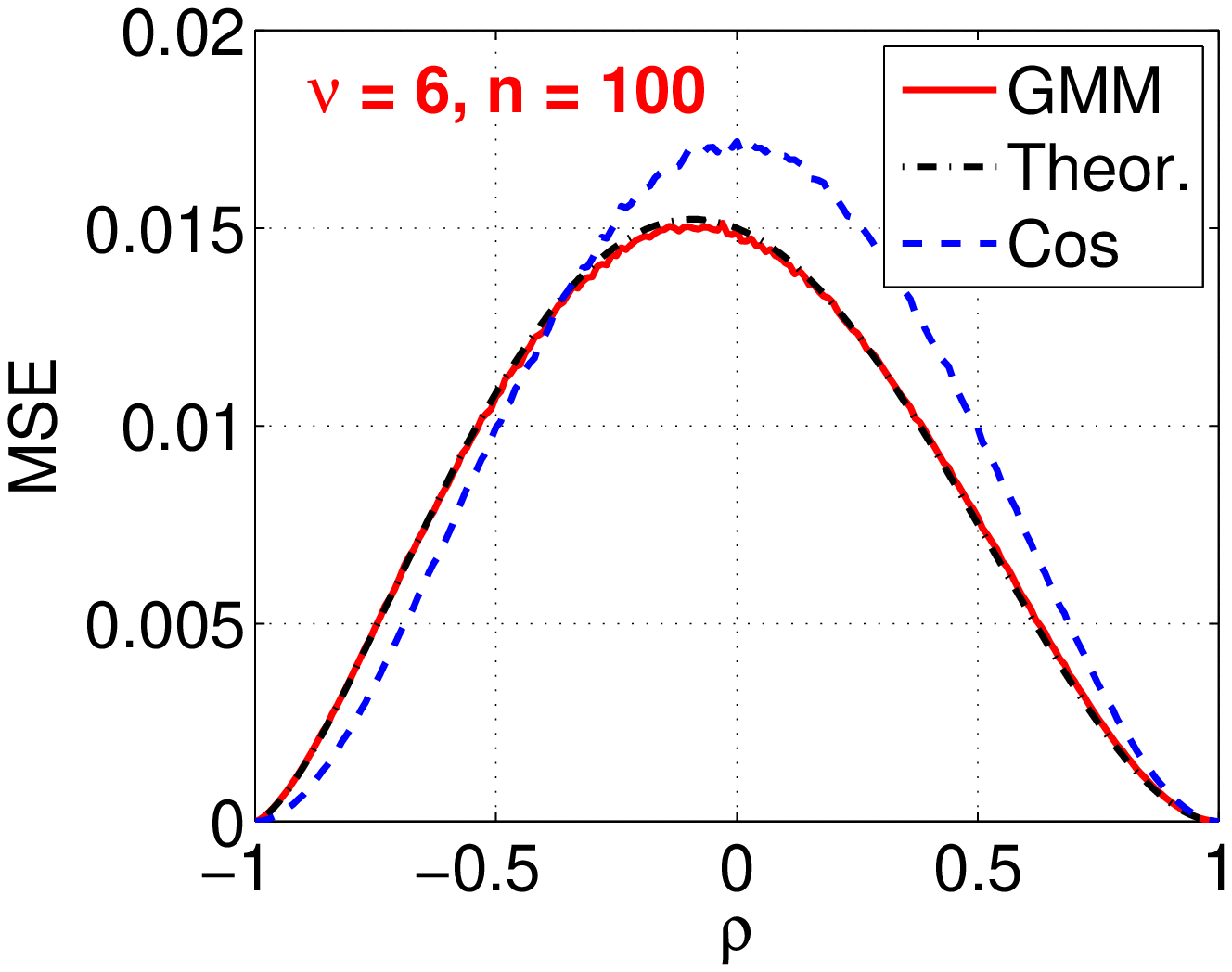}
}

\mbox{
\includegraphics[width=2.2in]{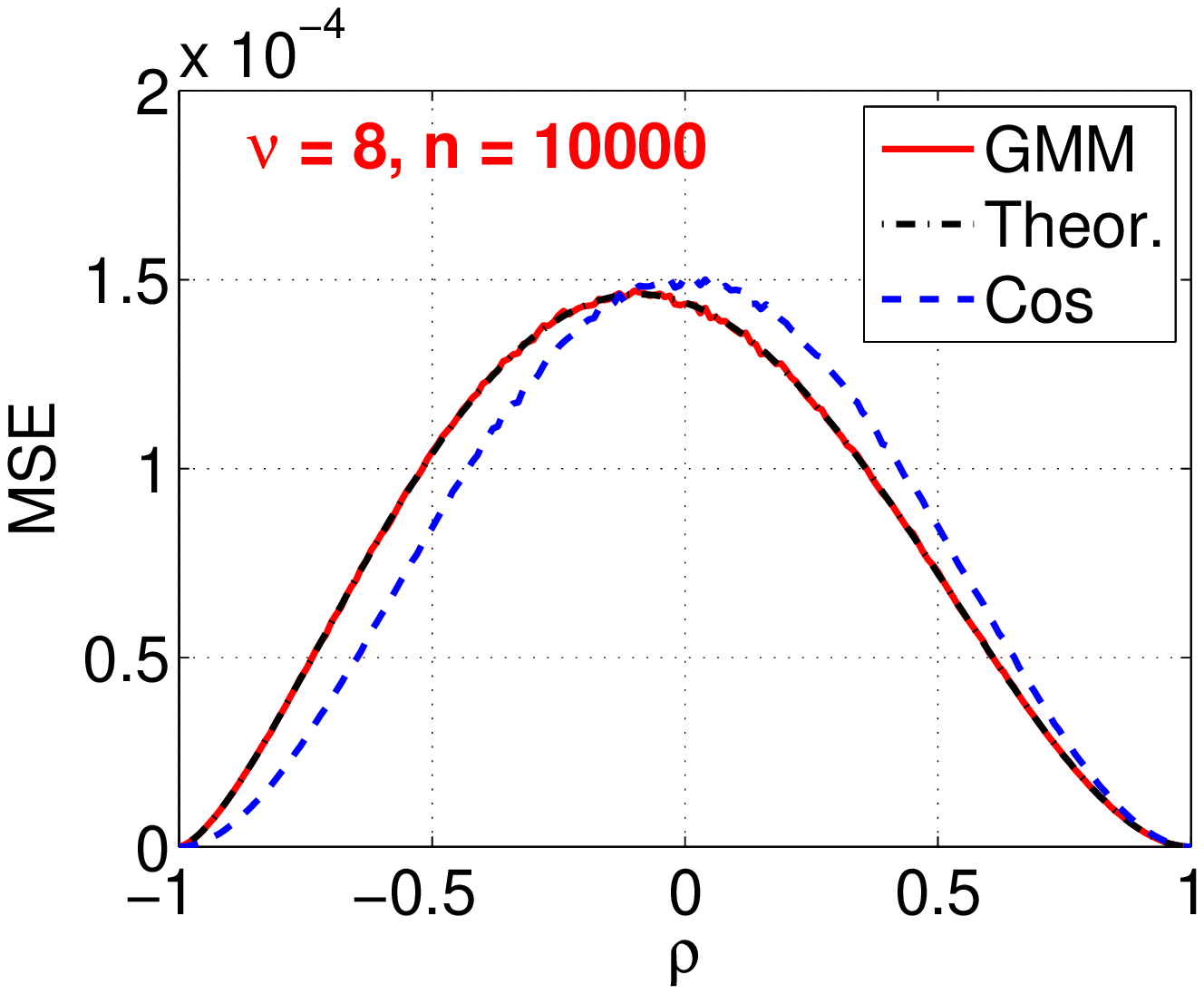}
\includegraphics[width=2.2in]{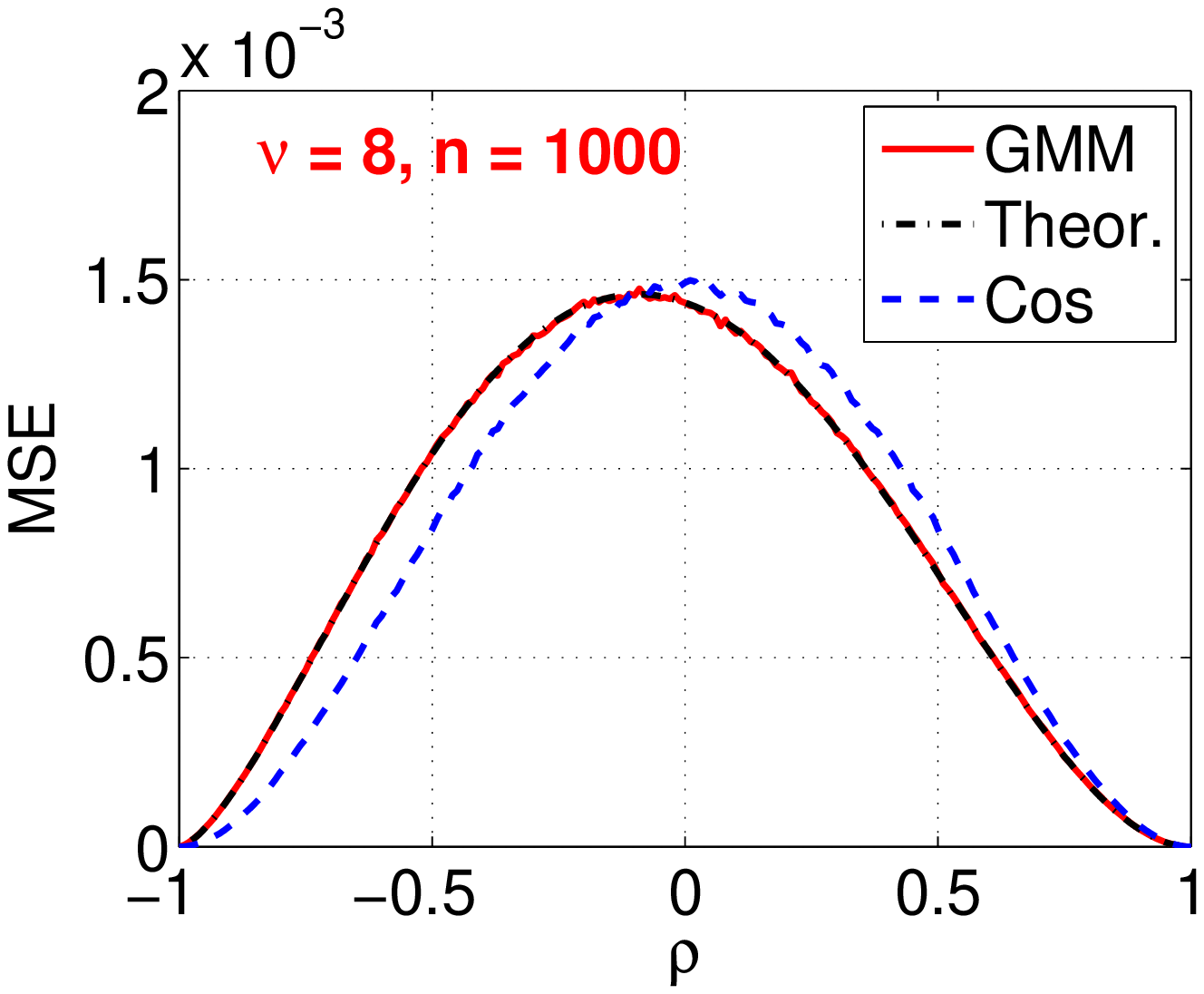}
\includegraphics[width=2.2in]{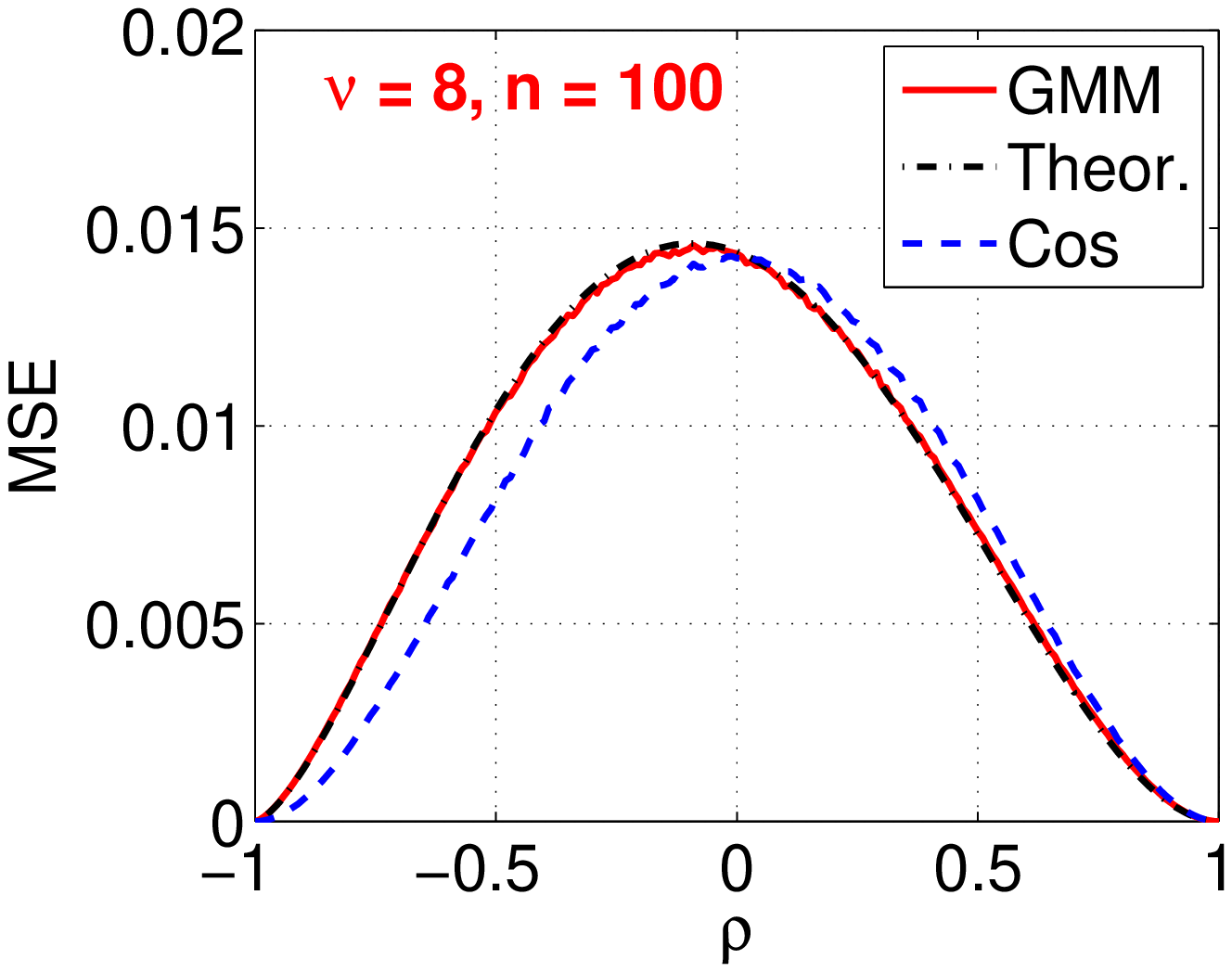}
}

\mbox{
\includegraphics[width=2.2in]{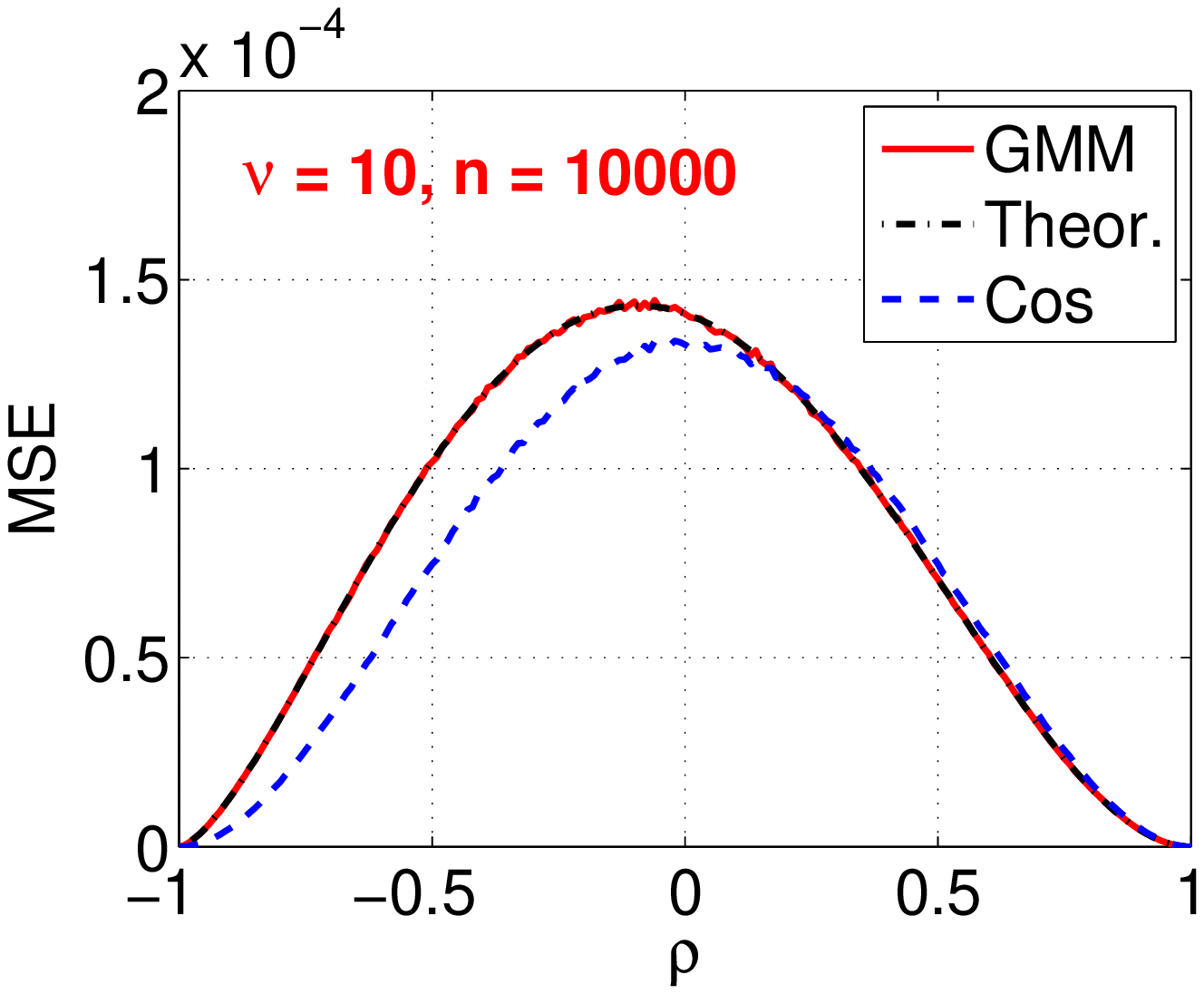}
\includegraphics[width=2.2in]{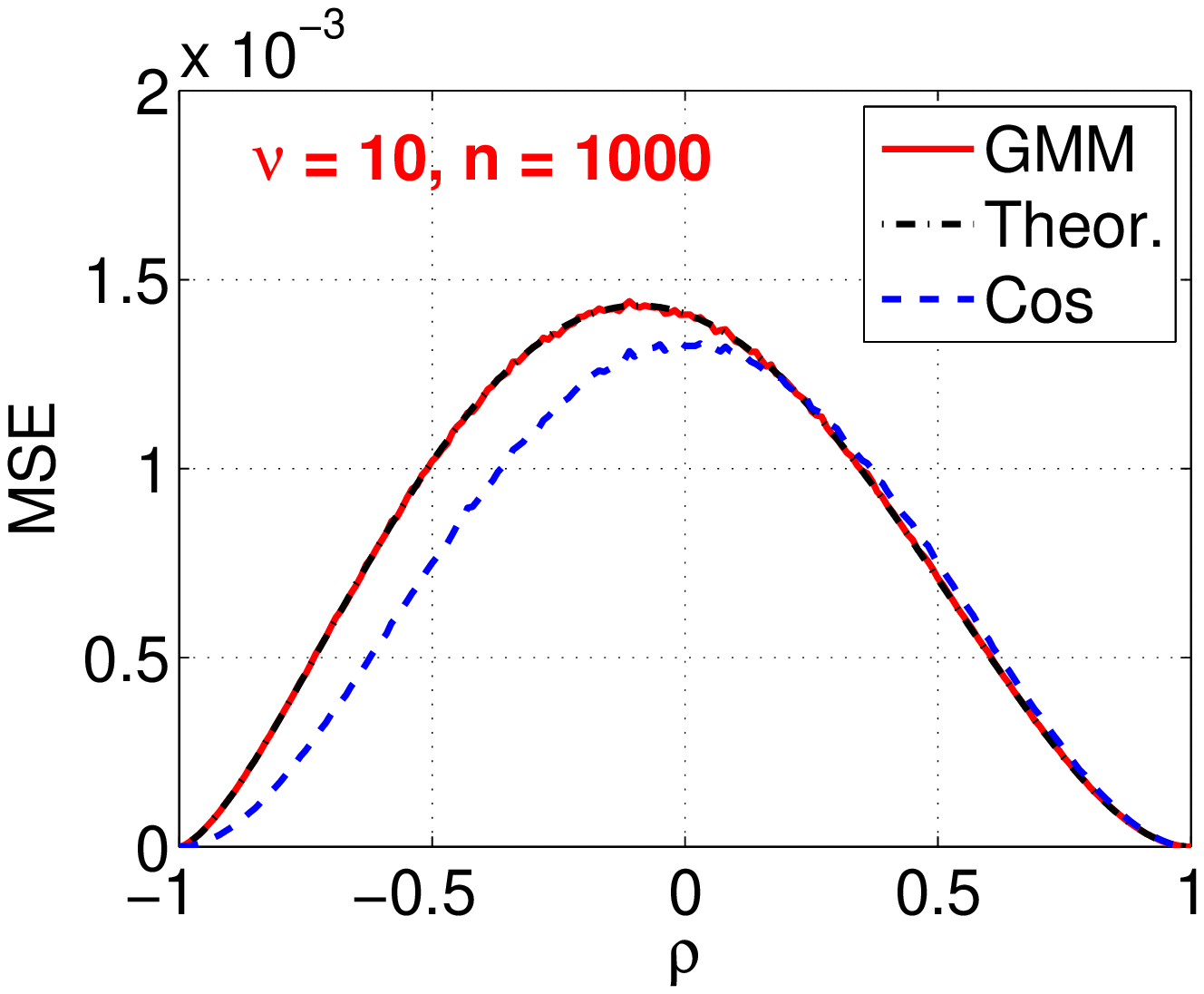}
\includegraphics[width=2.2in]{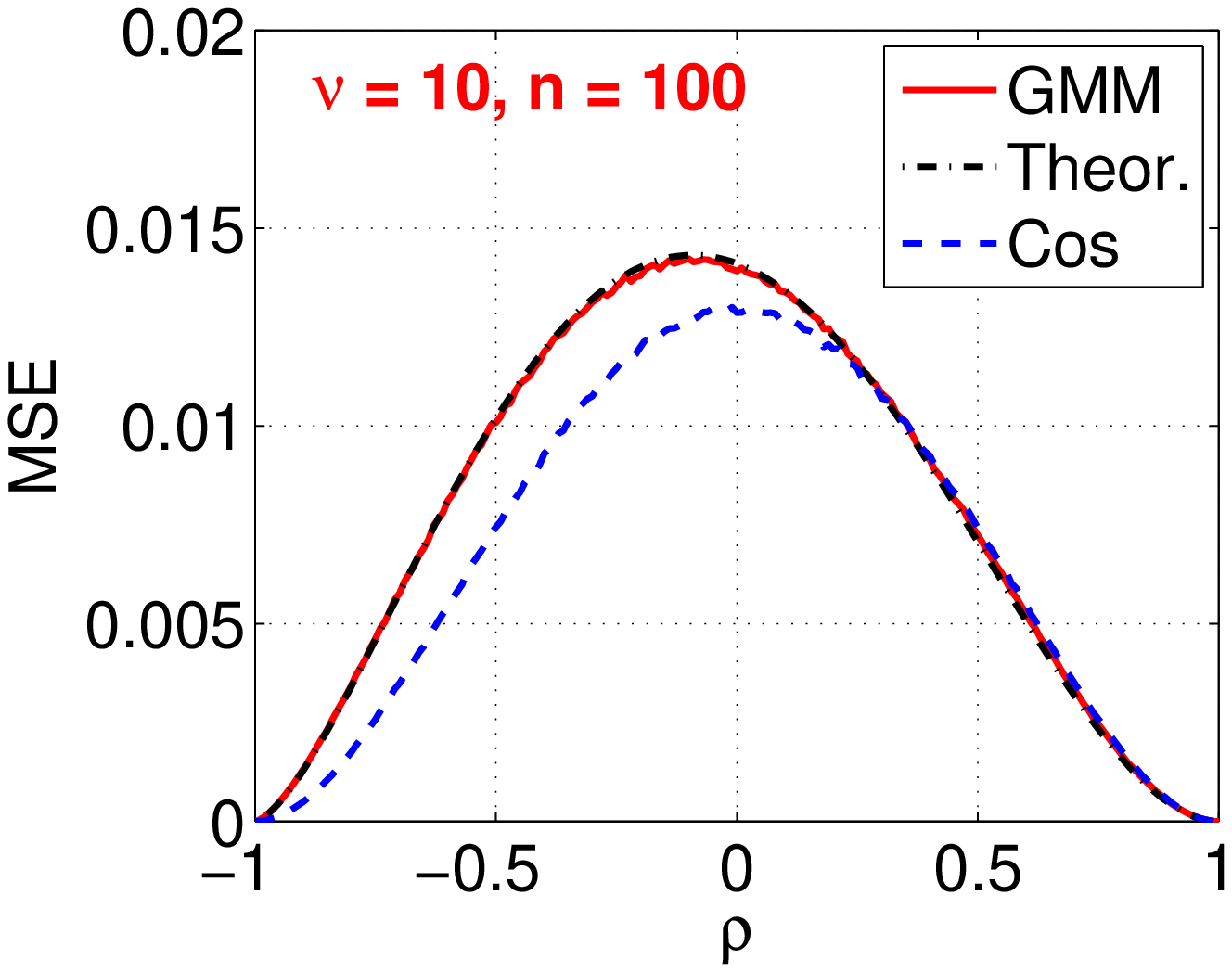}
}

\mbox{
\includegraphics[width=2.2in]{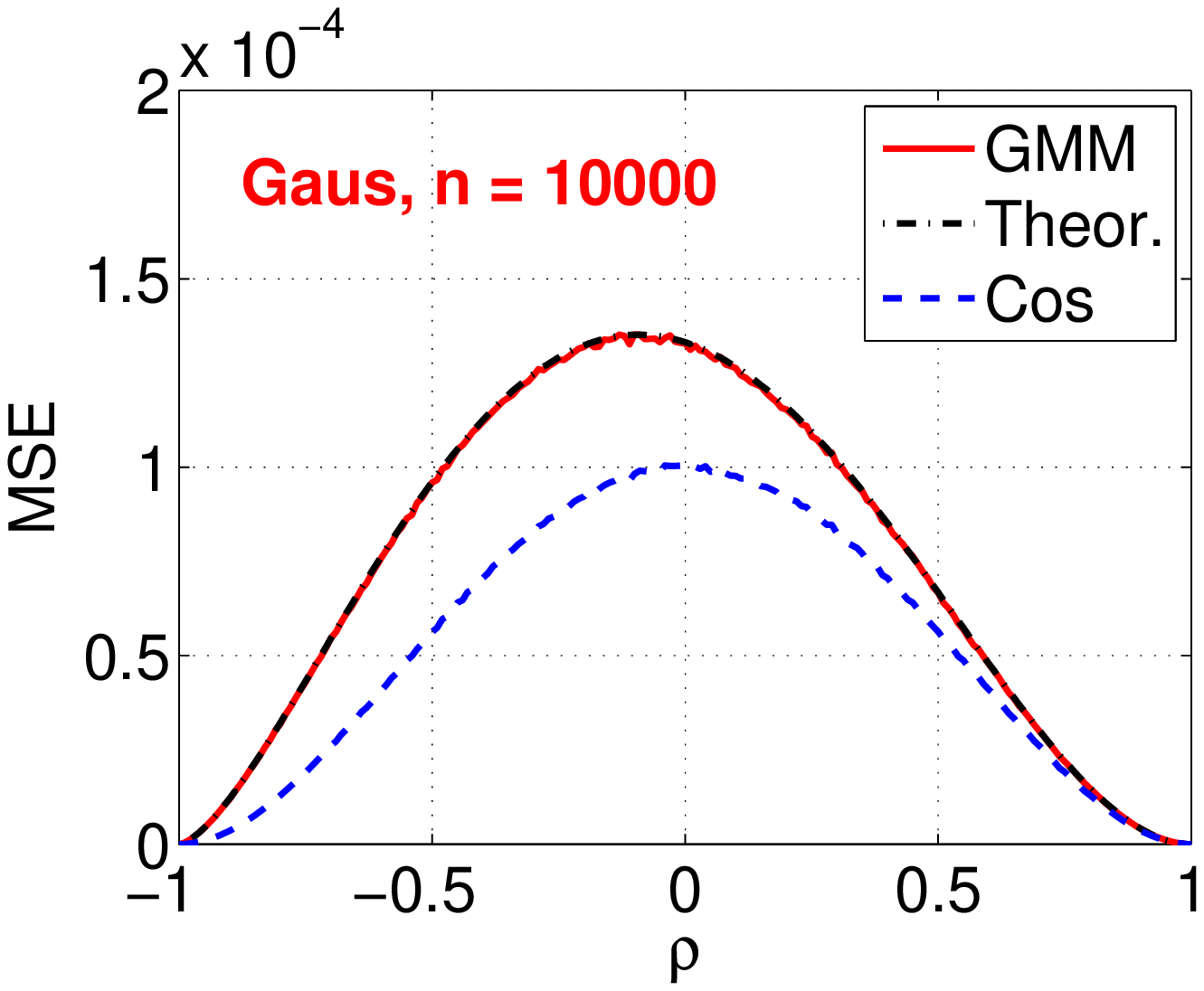}
\includegraphics[width=2.2in]{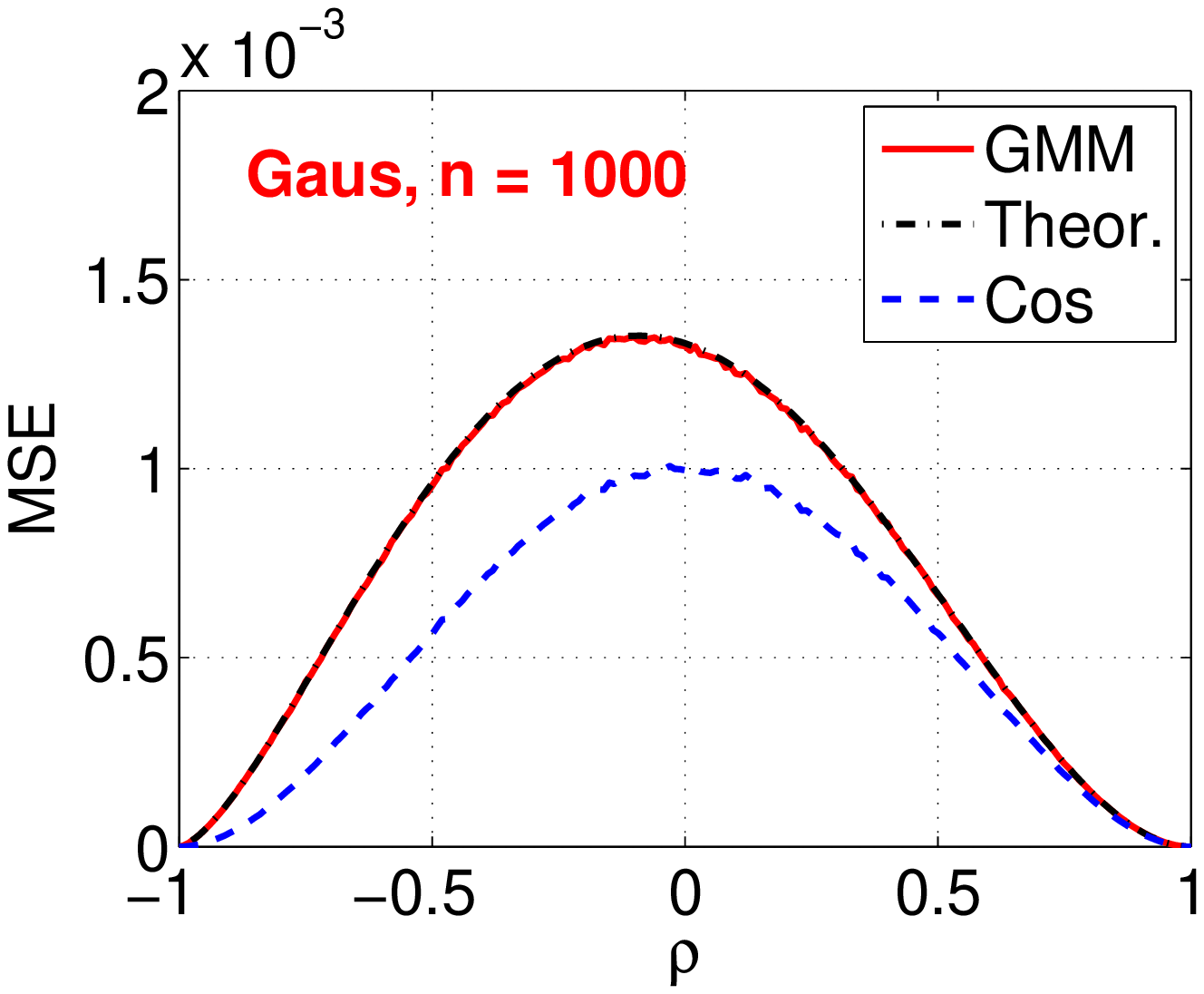}
\includegraphics[width=2.2in]{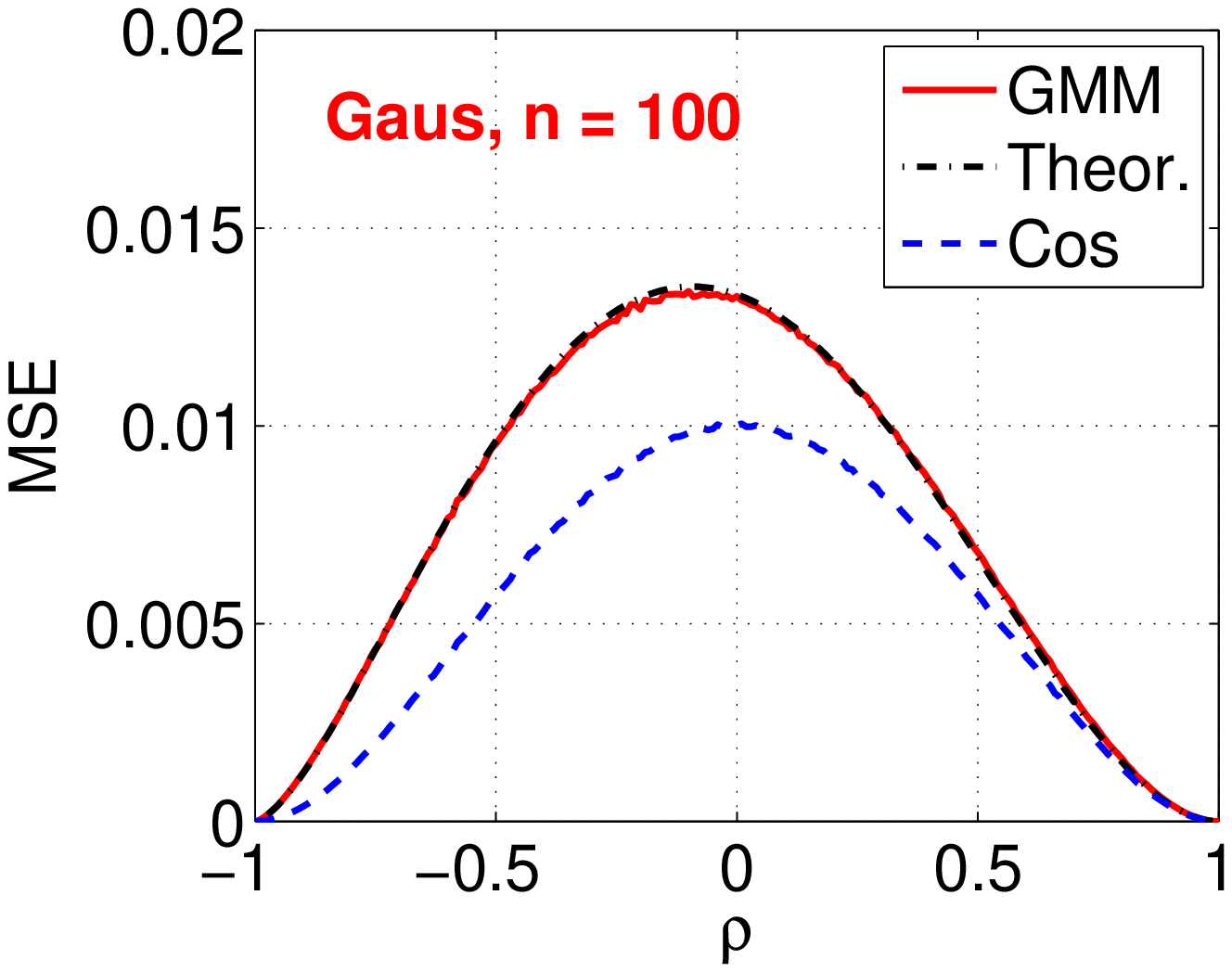}
}

\end{center}
\vspace{-0.2in}
\caption{Continued from Figure~\ref{fig_Mse1}. We present results for larger $\nu$ (5, 6, 8, 10) and $\nu=\infty$ (i.e., Gaussian data, the bottom row).  Roughly speaking, when $\nu<8$, it is preferable to use $\hat{\rho}_g$, the estimator based on GMM. In fact, even when data are perfectly Gaussian, using $\hat{\rho}_g$ does not result in too much loss of accuracy. }\label{fig_Mse2}
\end{figure}

\newpage\clearpage

\section{Concluding Remarks}

The ``cosine'' similarity commonly used in practice essentially assumes that the data are normally (or equivalently) distributed. The data in reality, however, are typically heavy-tailed and sparse. A concurrent line of work~\cite{Report:Li_GMM16,Report:Li_GMM_Nys16} has shown that the new measure named ``generalized min-max'' (GMM)  is particularly effective as a positive definite kernel and there is an efficient computational procedure to convert this nonlinear kernel into linear kernel. Extensive experiments on more than 50  datasets~\cite{Report:Li_GMM16,Report:Li_GMM_Nys16}  have demonstrated the promising performance in machine learning tasks. This motivates us to develop the theoretical results for analyzing GMM. \\

We show that, under  mild conditions,  GMM converges to a limit as long as the data have bounded first moment. In contrast, the cosine similarity requires that data to have bounded second moment. We derive the explicit expression for the limit and establish the asymptotic normality of GMM with explicit (and sophisticated) variance expressions. Those theoretical results will be useful for  further analyzing of GMM  in statistics, machine learning, and other applications. 

\bibliographystyle{abbrv}
\bibliography{../bib/mybibfile}

\begin{thebibliography}{10}

\bibitem{Book:Anderson03}
T.~W. Anderson.
\newblock {\em An Introduction to Multivariate Statistical Analysis}.
\newblock John Wiley \& Sons, Hoboken, New Jersey, third edition, 2003.

\bibitem{Book:Bingham87}
N.~H. Bingham, C.~M. Goldie, and J.~L. Teugels.
\newblock {\em Regular Variation}.
\newblock Cambridge University Press, 1987.
\newblock Cambridge Books Online.

\bibitem{Proc:Charikar}
M.~S. Charikar.
\newblock Similarity estimation techniques from rounding algorithms.
\newblock In {\em STOC}, pages 380--388, Montreal, Quebec, Canada, 2002.

\bibitem{Article:Crovella_power-law}
M.~E. Crovella and A.~Bestavros.
\newblock Self-similarity in world wide web traffic: Evidence and possible
  causes.
\newblock {\em IEEE/ACM Trans. Networking}, 5(6):835--846, 1997.

\bibitem{Proc:Faloutsos_Sigcomm99}
M.~Faloutsos, P.~Faloutsos, and C.~Faloutsos.
\newblock On power-law relationships of the \text{Internet} topology.
\newblock In {\em SIGMOD}, pages 251--262, Cambridge,MA, 1999.

\bibitem{Book:Gnedenko_54}
B.~V. Gnedenko and A.~N. Kolmogorov.
\newblock {\em Limit Distributions for Sum of Independent Random Variables}.
\newblock Addison Wesley, Reading, MA, 1954.

\bibitem{Proc:Ioffe_ICDM10}
S.~Ioffe.
\newblock Improved consistent sampling, weighted minhash and \text{L1}
  sketching.
\newblock In {\em ICDM}, pages 246--255, Sydney, AU, 2010.

\bibitem{Proc:Kleinberg_FOCS99}
J.~Kleinberg and E.~Tardos.
\newblock Approximation algorithms for classification problems with pairwise
  relationships: Metric labeling and \text{Markov} random fields.
\newblock In {\em FOCS}, pages 14--23, New York, 1999.

\bibitem{Article:Leland_power-law}
W.~E. Leland, M.~S. Taqqu, W.~Willinger, and D.~V. Wilson.
\newblock On the self-similar nature of \text{Ethernet} traffic.
\newblock {\em IEEE/ACM Trans. Networking}, 2(1):1--15, 1994.

\bibitem{Proc:Li_KDD15}
P.~Li.
\newblock 0-bit consistent weighted sampling.
\newblock In {\em KDD}, Sydney, Australia, 2015.

\bibitem{Report:Li_GMM16}
P.~Li.
\newblock Generalized min-max kernel and generalized consistent weighted
  sampling.
\newblock Technical report, arXiv:1605.05721, 2016.

\bibitem{Report:Li_GMM_Nys16}
P.~Li.
\newblock Nystrom method for approximating the gmm kernel.
\newblock Technical report, arXiv:1605.05721, 2016.

\bibitem{Report:Manasse_CWS10}
M.~Manasse, F.~McSherry, and K.~Talwar.
\newblock Consistent weighted sampling.
\newblock Technical Report MSR-TR-2010-73, Microsoft Research, 2010.

\bibitem{Article:Newman_05}
M.~E.~J. Newman.
\newblock Power laws, \text{P}areto distributions and \text{Z}ipf's law.
\newblock {\em Contemporary Physics}, 46(5):232--351, 2005.

\end{thebibliography}

\appendix

\section{Proof of Theorem~\ref{thm_mean}}

For a random vector $(X,Y)$, we are interested in quantities
\bes
\mu_1 = \E\, \frac{X_+\wedge Y_+ + X_-\wedge Y_-}{X_+\vee Y_+ + X_-\vee Y_-},\quad
\mu_\infty = \frac{\E(X_+\wedge Y_+ + X_-\wedge Y_-)}{\E(X_+\vee Y_+ + X_-\vee Y_-)}.
\ees
Without any assumption, we have
\bes
\mu_1 = \E\, \frac{X_+\wedge Y_+}{X_+\vee Y_+}+\E\, \frac{X_-\wedge Y_-}{X_-\vee Y_-}
= \E\, \frac{|X|\wedge|Y|}{|X|\vee |Y|}I\{XY>0\}
= \E\, \frac{|X/Y|\wedge 1}{|X/Y|\vee 1}I\{X/Y>0\}.
\ees
When $\E(|X|\wedge |Y|)<\infty$,
\bes
\mu_\infty = \frac{\E(|X|\wedge |Y|)I\{XY>0\}}{\E\big[(|X|+|Y|)I\{XY\le 0\}+(|X|\vee|Y|)I\{XY>0\}\big]}.
\ees

If $(X,Y)$ is symmetric in the sense of $(X,Y)\sim (-X,-Y)$, then
\bes
\mu_1 = 2\, \E\, \frac{X_+\wedge Y_+}{X_+\vee Y_+}
\ees
and
\bes
\mu_\infty = \frac{\E(X_+\wedge Y_+)}{\E(X_+\vee Y_+)}.
\ees

The vector $(X,Y)$ has an elliptical distribution if
\bel{elliptical}\notag
(X,Y)^T = AU T = {a_1^TU T\choose a_2^TUT}
\eel
where $A = (a_1,a_2)^T$ is a deterministic $2\times 2$ matrix,
$U$ is a vector uniformly distribution in the unit circle and $T$ is a positive random variable
independent of $U$.
In this case, $U\sim -U$, so that $(X,Y)$ is symmetric.
If $T$ has a finite expectation, then $T$ can be can cancelled in the calculation of
$\mu_1$ and $\mu_\infty$, so that
\bes
\mu_1 = 2\, \E\, \frac{(a_1^TU)_+\wedge (a_2^TU)_+}{(a_1^TU)_+\vee (a_2^TU)_+}
\ees
and
\bes
\mu_\infty = \frac{\E\{(a_1^TU)_+\wedge (a_2^TU)_+\}}{\E\{(a_1^TU)_+\vee (a_2^TU)_+\}}.
\ees
Since a bivariate Gaussian distribution is elliptical with $T^2\sim \chi^2_2$, the elliptical case with
finite $\E T$ is equivalent to the bivariate Gaussian case
\bes
(X,Y) \sim N(0,\Sigma)\ \hbox{ with }\ \Sigma = AA^T =
\begin{pmatrix} 1 & \sigma \rho \cr \sigma\rho & \sigma^2 \end{pmatrix}.
\ees
Note that we set $\Var(X)=1$ due to scale invariance of $\mu_1$ and $\mu_\infty$.

For $\sigma >0$ and $\rho\in [-1,1]$, let $\alpha = \sin^{-1}\big(\sqrt{1/2 - \rho/2}\big)\in [0,\pi/2]$,
and $\tau \in [-\pi/2+2\alpha,\pi/2]$ be the solution of $\cos(\tau - 2\alpha)/\cos\tau = \sigma$.
We have $\tau = \arctan(\sigma/\sin(2\alpha) - \cot(2\alpha))$. Define
\bes
f_1(\rho,\sigma) &=& \frac{1}{\sigma\pi}\Big((\tau+\pi/2-2\alpha)\cos(2\alpha)+\sin(2\alpha)
\log\frac{\cos(2\alpha-\pi/2)}{\cos\tau}\Big)
\cr && + \frac{\sigma}{\pi} \Big((\pi/2-\tau)\cos(2\alpha)+\sin(2\alpha)
\log\frac{\cos(2\alpha-\pi/2)}{\cos(2\alpha-\tau)}\Big),
\ees
and
\bel{f_infty}\notag
f_\infty(\rho,\sigma) =
\frac{1-\sin(2\alpha-\tau) +\sigma(1 - \sin\tau)}
{\sigma(1+\sin\tau) + 1+\sin(2\alpha-\tau)}.
\eel
We note that $\sin(2\alpha) = 2\sin\alpha\cos\alpha = 2\sqrt{1/2-\rho/2}\sqrt{1/2+\rho/2} = \sqrt{1-\rho^2}$,
$\cos(2\alpha)=\rho$, and $\cos(2\alpha - \pi/2) = \sin(2\alpha)= \sqrt{1-\rho^2}$.
Moreover, $\tan\tau = \{\sigma - \cos(2\alpha)\}/\sin(2\alpha) = (\sigma - \rho)/\sqrt{1-\rho^2}$,
so that $1/\cos^2\tau = 1+\tan^2\tau = 1 + (\sigma - \rho)^2/(1-\rho^2)
= (1 - \rho^2 + \sigma^2-2\sigma\rho + \rho^2)/(1-\rho^2)= (1+ \sigma^2-2\sigma\rho)/(1-\rho^2)$.
Thus,
\bes
\frac{\cos^2(2\alpha-\pi/2)}{\cos^2\tau} =1+ \sigma^2-2\sigma\rho,\
\frac{\cos^2(2\alpha-\pi/2)}{\cos^2(2\alpha-\tau)}=\frac{\cos^2(2\alpha-\pi/2)}{\sigma^2\cos^2 \tau}
= 1+ \sigma^{-2}-2\rho/\sigma.
\ees

Consider the Gaussian case
\bes
(X,Y) \sim N(0,\Sigma)\ \hbox{ with }\ \Sigma = AA^T =
\begin{pmatrix} 1 & \sigma \rho \cr \sigma\rho & \sigma^2 \end{pmatrix}.
\ees
Let
\bes
A = \begin{pmatrix} \cos\alpha & \sin\alpha \cr \sigma \cos\alpha & - \sigma \sin\alpha \end{pmatrix}.
\ees
We have
\bes
AA^T = \begin{pmatrix} 1 & \sigma(\cos^2\alpha - \sin^2\alpha)
\cr \sigma(\cos^2\alpha - \sin^2\alpha)  & \sigma^2 \end{pmatrix}
= \begin{pmatrix} 1 & \sigma \rho \cr \sigma\rho & \sigma^2 \end{pmatrix}.
\ees
Let $\theta$ be a uniform variable in $(-\pi,\pi)$. Since $U\sim (\cos\theta,\sin\theta)^T$,
\bes
{X\choose Y}
\sim \begin{pmatrix} \cos\alpha\cos\theta + \sin\alpha\sin\theta
\cr \sigma(\cos\alpha\cos\theta - \sin\alpha\sin\theta) \end{pmatrix}T
= \begin{pmatrix} \cos(\theta - \alpha)
\cr \sigma \cos(\theta + \alpha) \end{pmatrix}T
\sim \begin{pmatrix} \cos(\theta - 2\alpha) \cr \sigma \cos\theta\end{pmatrix}T
\ees
As $\alpha\in (0,\pi/2)$ , it follows that
\bes
\mu_1 &=& \frac{2}{2\pi}\int_{-\pi}^\pi \frac{(\cos(\theta - 2\alpha))_+\wedge (\sigma \cos\theta)_+}
{(\cos(\theta - 2\alpha))_+\vee (\sigma \cos\theta)_+}d\theta
\cr &=& \frac{1}{\pi}\int_{-\pi/2+2\alpha}^{\pi/2} \frac{(\cos(\theta - 2\alpha)/\cos\theta) \wedge \sigma}
{(\cos(\theta - 2\alpha)/\cos\theta) \vee \sigma}d\theta.
\ees
As $\cos(\theta - 2\alpha)/\cos\theta = \cos(2\alpha)+\tan\theta \sin(2\alpha)$,
for $\cos\theta>0$ $\cos(\theta - 2\alpha)/\cos\theta = \sigma$ iff $\theta =\tau$
and $\tau \in [-\pi/2+2\alpha,\pi/2]$. Thus, with $t = 2\alpha - \theta$,
\bes
\mu_1 &=& \frac{1}{\pi}\int_{-\pi/2+2\alpha}^{\tau}\frac{\cos(\theta - 2\alpha)}{\sigma \cos\theta}d\theta
+\frac{1}{\pi}\int_{\tau}^{\pi/2} \frac{\sigma \cos\theta}{\cos(\theta - 2\alpha)}d\theta
\cr &=& \frac{1}{\sigma\pi}\int_{-\pi/2+2\alpha}^{\tau}\{\cos(2\alpha)+\tan\theta \sin(2\alpha)\}d\theta
+\frac{\sigma}{\pi} \int_{-\pi/2+2\alpha}^{2\alpha-\tau} \frac{\cos(t-2\alpha)}{\cos t}dt
\cr &=& f_1(\rho,\sigma).
\ees
We note that $\tau=\alpha$ when $\sigma=1$. Similarly,
\bes
\mu_\infty &=& \frac{(2\pi)^{-1}\int_{-\pi}^\pi (\cos(\theta - 2\alpha))_+\wedge (\sigma \cos\theta)_+d\theta}
{(2\pi)^{-1}\int_{-\pi}^\pi (\cos(\theta - 2\alpha))_+\vee (\sigma \cos\theta)_+d\theta }
\cr &=& \frac{\int_{-\pi/2+2\alpha}^{\tau} \cos(\theta - 2\alpha)d\theta
+\sigma \int_{\tau}^{\pi/2} \cos\theta d\theta}
{\sigma\int_{-\pi/2}^{\tau} \cos \theta d\theta
+ \int_{\tau}^{\pi/2+2\alpha} \cos(\theta - 2\alpha)d\theta}
\cr &=&\frac{1-\sin(2\alpha-\tau) +\sigma(1 - \sin\tau)}
{\sigma(1+\sin\tau) + 1+\sin(2\alpha-\tau)}.
\cr &=& f_\infty(\rho,\sigma).
\ees

It is well known~\cite{Book:Gnedenko_54,Book:Bingham87} that
\bes
\frac{\max_{i\le n}T_i}{T_1+\cdots+T_n} = o_{\P}(1)
\ees
if and only if
\bel{cond}
\lim_{t\to\infty} \frac{t\,\P(T>t)}{\E\min(T,t)} = 0.
\eel
Suppose (\ref{cond}) holds. Let $(X_i,Y_i)$ be a sequence of iid variables from $(X,Y)$.
Then,
\bel{th-1}\notag
\frac{\sum_{i=1}^n\{(X_i)_+\wedge (Y_i)_+ + (X_i)_-\wedge (Y_i)_-\}}
{\sum_{i=1}^n\{(X_i)_+\vee (Y_i)_+ + (X_i)_-\vee (Y_i)_-\}} = f_\infty(\rho,\sigma) + o_{\P}(1).
\eel
This can be seen as follows. Write
\bes
(X_i,Y_i)^T = AU_iT_i,\ \hbox{ with }\
A = \begin{pmatrix} \cos\alpha & \sin\alpha \cr \sigma \cos\alpha & - \sigma \sin\alpha \end{pmatrix}.
\ees
We have
\bes
\Var\Big(\frac{\sum_{i=1}^n (X_i)_+\wedge (Y_i)_+}{\sum_{i=1}^nT_i}\Big|T_1,\ldots,T_n\Big)
\le \frac{C_0 \sum_{i=1}^n T_i^2}{(\sum_{i=1}^nT_i)^2}
\le \frac{C_0\max_{i\le n}T_i}{T_1+\cdots+T_n} = o_{\P}(1).
\ees
After applying this argument to $(X_i)_-\wedge (Y_i)_-$,
$(X_i)_+\vee (Y_i)_+$ and $(X_i)_-\vee (Y_i)_-$, the conclusion follows from
\bes
\frac{\E\big[\sum_{i=1}^n\big((X_i)_+\wedge (Y_i)_+ + (X_i)_-\wedge (Y_i)_-\big)\big|T_1,\ldots,T_n\big]}
{\E\big[\sum_{i=1}^n\big((X_i)_+\vee (Y_i)_+ + (X_i)_-\vee (Y_i)_-\big)\big|T_1,\ldots,T_n\big]}
= f_\infty(\rho,\sigma).
\ees

Now consider the bivariate $t$-distribution as an example:
\bes
(X,Y)^T \sim N(0,\Sigma)\sqrt{\nu/\chi^2_\nu}.
\ees
where $\chi^2_\nu$ is independent of $N(0,\Sigma)$.
Since $N(0,\Sigma)$ can be written as $AU\sqrt{\chi^2_2}$,
the bivariate $t$-distribution can be written as
\bes
(X,Y)^T \sim AUT\ \hbox{ with } T \sim \sqrt{\chi^2_2 \nu /\chi^2_\nu} \sim \sqrt{2F_{2,\nu}}
\ees
with two independent chi-square variables, where $F_{2,\nu}$ denotes the
$F$ distribution. It can be shown that 
\bel{ET}\notag
\E T 
= \frac{\sqrt{\nu}\,\Gamma(\nu/2-1/2)\Gamma(1/2)}{2\,\Gamma(\nu/2)},\quad \nu > 1.
\eel
For example, $\E T = \pi/\sqrt{2}$ for $\nu=2$.
For $\nu=1$, we still have (\ref{th-1}), as (\ref{cond}) follows from
\bes
\frac{t\,\P(T>t)}{\E\min(T,t)}
= \frac{t (1+t^2)^{-1/2}}{\int_0^t (1+x^2)^{-1/2}dx}
=  \frac{1+o(1)}{\log t} \to 0.
\ees

\section{Proof of Theorem~\ref{thm_normality}}

Let $T$ be independent of $(\xi,\zeta)$ and
$(T_i,\xi_i,\zeta_i)$ be iid copies of $(T,\xi,\zeta)$.
Assume that $\E T^2+ \E(\xi\E\zeta - \zeta\E\xi)^2<\infty$.  Then,
\bes
n^{1/2}\left(\frac{\sum_{i=1}^n T_i\xi_i}{\sum_{i=1}^n T_i\zeta_i} - \frac{\E\xi}{\E\zeta}\right)
= n^{1/2}\frac{\sum_{i=1}^n T_i(\xi_i\E\zeta - \zeta_i\E\xi)}{{\E\zeta\sum_{i=1}^n T_i\zeta_i}}
\toD N\left(0,\frac{ V\E T^2}{(\E T)^2(\E\zeta)^4}\right).
\ees
with
\bes
V = \E(\xi\E\zeta - \zeta\E\xi)^2.
\ees
Alternatively, if the condition $\E T^2<\infty$ is replaced by
\bel{cond-2}
\lim_{t\to\infty} \frac{t\,\P(T^2>t)}{\E\min(T^2,t)} = 0,
\eel
then,
\bes
&& \frac{\sum_{i=1}^n T_i}{(\sum_{i=1}^n T_i^2)^{1/2}}
\left(\frac{\sum_{i=1}^n T_i\xi_i}{\sum_{i=1}^n T_i\zeta_i} - \frac{\E\xi}{\E\zeta}\right)
\cr &=& \left(\frac{\sum_{i=1}^n T_i}{\E\zeta\sum_{i=1}^n T_i\zeta_i}\right)
\frac{\sum_{i=1}^n T_i(\xi_i\E\zeta - \zeta_i\E\xi)}{(\sum_{i=1}^n T_i^2)^{1/2}}
\cr &=& \left(1+o(1)\right)
\frac{\sum_{i=1}^n T_i(\xi_i\E\zeta - \zeta_i\E\xi)}{(\E\zeta)^2(\sum_{i=1}^n T_i^2)^{1/2}}
\cr &\toD& N\left(0,\frac{V}{(\E\zeta)^4}\right).
\ees

Suppose $(X,Y)$ is elliptical  and
\bes
\xi = \{(X)_+\wedge (Y)_+ + (X)_-\wedge (Y)_-\}/T,\quad
\zeta = \{(X)_+\vee (Y)_+ + (X)_-\vee (Y)_-\}/T.
\ees
As in the computation of $f_\infty$, we have
\bes
\E\xi &=& \frac{2}{2\pi} \int_{-\pi}^\pi (\cos(\theta - 2\alpha))_+\wedge (\sigma \cos\theta)_+d\theta
\cr &=& \frac{1}{\pi}\left\{\int_{-\pi/2+2\alpha}^{\tau} \cos(\theta - 2\alpha)d\theta
+\sigma \int_{\tau}^{\pi/2} \cos\theta d\theta\right\}
\cr &=& \frac{1}{\pi}\left\{1-\sin(2\alpha-\tau) +\sigma(1 - \sin\tau)\right\}
\cr &\overset{\sigma=1}{=}& \frac{2}{\pi}\left\{1-\sin\alpha \right\}
\ees
and
\bel{Ezeta}\notag
\E\zeta &=& \frac{2}{2\pi} \int_{-\pi}^\pi (\cos(\theta - 2\alpha))_+\vee (\sigma \cos\theta)_+d\theta
\cr &=& \frac{1}{\pi}\left\{\sigma\int_{-\pi/2}^{\tau} \cos \theta d\theta
+ \int_{\tau}^{\pi/2+2\alpha} \cos(\theta - 2\alpha)d\theta\right\}
\cr &=& \frac{1}{\pi}\left\{\sigma(1+\sin\tau) + 1+\sin(2\alpha-\tau)\right\}
\cr &\overset{\sigma=1}{=}& \frac{2}{\pi}\left\{1+\sin\alpha \right\}
\eel

Moreover

\bes
\E \xi^2 &=& \E\left[\{(X)_+\wedge (Y)_+ + (X)_-\wedge (Y)_-\}/T\right]^2
\cr &=& \frac{2}{2\pi} \int_{-\pi}^\pi \left\{(\cos(\theta - 2\alpha))_+\wedge (\sigma \cos\theta)_+\right\}^2d\theta
\cr &=& \frac{1}{\pi}\left\{\int_{-\pi/2+2\alpha}^{\tau} \cos^2(\theta - 2\alpha)d\theta
+\sigma^2 \int_{\tau}^{\pi/2} \cos^2\theta d\theta\right\}
\cr &=&\frac{1}{2\pi}\left\{\left.\left(\theta+\frac{1}{2}\sin(2\theta-4\alpha)\right)\right|_{-\pi/2+2\alpha}^{\tau}+\sigma^2\left.\left(\theta+\frac{1}{2}\sin(2\theta)\right)\right|_{\tau}^{\pi/2}
\right\}
\cr &=&\frac{1}{2\pi}\left\{\left(\tau+\frac{1}{2}\sin(2\tau-4\alpha)\right) - \left(-\pi/2+2\alpha\right)+\sigma^2\left(\pi/2-\tau-\frac{1}{2}\sin(2\tau)\right)\right\}
\cr &=&\frac{1}{2\pi}\left\{\tau+\pi/2-2\alpha+\frac{1}{2}\sin(2\tau-4\alpha)+\sigma^2\left(\pi/2-\tau-\frac{1}{2}\sin(2\tau)\right)\right\}
\cr &\overset{\sigma=1}{=}&\frac{1}{2\pi}\left(\pi-2\alpha-\sin2\alpha\right)
\ees

\bes
\E \zeta^2 &=& \E\left[\{(X)_+\vee (Y)_+ + (X)_-\vee (Y)_-\}/T\right]^2
\cr &=& \frac{2}{2\pi} \int_{-\pi}^\pi \left\{(\cos(\theta - 2\alpha))_+\vee (\sigma \cos\theta)_+\right\}^2d\theta
 +\frac{4}{2\pi} \int_{-\pi}^\pi \left\{(\cos(\theta - 2\alpha))_+\times (\sigma \cos\theta)_-\right\} d\theta
\cr &=& \frac{1}{\pi}\left\{\sigma^2\int_{-\pi/2}^{\tau} \cos^2 \theta d\theta
+ \int_{\tau}^{\pi/2+2\alpha} \cos^2(\theta - 2\alpha)d\theta\right\}
\cr && -\frac{\sigma}{\pi} \int_{\pi/2}^{\pi/2+2\alpha}\left\{\cos(2\alpha)+\cos(2\theta - 2\alpha)\right\} d\theta
\cr &=& \frac{1}{\pi}\left\{\sigma^2 \left.\left(\frac{\theta}{2}+\frac{1}{4}\sin(2\theta)\right)\right|_{-\pi/2}^{\tau}
+ \left.\left(\frac{\theta}{2}+\frac{1}{4} \sin(2\theta - 4\alpha)\right)\right|_{\tau}^{\pi/2+2\alpha}\right\}
\cr && -\frac{\sigma}{\pi} \left.\left(\cos(2\alpha)\theta+\frac{1}{2}\sin(2\theta - 2\alpha)\right)\right|_{\pi/2}^{\pi/2+2\alpha}
\cr &=& \frac{1}{\pi}\left\{\sigma^2 \left(\frac{\tau}{2}+\frac{1}{4}\sin(2\tau) + \frac{\pi}{4}\right)
+ \left(\frac{\pi}{4}+\alpha-\frac{\tau}{2}-\frac{1}{4} \sin(2\tau - 4\alpha)\right)\right\} +\frac{\sigma}{\pi} \left(\sin2\alpha-2\alpha\cos2\alpha\right)
\cr &\overset{\sigma=1}{=}&\frac{1}{\pi}\left(\frac{\pi}{2}+\alpha+\frac{3}{2}\sin2\alpha-2\alpha\cos2\alpha\right)
\ees

and
\bes
\E(\xi\zeta) &=& \E\left[\{(X)_+\wedge (Y)_+ + (X)_-\wedge (Y)_-\}
\{(X)_+\vee (Y)_+ + (X)_-\vee (Y)_-\}/T^2\right]
\cr &=& \frac{2}{2\pi}\int_{-\pi}^\pi \left\{(\cos(\theta - 2\alpha))_+\times (\sigma \cos\theta)_+\right\} d\theta
\cr &=& \frac{\sigma}{2\pi} \int_{-\pi/2+2\alpha}^{\pi/2}\left\{\cos(2\alpha)+\cos(2\theta - 2\alpha)\right\} d\theta
\cr &=& \frac{\sigma}{2\pi}\left\{\theta \cos(2\alpha)+2^{-1}\sin(2\theta)\right\}\Big|_{-\pi/2+\alpha}^{\pi/2-\alpha}
\cr &=& \frac{\sigma}{2\pi} \left((\pi-2\alpha)\cos2\alpha + \sin2\alpha\right)
\cr &\overset{\sigma=1}{=}& \frac{1}{2\pi} \left((\pi-2\alpha)\cos2\alpha + \sin2\alpha\right)
\ees
Consequently,
\begin{align}\notag
V =& \E\xi^2(\E\zeta)^2 + \E\zeta^2(\E\xi)^2 - 2\E\xi\E\zeta \E(\xi\zeta)\\\notag
=&\frac{1}{4\pi^3}\left\{2\tau+\pi-4\alpha+\sin(2\tau-4\alpha)+\sigma^2\left(\pi-2\tau-\sin(2\tau)\right)\right\}\left\{\sigma(1+\sin\tau) + 1+\sin(2\alpha-\tau)\right\}^2\\\notag
+&\frac{1}{4\pi^3}\left\{\sigma^2 \left(2{\tau}+\sin(2\tau) + {\pi}\right)
+ \left({\pi}+4\alpha-2{\tau}- \sin(2\tau - 4\alpha)\right) +4{\sigma} \left(\sin2\alpha-2\alpha\cos2\alpha\right)\right\}\\\notag
&\hspace{0.2in}\times\left\{1-\sin(2\alpha-\tau) +\sigma(1 - \sin\tau)\right\}^2\\\notag
-&\frac{\sigma}{\pi^3} \left((\pi-2\alpha)\cos2\alpha + \sin2\alpha\right)\left\{1-\sin(2\alpha-\tau) +\sigma(1 - \sin\tau)\right\}\left\{\sigma(1+\sin\tau) + 1+\sin(2\alpha-\tau)\right\}
\end{align}
The expression can be simplified when $\sigma=1$:
\bes
\E\xi &\overset{\sigma=1}{=}& \frac{2}{\pi}\left\{1-\sin\alpha \right\}\\
\E\zeta &\overset{\sigma=1}{=}& \frac{2}{\pi}\left\{1+\sin\alpha \right\}\\
\E \xi^2 &\overset{\sigma=1}{=}&\frac{1}{2\pi}\left(\pi-2\alpha-\sin2\alpha\right)\\
\E \zeta^2  &\overset{\sigma=1}{=}&\frac{1}{\pi}\left(\frac{\pi}{2}+\alpha+\frac{3}{2}\sin2\alpha-2\alpha\cos2\alpha\right)\\
\E(\xi\zeta) &\overset{\sigma=1}{=}& \frac{1}{2\pi} \left((\pi-2\alpha)\cos2\alpha + \sin2\alpha\right)
\ees
Thus, when  $\sigma=1$, we have
\begin{align}\notag
V =& \E\xi^2(\E\zeta)^2 + \E\zeta^2(\E\xi)^2 - 2\E\xi\E\zeta \E(\xi\zeta)\\\notag
=&\frac{1}{2\pi}\left(\pi-2\alpha-\sin2\alpha\right)\left[\frac{2}{\pi}\left(1+\sin\alpha \right)\right]^2
+\frac{1}{\pi}\left(\frac{\pi}{2}+\alpha+\frac{3}{2}\sin2\alpha-2\alpha\cos2\alpha\right)\left[\frac{2}{\pi}\left(1-\sin\alpha \right)\right]^2\\\notag
&-2\frac{1}{2\pi} \left((\pi-2\alpha)\cos2\alpha + \sin2\alpha\right)\frac{2}{\pi}\left(1-\sin\alpha \right) \frac{2}{\pi}\left(1+\sin\alpha \right)\\\notag
=&\frac{4}{\pi^3}\sin^2\alpha\left(3\pi-8\cos\alpha+2\sin2\alpha+\pi\cos2\alpha
-8\alpha\sin\alpha-4\alpha\cos2\alpha\right)
\end{align}

For a bivariate $t$-distribution with $\nu$ degrees of freedom, we have $T \sim \sqrt{\chi^2_2 \nu /\chi^2_\nu} \sim \sqrt{2F_{2,\nu}}$,  
and
\bes
\E T^2  = 2\E \left\{F_{2,\nu}\right\} = \frac{2\nu}{\nu-2},\hspace{0.5in} 
\E T =\frac{\sqrt{\pi}}{2}\frac{\Gamma(\nu/2-1/2)\sqrt{\nu}}{\Gamma(\nu/2)}
\ees

Thus, when $\E T^2<\infty$, we have the asymptotic normality
\bel{normality-1}\notag
n^{1/2}\left(\frac{\sum_{i=1}^n T_i\xi_i}{\sum_{i=1}^n T_i\zeta_i} - f_\infty(\rho,\sigma)\right)
\toD N\left(0,\frac{ V\E T^2}{(\E T)^2(\E\zeta)^4}\right).
\eel

For $t$-distribution with $\nu=2$, condition (\ref{cond-2}) holds as
\bes
\frac{t\,\P(T^2>t)}{\E\min(T^2,t)} = \frac{t (1+t/2)^{-1}}{\int_0^t(1+x/2)^{-1}dx}
\asymp \frac{1}{\log t} \to 0.
\ees
Moreover, $\P(\max_{i\le n}T_i^2>n/\eps) = O(\eps)$, $\E(T^2\wedge(n/\eps))\approx 2\log n$
and $\E(T^2\wedge(n/\eps))^2=O(n)$, so that
\bes
\frac{\sum_{i=1}^n T_i^2}{2n\log n} = 1 + O_P(1/\log n).
\ees
Thus, for $\nu =2$,
\bel{normality-2}\notag
&& \left(\frac{n}{\log n}\right)^{1/2}
\left(\frac{\sum_{i=1}^n T_i\xi_i}{\sum_{i=1}^n T_i\zeta_i} - f_\infty(\rho,\sigma)\right)
\toD N\left(0,\frac{4V}{\pi^2(\E\zeta)^4}\right).
\eel

\section{Proof of Theorem~\ref{thm_cosine}}

\cite{Book:Anderson03} provides the result for the normal case. We extend  the results of~\cite{Book:Anderson03} to the general elliptical family. Again,  a vector $(X,Y)$ has an elliptical distribution if
\bel{elliptical}\notag
(X,Y) = T(\xi,\zeta),\ (\xi,\zeta)^T =  AU  = {a_1^TU \choose a_2^TU}
\eel
where $A = (a_1,a_2)^T$ is a deterministic $2\times 2$ matrix,
$U$ is a vector uniformly distribution in the unit circle and $T$ is a positive random variable
independent of $U$. We want to compute the asymptotic variance
of the sample correlation
\bes
\hrho_n = \frac{\sum_{i=1}^n X_iY_i}
{\sqrt{\sum_{i=1}^n\sum_{j=1}^n  X_i^2 Y_j^2}}
\ees
Due to scale invariance, it suffices to consider the case of $\E X^2  = \E Y^2 =1$.

\bes
\hrho_n - \rho
 &=& \frac{\sum_{i=1}^n X_iY_i/n - \rho}{\sqrt{\sum_{i=1}^n\sum_{j=1}^n  X_i^2Y_j^2/n^2}}
+ \rho \frac{1 - \sqrt{\sum_{i=1}^n\sum_{j=1}^n  X_i^2Y_j^2/n^2}}
{\sqrt{\sum_{i=1}^n\sum_{j=1}^n  X_i^2Y_j^2/n^2}}
\cr &=& \sum_{i=1}^n \frac{X_iY_i}{n}  - \rho
+ \rho \frac{1 - \sum_{i=1}^n\sum_{j=1}^n  X_i^2Y_j^2/n^2}{2} + O_P(1/n)
\cr &=& \sum_{i=1}^n \frac{X_iY_i}{n} - \rho + \frac{\rho}{2}\Big(1- \sum_{i=1}^n\frac{X_i^2}{n}\Big)
+ \frac{\rho}{2}\Big(1- \sum_{i=1}^n\frac{Y_i^2}{n}\Big) + O_P(1/n).
\ees
Thus, the asymptotic variance of $\hrho$ is
\bes
V &=& \E\Big(XY - \rho - (\rho/2)(X^2 + Y^2 -2)\Big)^2
\cr &=& \E\Big(XY - (\rho/2)(X^2 + Y^2)\Big)^2
\cr &=& \E\Big(T^2\{\xi\zeta - (\rho/2)(\xi^2 + \zeta^2)\}\Big)^2.
\cr &=& \E T^4 \ \E \Big(\xi\zeta - (\rho/2)(\xi^2 + \zeta^2)\Big)^2.
\ees
Let $\E_0$ be the expectation in the Gaussian case. We have
$T^2 \sim \chi^2_2$ under $\E_0$, $\E_0 T^2 = 2$, $\E_0 T^4 = \Var_0(T^2)+(\E T_0^2)^2 = 4 + 4 = 8$,
and $V_0 = (1-\rho^2)^2$. A comparison with the solution in the Gaussian case yields
\bes
V &=& \frac{\E T^4 (\E_0T^2)^2}{(\E T^2)^2 \E_0 T^4}
\bigg\{\frac{\E_0 T^4}{(\E_0 T^2)^2}\E \Big(\xi\zeta - (\rho/2)(\xi^2 + \zeta^2)\Big)^2\bigg\}
\cr &=& \frac{ 4 \E T^4}{ 8 (\E T^2)^2}
\bigg\{\frac{\E_0 T^4}{(\E_0 T^2)^2}\E_0 \Big(\xi\zeta - (\rho/2)(\xi^2 + \zeta^2)\Big)^2\bigg\}
\cr &=& \frac{\E T^4}{2(\E T^2)^2}\E_0\Big(XY - \rho - (\rho/2)(X^2 + Y^2 -2)\Big)^2\bigg|_{\E_0 X^2 =\E_0 Y^2=1}
\cr &=& \frac{\E T^4}{2(\E T^2)^2}(1-\rho^2)^2.
\ees

\end{document}